\UseRawInputEncoding
\documentclass[twocolumn,aps,preprint,10pt,superscriptaddress,floatfix,showpacs]{revtex4-1}
\usepackage{graphicx}
\usepackage{epsfig,rotate,latexsym}
\usepackage{color, bm}
\usepackage{graphics}
\usepackage{dcolumn}
\usepackage{bm}
\usepackage{epsfig}
\usepackage{color}
\usepackage{epic}
\usepackage{amsmath}
\usepackage{amsfonts}
\usepackage{hyperref}
\usepackage{float}

\begin{document}
\title{Coexisting magnetic structures and spin-reorientation  in Er$_{0.5}$Dy$_{0.5}$FeO$_{3}$: Bulk magnetization,  neutron scattering, specific heat, and \emph{Ab-initio} studies}
\author{Sarita Rajput}
\affiliation{Department of Physics, Indian Institute of Technology Roorkee, Roorkee 247 667, India
}
\author{Padmanabhan Balasubramanian}
\affiliation{Department of Physics, Graphic Era University, Dehra Dun, Uttarakhand 248 002, India
}
\author{Ankita Singh}
\affiliation{Department of Physics, Indian Institute of Technology Roorkee, Roorkee 247 667, India
}
\author{Francoise Damay}
\affiliation{Laboratoire L\'{e}on Brillouin, CEA-CNRS, CEA/Saclay, 91191 Gif sur Yvette, France
}
\author{C.M.N. Kumar}
\affiliation{Institute of Solid State Physics, Vienna University of Technology, Wiedner Hauptstrasse 8-10, 1040 Wien, Austria
}
\affiliation{AGH University of Science and Technology, Faculty of Physics and Applied Computer Science, 30-059 Krak{\'{o}}w, Poland}
\author{W. Tabis}
\affiliation{AGH University of Science and Technology, Faculty of Physics and Applied Computer Science, 30-059 Krak{\'{o}}w, Poland}
\author{T. Maitra}
\affiliation{Department of Physics, Indian Institute of Technology Roorkee, Roorkee 247 667, India
}
\author{V. K. Malik}
\email{vivek.malik@ph.iitr.ac.in}
\affiliation{Department of Physics, Indian Institute of Technology Roorkee, Roorkee 247 667, India
}
\date{\today}

\begin{abstract}
The complex magnetic structures, spin-reorientation and associated exchange interactions have been investigated in Er$_{0.5}$Dy$_{0.5}$FeO$_3$ using bulk magnetization, neutron diffraction, specific heat measurements and density functional theory calculations.
The Fe$^{3+}$ spins order as G-type antiferromagnet structure depicted by ${\Gamma}_{4}$($G_{x}$,\,$A_{y}$,\,$F_{z}$)  irreducible representation below 700\,K, similar to its end compounds. The  bulk magnetization data indicate occurrence of the spin-reorientation and rare-earth magnetic ordering below $\sim$75\,K and 10\,K, respectively. The neutron diffraction studies confirm an ``incomplete" ${\Gamma}_{4}$${\rightarrow}$ ${\Gamma}_{2}$($F_{x}$,\,$C_{y}$,\,$G_{z}$) spin-reorientation initiated $\leq$75\,K. Although, the relative volume fraction of the two magnetic structures varies with decreasing temperature, both co-exist even at 1.5\,K. 
Below 10\,K, magnetic ordering of Er$^{3+}$/Dy$^{3+}$ moments in $c_{y}^R$ arrangement develops, which gradually increases in intensity with decreasing temperature. At 2\,K,  magnetic structure associated with $c_{z}^R$ arrangement of Er$^{3+}$/Dy$^{3+}$ moments also appears.
 At 1.5\,K the magnetic structure of Fe$^{3+}$ spins is represented by a combination of ${\Gamma}_{2}$+${\Gamma}_{4}$+${\Gamma}_{1}$, while the rare earth magnetic moments' order coexists as $c_{y}^R$ and $c_{z}^R$ corresponding to ${\Gamma}_{2}$ and ${\Gamma}_{1}$ representation, respectively. 
The observed Schottky anomaly at 2.5\,K suggests that the ``rare-earth ordering" is induced by polarization due to Fe$^{3+}$ spins. 
 The Er$^{3+}$-Fe$^{3+}$ and Er$^{3+}$-Dy$^{3+}$ exchange interactions, obtained from first principle calculations, indicate that these interactions primarily cause the complicated spin-reorientation and $c_{y}^R$ rare-earth ordering in the system, respectively, while the dipolar interactions between rare-earth moments, result in the $c_{z}^R$ type rare-earth ordering at 2\,K.
\end{abstract}

\pacs{75.25.-j, 75.47.Lx, 75.30.Gw, 71.20.-b} 
\maketitle

\section{Introduction}
\label{intro}
 Rare-earth orthoferrites materials $R$FeO$_{3}$ ($R$$=$ rare-earth ion), have shown potential for technologically relevant applications via observation of spin switching, spontaneous exchange bias and optically controlled ultrafast spin dynamics\cite{Guo2020,Mikhaylovskiy2015, Mikhaylovskiy2015b,Mikhaylovskiy2014,Yamaguchi2013,Ding2019,Kimel2004}. Additional important properties include large linear magneto-dielectric effect, spontaneous ferroelectric polarization, mutilferroicity, and magneto-caloric effect\cite{Deng2015,Du2010,Tokunaga2008,Yokota2015,Ankita2019,Ke2016}. \par
 The orthoferrites materials belong to the family of perovskites and crystallize in the structure represented by orthorhombic space group $D_{16}^{2h}$:$Pbnm$ symmetry, as observed in case of manganite and cobaltate materials\cite{Chakraborty2016, Karel2009}. Unlike manganites, long-range static Jahn-Teller effect is absent in orthoferrites \cite{Chakraborty2016}. The FeO$_{6}$ octahedra in rare-earth orthoferrites possess a GdFeO$_{3}$ type distortion with nearly equal Fe-O bond lengths\cite{Slawinski2005,Ross2004,Glazer1972,Woodward1997a,Woodward1997b}. The structural distortion increases with atomic number of $R$\cite{Marezio70}.\par
The rare-earth orthoferrites ($R$FeO$_{3}$) are anti-ferromagnets with a N\'eel temperature($T_\mathrm{N1}$)  in the temperature range of 650-760\,K\cite{KOEHLER1960,White1969}. Increase in the atomic number of $R$ results in systematic reduction of $T_\mathrm{N1}$\cite{White1969}.
At $T_\mathrm{N1}$, the orthoferrites undergo transition from paramagnetic to G-type anti-ferromagnetic state with ordering wave-vector $\vec {k}$=(0,\,0,\,0). The magnetic structure belongs to ${\Gamma}_{4}$ irreducible representation, which  can be written as ($G_{x}$,\,$A_{y}$,\,$F_{z}$) in Bertaut notation\cite{bertaut1963magnetism}. The Fe$^{3+}$ spins order primarily in antiferromagnetic G-type configuration along the crystallographic $a$-axis($G_{x}$), while $A_{y}$ and $F_{z}$ correspond to the  A-type antiferromagnetic and ferromagnetic arrangement of the spins along the crystallographic $b$ and $c$-axes due to covert and overt canting, respectively\cite{White1969,Yamaguchi1974,bartolome1997single}.\par The dominant interaction in all the orthoferrite compounds is an isotropic Fe$^{3+}$-Fe$^{3+}$ super-exchange interaction which is denoted as $J_\mathrm{Fe-Fe}$. The additional isotropic interactions are the $R^{3+}-$Fe$^{3+}$ and $R^{3+}-R^{3+}$ interactions\cite{Yamaguchi1974}.
The strength of the isotropic interactions are in the following order, $J_\mathrm{Fe-Fe}$\,$>$\,$J_{R-\mathrm{Fe}}$\,$>$\,$J_{R-R}$. 
The anti-symmetric Dzhyloshinski-Moriya interaction is responsible for the covert and overt magnetic orderings ($A_{y}$,\,$F_{z}$) due to small canting of the Fe$^{3+}$ spins. Yamaguchi {\it et al.}\cite{Yamaguchi1974} also discussed the role of anisotropic parts of $R^{3+}$-Fe$^{3+}$ exchnage interactions, which are responsible for the spin reorientation in orthoferrites.
\par
NdFeO$_{3}$ is one of the most studied orthoferrite compound for magnetic ordering and spin reorientation process\cite{KOEHLER1960,White1969,Pinto1972,Yamaguchi1974,SOSNOWSK1986,Przeniosto1996,bartolome1997single,Slawinski2005,Yuan2011,SONG2015,Jiang2016} in which the Fe$^{3+}$ spins undergo a continuous ${\Gamma}_{4}$${\rightarrow}$${\Gamma}_{2}$($F_{x}$,\,$C_{y}$,\,$G_{z}$) reorientation between 200 and 150\,K \cite{SOSNOWSK1986,Yuan2011, Chen2012}. In the ${\Gamma}_{2}$ magnetic structure, the crystallographic $c$-axis is the easy axis of the ordered Fe$^{3+}$ spins in basic G-type antiferromagnetic configuration along with ferromagnetic $F_{x}$ and antiferromagnetic $C_{y}$ configurations due to overt and covert canting of the spins\cite{bertaut1963magnetism,Yamaguchi1974}. The Nd$^{3+}-$Fe$^{3+}$ interactions leads to polarization of the Nd$^{3+}$ magnetic moments, which result in their long range ordering at liquid He temperatures\cite{bartolome1997single, Przeniosto1996,PRZENIOSLO1995}. In most of the orthoferrites (except DyFeO$_{3}$), the $R^{3+}$-moments usually order in a structure which is symmetry-compatibile with the G-type arrangement of Fe$^{3+}$ spins\cite{Yamaguchi1974}. For instance, in NdFeO$_{3}$, the Nd$^{3+}$ moments order  as $c_{y}^R$, compatible to the ${\Gamma}_{2}$ representation of Fe$^{3+}$ spins\cite{bartolome1997single,Przeniosto1996,PRZENIOSLO1995}. At the lowest temperatures (usually below 2\,K) the Nd$^{3+}-$Nd$^{3+}$ interactions begin to supercede the Nd$^{3+}-$Fe$^{3+}$ interactions, resulting in long-ranged independent ordering of the Nd$^{3+}$ moments, which can be considered as Landau-type second order phase transition \cite{Bartolome1994}.
\par
Another isostructural,  but relatively less studied, orthoferrite, ErFeO$_{3}$,  shows magnetic properties similar to NdFeO$_{3}$. Fe$^{3+}$ spins in ErFeO$_{3}$ order as canted G-type antiferromagnet in ${\Gamma}_{4}$ representation below 620\,K ($T_\mathrm{N1}$)\cite{KOEHLER1960}. Between $\sim$100/110\,K and  $\sim$80/90\,K, the Fe$^{3+}$ spins undergo a ${\Gamma}_{4}{\rightarrow}{\Gamma}_{2}$ type gradual spin reorientation \cite{Bozorth58,Grant1969, Bazaliy2004,Pinto1971,Tsymbal2007}.
The isotropic and anisotropic Er$^{3+}$-Fe$^{3+}$ exchange interactions result in polarization of the Er$^{3+}$ moments which causes the spin reorientation of Fe$^{3+}$ spins.
\par
 In the low temperature phase (below 20 \,K), the magnetic arrangement of Fe$^{3+}$ spins is more complex due to the influence of Er$^{3+}$ ordering on Fe$^{3+}$ magnetic structure. Koehler {\it et al.}\cite{KOEHLER1960} proposed that the magnetic ordering of Fe$^{3+}$ spins, below rare-earth ordering temperature, could be given by a G-type antiferromagnetic structure with Fe$^{3+}$ spins confined in $a$-$b$ plane. At 1.3\,K,  Er$^{3+}$ moments ordered in a magnetic structure given by $C_{z}$ configuration\cite{KOEHLER1960}. 
Based on neutron diffraction data, Gorodetsky {\it et al.}\cite{Gorodetsky1973} proposed the possibility of two the Fe$^{3+}$ magnetic structure below the  Er$^{3+}$ ordering temperature. First magnetic structure was given by mixed representation of $G_{xy}$ with Fe$^{3+}$ spins aligned in $ab$ plane at an angle of 33$^{\circ}\pm$4$^{\circ}$ from the $b$ axis. As per the second proposed magnetic structure,  Fe$^{3+}$ spins should order as G -type structure in $bc$ plane ($G_{yz}$) with the spins at an angle of 51$^{\circ}\pm$8$^{\circ}$ from the $b$ axis. Based on magnetization and torque measurements of the same study\cite{Gorodetsky1973},  Gorodetsky {\it et al.} confirmed the existence of $G_{yz}$ magnetic structure for Fe$^{3+}$ spins at $T<4.5$\,K.  The low temperature magnetic structure of Fe$^{3+}$ moments remained unresolved for long time.  Recently, Deng {\it et al.}\cite{Deng2015} confirmed with the help of neutron diffraction and symmetry analysis that the low temperature magnetic structure of Fe$^{3+}$ remains ${\Gamma}_{2}$ below spin reorientation temperature. Additionally Deng {\it et al.} also detected $C_y$ ($C_x$ in $Pnma$ space group)  mode of  ${\Gamma}_{2}$($F_x,C_y,G_z$) magnetic structure below 5\,K, at which, the Er$^{3+}$ moments start to arrange in the long range antiferromagnetic ordering given by $c_{z}^R$($c_{y}^R$ in $Pnma$ space group) configuration\cite{Gorodetsky1973, Deng2015}. The concurrence  of centero-symmetric space group of Fe$^{3+}$ sublattice along with centero-asymmetric space group of Er$^{3+}$  opens the possibility of improper polarization along with magnetic ordering and hence multiferroicity in ErFeO$_3$\cite{Deng2015}. Yokota {\it et al.} \cite{Yokota2015} observed the coexistence of ferroelectricity and magnetism in ErFeO$_3$ thin films by  modification of structure from orthorhombic to hexagonal via  yttria stabilized zirconia substrate. \par
 In the family of orthoferrites, the DyFeO$_{3}$ exhibits an exceptional trend in spin-reorientation and rare-earth ordering\cite{Bozorth58,Gorodetsky1968, Prelorendjo1980, Zhao2014, Wang2016,Tokunaga2008}. Below $T_\mathrm{N1}$(${\sim}$650\,K), the Fe$^{3+}$ moments order with a structure given by  ${\Gamma}_{4}$ representation. However, near ${\sim}$35\,K, a ${\Gamma}_{4}$${\rightarrow}$${\Gamma}_{1}$($A_{x}$,\,$G_{y}$,\,$C_{z}$)  type spin reorientation occurs where  antiferromagnetic axis ($b$ axis) becomes parallel to Dzyaloshinsky-Moriya vector and weak ferromagnetic component cease to exist due to a pure uncanted antiferromagnetic structure\cite{Gorodetsky1968, Prelorendjo1980}. This transition, also known as Morin transition, is unique for DyFeO$_3$ in the family of rare-earth orthoferrites\cite{Morin50,Bozorth58,Gorodetsky1968,White1969,Yamaguchi1973}.
 \par
  Below $T_\mathrm{N2}$ ${\sim}$ 4\,K, due to the Dy$^{3+}$-Dy$^{3+}$ exchange and dipole interactions, the Dy$^{3+}$ moments arrange in the ${\Gamma}_{5}$($G_{x}^{R}$,$A_{y}^{R}$) configuration making an angle of 30$^{o}$ with the $b$ axis \cite{Berton1968,NOWIK1966,Belov1968,Gorodetsky1968,White1969}.  Iso-structural Dy-based compounds viz. DyAlO$_{3}$ \cite {Holmes1972} and DyCrO$_{3}$ \cite{ Krynetskii1997}, also arrange in the same configuration. Due to large single ion anisotropy, the Dy$^{3+}$ moments are confined to the $a$-$b$ plane\cite{Wu2017}.
  \par
    Prelorendjo {\it et al.}\cite{Prelorendjo1980} also studied magnetic field induced spin reorientation.  At 4.2\,K, ${\Gamma}_{1}\rightarrow{\Gamma}_{4}$, ${\Gamma}_{1}\rightarrow{\Gamma}_{2}$, and ${\Gamma}_{1}\rightarrow{\Gamma}_{4}$ type spin reorientations of the Fe$^{3+}$ sublattice were observed in DyFeO$_3$ on application of external magnetic field along $b$, $a$, and $c$-axis of the crystal, respectively\cite{Prelorendjo1980}. The ($G_{x}^{R}$,\,$A_{y}^{R}$) arrangement of Dy$^{3+}$ moments, is not symmetry compatible with the field induced ${\Gamma}_{2}$ magnetic structure of the Fe$^{3+}$ sub lattice\cite{Yamaguchi1973}. Co-existence of two incompatible magnetic structures breaks the inversion symmetry which is essential for linear magneto-electric effect\cite{Yamaguchi1973, Zvezdin2009}. Experimentally, a linear magneto-electric tensor component with a value as large as 2.4x10$^{-2}$ esu is observed below the Dy$^{3+}$ antiferromagnetic ordering temperature\cite{Tokunaga2008}. Interestingly, large ferroelectric polarization along with magnetic ordering also is achieved by application of an external magnetic field along the $c$ axis of DyFeO$_{3}$ \cite{Tokunaga2008,Rajeswaran2013, Zhao2014}. Rajeswaran {\it et al.}\cite{Rajeswaran2013} claimed to observe simultaneous ferroelectricity and weak ferromagnetism above Morin transition temperature in polycrystalline DyFeO$_3$. Recent study by Wang {\it et al.}\cite{Wang2016} observed a long- to short-range ordering transition of Dy$^{3}$ concurrent to the  magnetic field (along the $c$-axis) induced  spin reorientation of  the Fe$^{3+}$ sublattice. The magnetic field induced short range ordering of the  Dy$^{3+}$ moments  is responsible for the observed  multiferroic phase induced by external magnetic field \cite{Wang2016,Tokunaga2008}
\par
In addition to the pure orthoferrites, doping and/or substitution at the Fe and/or $R$ sites shows interesting variations in the structural, magnetic, and electronic properties, while the fundamental characteristic of orthoferrites is still retained\cite{Ankita2017,Ankita2019,Tokunaga2014,MIHALIK2013,CHAKRABORTY2018,Wu2014,Lazurova2015}. For instance, doping at the Fe-site with Mn has been studied and resulted in a systematic decrease of the N\'eel temperature\cite{Ankita2017,Lazurova2015}, structural distortion due to Jahn-Teller effect\cite{Tirtha2016}, and modifies the preferred direction of  Fe$^{3+}$ spins due to single ion anisotropy of Mn$^{3+}$ ions\cite{Ankita2017, Harikrishnan2016}. Substitution at A-site by another rare-earth though does not affect the structural behaviour and N\'eel temperature, but however affects the spin-reorientation and the rare-earth ordering{\color{red}\cite{}}.
For instance, two-fold spin-reorientations are observed in single crystals of Dy$_{0.5}$Pr$_{0.5}$FeO$_{3}$ and Ho$_{0.5}$Dy$_{0.5}$FeO$_{3}$ with varying temperature and magnetic field\cite{Wu2014, Tirtha2018}. However, in these cases, the nature of the rare-earth ordering and its effect on the spin reorientation are not known. 
\par
Considering the striking contrast in the nature of Fe$^{3+}$  and $R$$^{3+}$-orderings in ErFeO$_{3}$ and DyFeO$_{3}$, it would be interesting to explore the properties of Er$_{0.5}$Dy$_{0.5}$FeO$_{3}$(EDFO). 
A very complex interplay of the various exchange interactions between the Er$^{3+}$/Dy$^{3+}$ and Fe$^{3+}$ sub lattices along with the $R^{3+}$$-$$R^{3+}$ exchange is expected. Additionally, due to the large magnetic moments of both the rare-earths, the classical dipole interactions are expected to play a prominent role in determining the complex ground state magnetic orders at the lowest temperatures. 
 Due to large differences in the nature of single ion anisotropy of both the rare-earth ions (Er$^{3+}$ and Dy$^{3+}$) which can compete with the exchange and dipolar interactions, it would also be interesting to establish, whether the rare-earth ordering in the system is long-ranged, or magnetic ground state turns into a spin-glass. Additionally, the possibilty of a magneto-electric effect is also worth exploring.
\par
Thus, we have experimentally studied the bulk magnetization, heat capacity, neutron diffraction and magneto-dielectric measurements of polycrystalline EDFO. Theoretically, density functional theory calculations are performed to understand the ground state electronic structure and evaluate the various exchange interactions from the total energies of various possible magnetic configurations.

\begin{figure}[h!] \center
       \begin{picture}(240,220)
        \put(-5,-5){\includegraphics[height=220\unitlength,width=250\unitlength,]{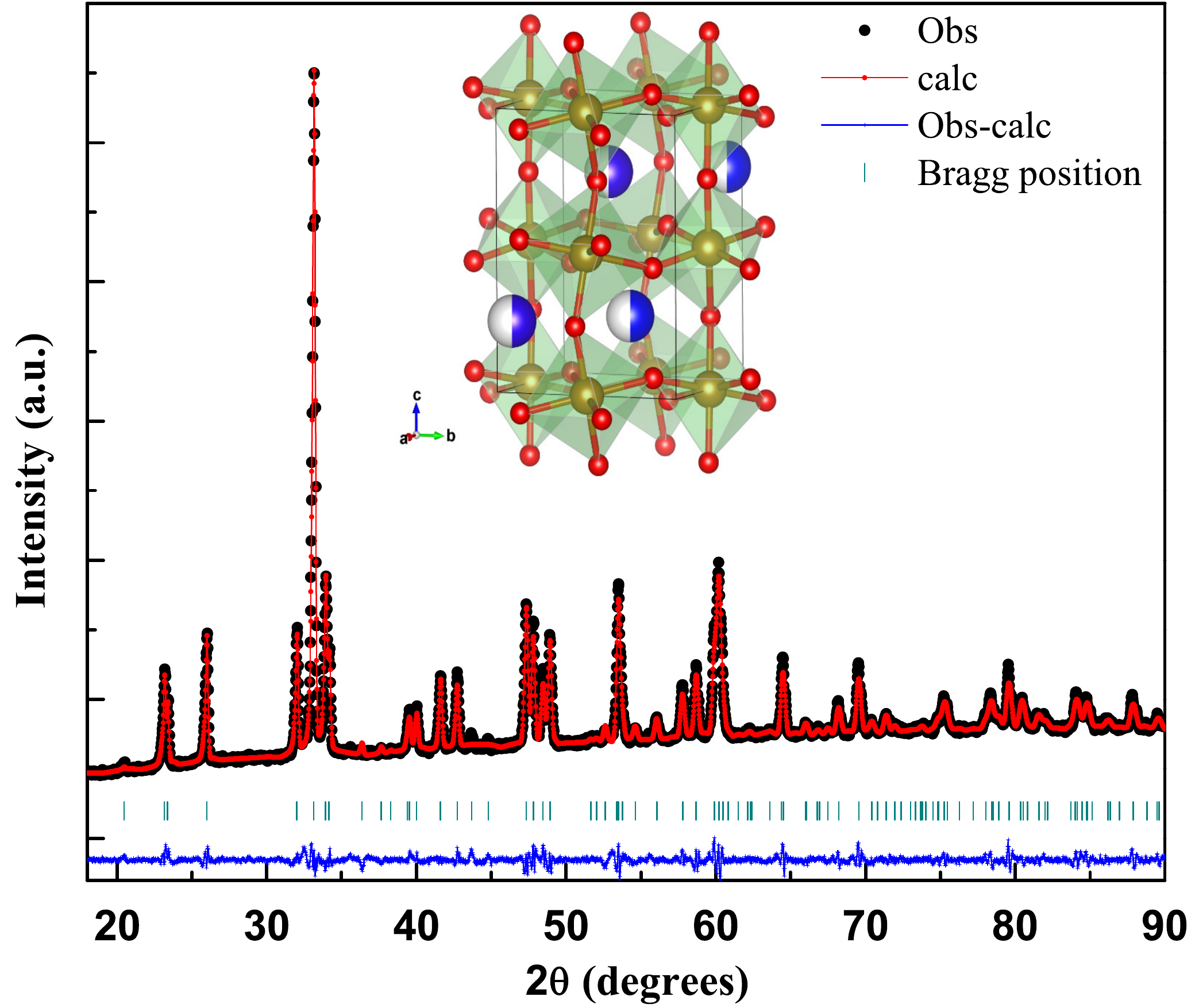}}
      \end{picture}
\caption{(color online) Observed and reifned x-ray diffraction pattern of EDFO at room temperature. We also show the unit cell of EDFO, wherein the Er and Dy atoms occupy same crystallographic site.The Er/Dy, Fe and O atoms are represented by Silver/blue, green and red spheres, respectively, in the unit cell.}
\label{xrd}
\end{figure}
\section{Methods}
\label{methd}
\subsection{Experimental}
\label{Expmethd}
Polycrystalline samples of EDFO were synthesized using solid state reaction method. Er$_{2}$O$_{3}$, Dy$_{2}$O$_{3}$, and Fe$_2$O$_3$ precursor powders were weighed in appropriate stoichiometry and ground in an agate mortar for 12 hours. The steps involved calcination at 1200\,$^{\circ}$C for 24 hrs, followed by heating at 1350\,$^{\circ}$C for 24 hrs with intermediate grinding.
Crystal structure of the sample was identified using a Bruker D8 two circle powder x-ray diffractometer with Cu $K_{\alpha}$ source. 
Bulk magnetization measurements were performed using SQUID magnetometer of Quantum Design Inc.'s Magnetic Properties Measurement System-XL (QD-MPMS-XL) and vibrating sample magnetometer option of Quantum Design Inc.'s Dynacool Physical Properties Measurement System (QD-PPMS). 
 Zero field cooled (ZFC) and field cooled (FC) magnetization measurements were carried out from 300\,K to 1.5\,K in the presence of 0.01 and 0.1\,T magnetic field. $M$-$H$ isotherms were measured at various temperatures between 300 and 1.5\,K. Using custom designed probe for QD-PPMS and Hioki EIM3536 LCR meter, magneto-dielectric studies were carried out in the temperature range of 300-2\,K, at frequencies ranging from 1\,kHz to 500\,kHz in the external magnetic field of 0, 0.1 and 1\,Tesla. Heat capacity measurements were performed using the QD-PPMS with $^{3}$He option in the temperature range 20-0.4\,K and magnetic field values of 0, 2 and 5\,T.
Powder Neutron diffraction studies in zero magnetic field were carried out at various temperatures between 300-1.5\,K to identify the crystal as well as magnetic structure and their evolution as a function of temperature. The neutron diffraction measurements were performed at powder diffractometer G-41-I 
(${\lambda}$ = 2.4206 ${\mathrm{\AA}}$), at LLB, Saclay in France.  The Rietveld analysis of the diffraction data was performed using FullProf suite of programs\cite{rietveld1969profile,rodriguez1990fullprof}. Magnetic structures were determined using the irreducible representations from BasIreps \cite{Hovestreydt92}. 
\begin{figure*}[tbh]
\begin{minipage}{1.0\linewidth}
\centering
\includegraphics[width=16.5cm]{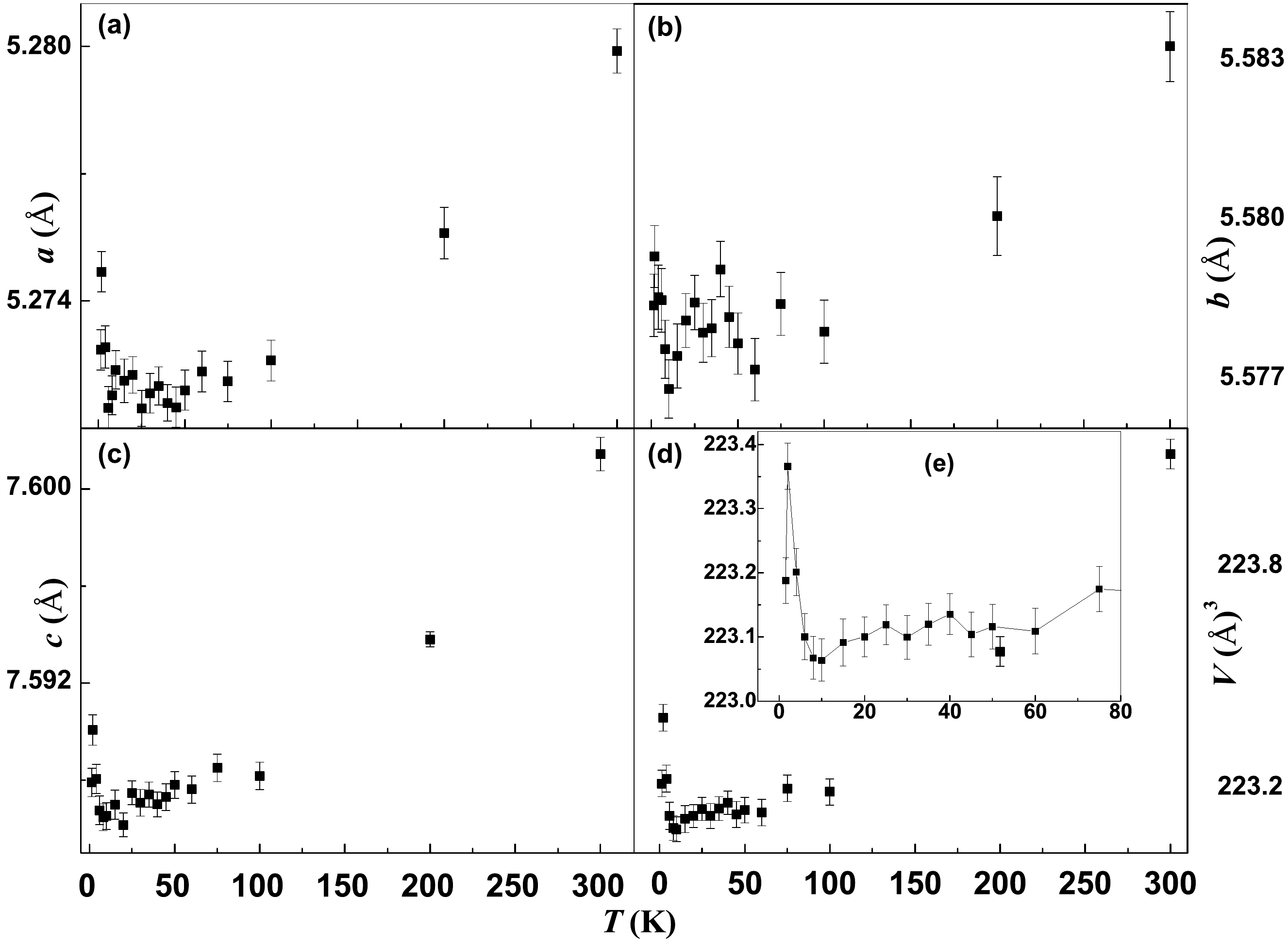}
\end{minipage}
\caption{(a)-(c):Temperature variation of lattice parameters, (d) unit cell volume of EDFO and (e) shows anomalous rise in volume below 10\,K.}
\label{latticeparams} 
\end{figure*}
\subsection{Theoretical}
\label{Theomethd}
Electronic structure of EDFO was studied using density functional theory implemented in the Vienna Ab-initio simulation program (VASP) which uses the projector augmented wave(PAW) method\cite{kresse1996efficient}. Calculations were performed using Perdew-Burke-Ernzerhof(PBE) based generalized gradient approximation (GGA)\cite{perdew1996} and GGA+$U$\cite{anisimov}. A cut-off energy of 500\,eV was used in the expansion of the plane waves. The structure was relaxed keeping the Er/Dy $4f$ electrons as core electrons. Ionic positions were relaxed until the forces on the ions are less than 0.1 meV\,$\mathrm{\AA}^{-1}$. For the electronic self-consistent calculations, the Er/Dy $4f$ electrons were treated as valence electrons. We have considered following orbitals in the valence band for each atom; Fe: $3d$, $4s$, O: $2s$, $2p$ and Er/Dy: $4f$, $5p, 5d$, $6s$. A 6$\times$6$\times$6 Monkhorst-Pack $k$-mesh centered at ${\Gamma}$ point in Brillouin zone was used for performing the Brillouin zone integrations.
\section{Experimental results}
\subsection{Structural Characterization}
\label{stuct_char}
Fig.~\ref{xrd} shows the room temperature powder x-ray diffraction pattern for EDFO. The pattern is refined by Rietveld method using Fullprof program. The pattern is refined to a single phase, with no trace of any impurity or unreacted phases. The compound crystallizes in the orthorhombic $Pbnm$ space group.  At room temperature, the estimated lattice parameters are $a$$=$5.2793$\mathrm{\AA}$, $b$$=$ 5.5835$\mathrm{\AA}$ and $c$$=$7.6011$\mathrm{\AA}$. The lattice(structural) parameters were also extracted from neutron powder diffraction patterns collected between 300 to 1.5\,K. 
In Fig.~\ref{latticeparams}(a-c), the temperature variation of the three lattice parameters is shown. As expected, $a$, $b$, and $c$ continuously decrease from 300 till 100\,K. Further till 10 \,K, $a$ and $c$ decrease gradually with temperature, while $b$ shows a ``hump" like feature between 50 and 10\,K. Below 5 \,K we observe a sharp increase in all the three lattice parameters with a maximum around 2\,K. 
The temperature variation of unit cell volume $V$ (Fig.~\ref{latticeparams}(d)) is similar to that of $a$ and $c$, with a slope change below 100 \,K.    
 In the Fig.~\ref{latticeparams}(e), we observe a sharp increase in volume with a maximum value at 2 \,K. In the absence of structural transformation, the ``isotropic" negative thermal expansion effects may be considered as magneto-elastic or magneto-volume effect\cite{Naveen2015}.
The anomalous negative thermal expansion is found to be associated with rare-earth ordering.
\begin{figure}[tbh] \center
       \begin{picture}(240,190)
        \put(-5,-5){\includegraphics[width=250\unitlength,]{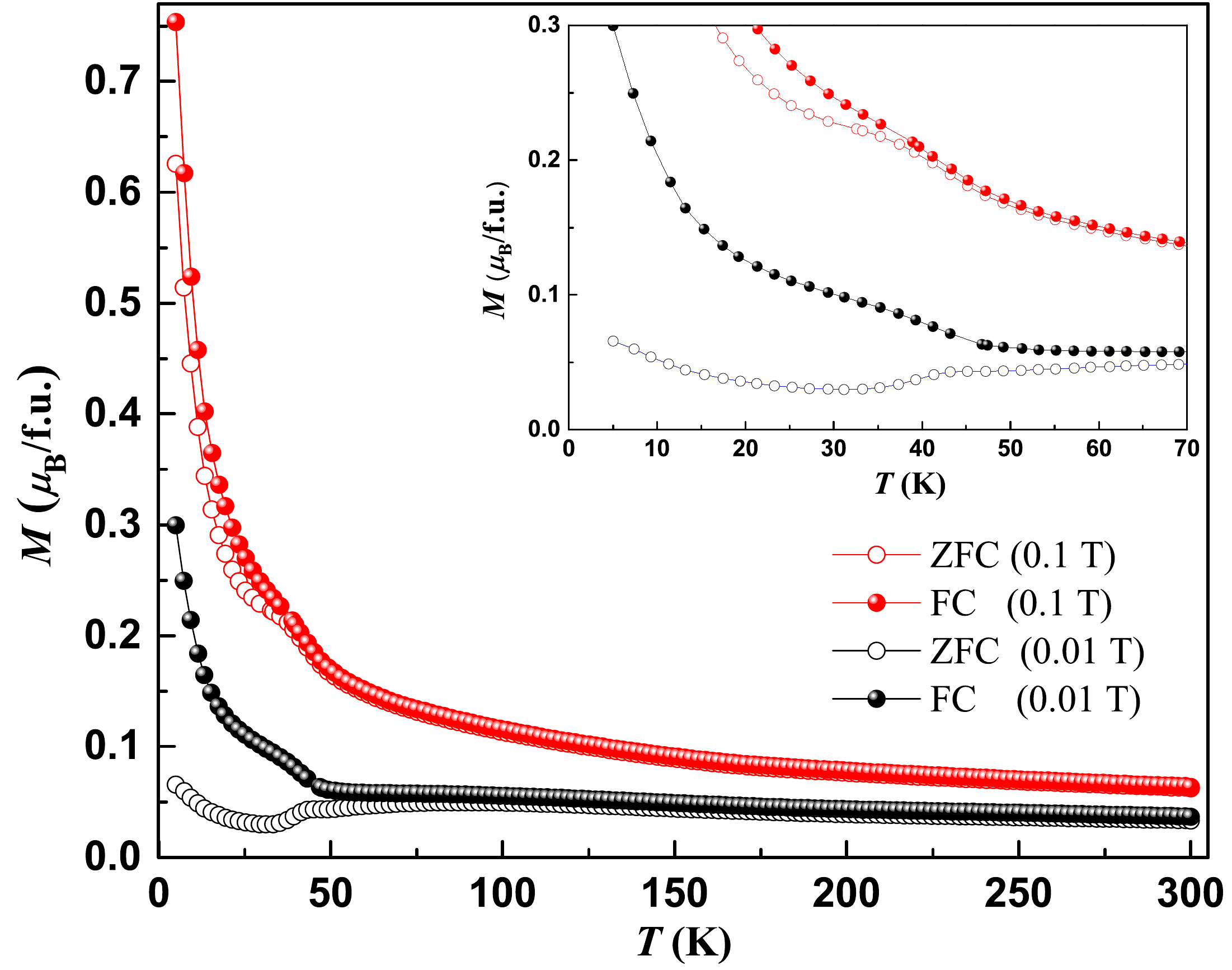}}
      \end{picture}
\caption{(color online) ZFC-FC magnetization of EDFO at (a) 0.01\,T and (b) 0.1\,T. The inset shows enlarged portion of the graph below 70\,K.}
\label{ZFC-FC}
\end{figure}
\begin{figure}[tbh] \center
       \begin{picture}(240,180)
        \put(-5,-5){\includegraphics[width=255\unitlength,]{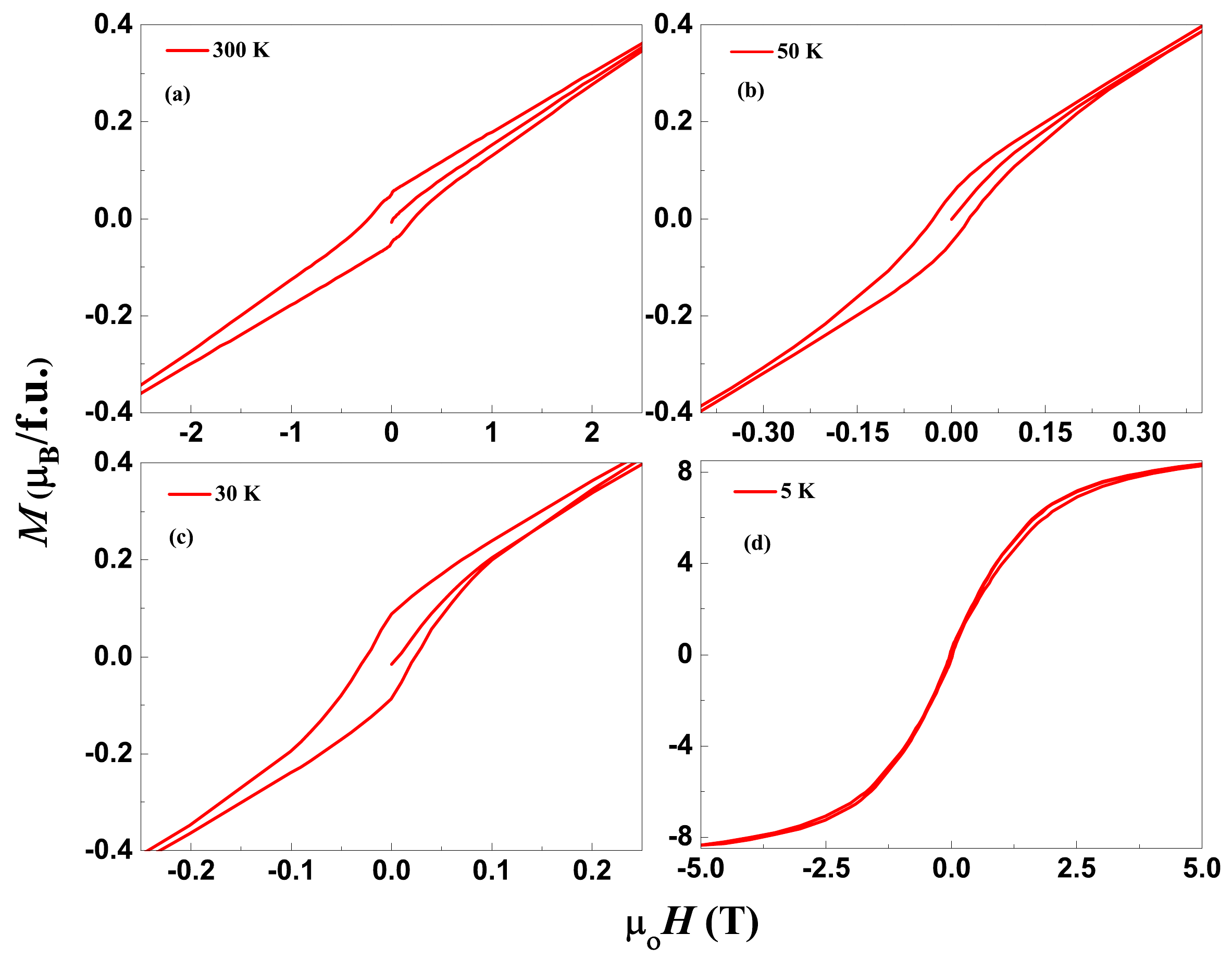}}
      \end{picture}
\caption{(color online) $M$-$H$ isotherms of EDFO at various temperatures.}
\label{MH}
\end{figure}
\subsection{Magnetic properties}
\subsubsection{DC Magnetization}
\label{avg_mag}
Fig.~\ref{ZFC-FC}, shows the temperature dependence zero field cooled (ZFC) and  field cooled (FC) magnetization measurements for EDFO from 2\,K to 300\,K at magnetic fields of 0.01\,T and 0.1\,T. 
At 300\,K, EDFO is expected to be  antiferromagnetically ordered in ${\Gamma}_{4}$($G_{x}$,\,$A_{y}$,\,$F_{z}$) magnetic structure, due to which we observe a small difference in the ZFC and FC magnetization. 
 The spin reorientation of Fe$^{3+}$ spins begins below $\sim$100/110\,K in ErFeO$_{3}$\cite{Bozorth58,Grant1969, Bazaliy2004,Pinto1971,Tsymbal2007}. Unlike ErFeO$_{3}$, the signature of spin reorientation in EDFO appears below 70\,K, indicated by the increase in bifurcation between ZFC-FC magnetization (in 0.01\,T) as seen in Fig.~\ref{ZFC-FC}. 
Near T$\sim$\,45\,K, the ZFC magnetization shows a drop  whereas the FC magnetization increases in a continuous manner with a change in slope. Below 25\,K, the ZFC magnetization rises again, similar to ErFeO$_{3}$ \cite{Huishen2013}.
Unlike DyFeO$_{3}$, signature of a clear Morin-like transition is not observed in EDFO\cite{Zhao2014}. The ZFC-FC measurements for 0.1\,T is also shown in Fig.~\ref{ZFC-FC}. With decrease in temperature, the ZFC and FC curves for 0.1\,T show a continuous increase with a small slope change near 45\,K.
\par
The isothermal field variation of magnetization for various temperatures are shown in Fig.~\ref{MH}. At 300\,K, the $M$-$H$ curve shows typical hysteresis loop of a canted antiferromagnet with a coercivity of nearly 0.2\,T, confirming the weak ferromagnetism in EDFO.
At 50\,K and 30\,K, the $M$-$H$ isotherm loops have relatively narrower hysteresis loops with an almost linear magnetization at higher magnetic fields. 
The slope of linear magnetization at higher fields, which increases with decreasing temperature, is due to the development of paramagnetic moment of both the rare-earth ions\cite{Huishen2013}. At 5\,K, the non-linear behavior of the magnetization, at higher magnetic field, suggests polarization of the $R^{3+}$ moments.
In a magnetic field of 5\,T, the magnetization attains a near-saturation ($M_\mathrm{sat}$) value of nearly 8.4\,${\mu}_\mathrm{B}$. 
Both Er$^{3+}$ and Dy$^{3+}$ ions have a ground state angular momentum quantum number $J$=15/2 and $\sim$10\,${\mu}_\mathrm{B}$ magnetic moment. 
 Magnetization studies on single crystals of DyFeO$_{3}$ reveal that the magnetic moment along $b$ axis attains a maximum value of nearly 9\,${\mu}_\mathrm{B}$, while a total magnetization of 10.6\,${\mu}_\mathrm{B}$ is obtained.
Similarly, in ErFeO$_{3}$, the total magnetic moment reaches a value of 7.6\,${\mu}_\mathrm{B}$ in a magnetic field of 5\,T\cite{Zhang2019}.
Thus the value of $M_\mathrm{sat}$ in EDFO is close to the total magnetic moment observed in ErFeO$_{3}$ and thus smaller than the expected average magnetic moment of $\sim$10\,${\mu}_\mathrm{B}$.
\begin{figure*}[tbh]
\begin{minipage}{0.5\linewidth}
\centering
\includegraphics[width=9.0cm]{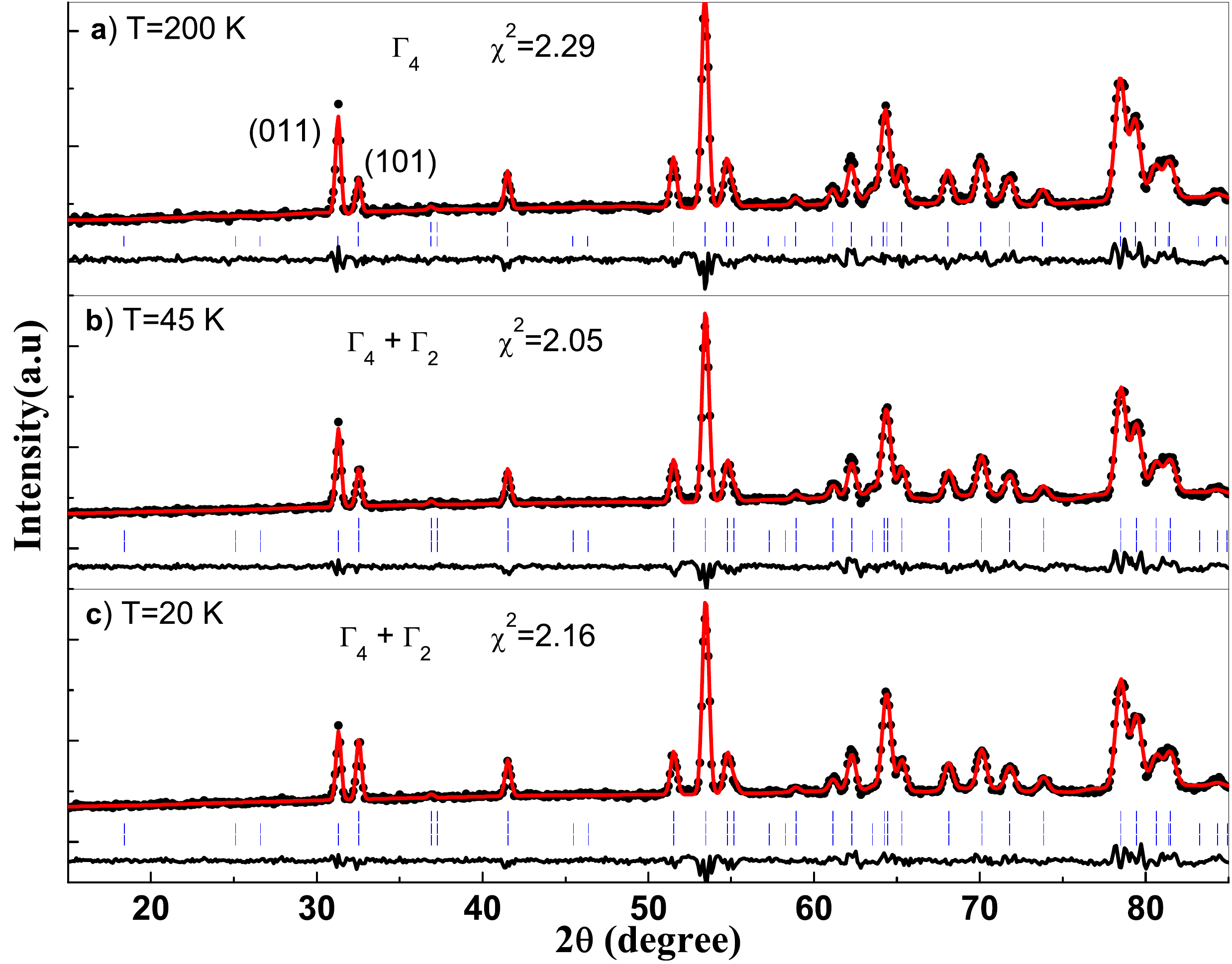}
\end{minipage}%
\begin{minipage}{0.5\linewidth}
\hspace*{-0.3cm}
\centering
\includegraphics[width=9.0cm]{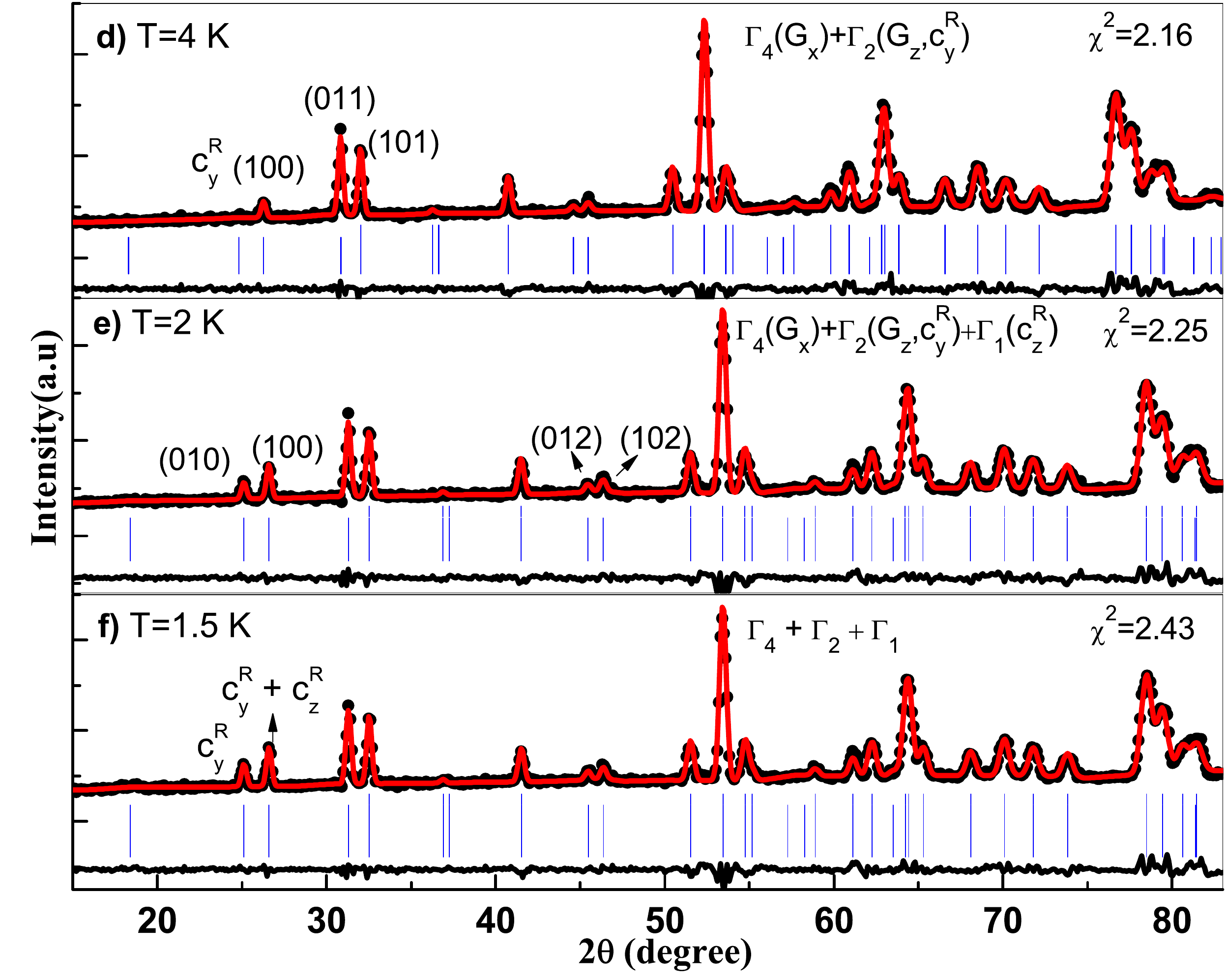}
\end{minipage}
\caption{Neutron powder diffraction pattern and refinements of EDFO at 200, 45 and 20\,K (left panel) showing the systematic evolution of intensity of (011) and (101) magnetic peaks. In the right panel, patterns and refinement for 4, 2 and 1.5\,K are shown. Additional magnetic peaks arising due to ordering of Er$^{3+}$/Dy$^{3+}$ moments are marked.}
\label{neutron_diff}
\end{figure*}
\subsubsection{Magnetic Neutron Diffraction}
In this section, the magnetic structures of EDFO obtained from neutron diffraction data is discussed. At 300\,K(pattern not shown), we observe the structurally forbidden (101) and (011) magnetic peaks associated with G-type magnetic ordering of the Fe$^{3+}$ spins. The neutron diffraction patterns obtained at 200\,K, 45\,K and 20\,K are shown in Fig.~\ref{neutron_diff}a), b), and c). The magnetic peaks (011) and (101) correspond to ordering vector, $\vec{k}$=(0,\,0,\,0). A ratio of nearly 1/3 between the intensities of (101) and (011) peaks confirms that the Fe$^{3+}$ spins are arranged as $G_{x}$ in ${\Gamma}_{4}$ magnetic structure at 200\,K\cite{Epstein1969}. Between 200 and 100 \,K, the ratio between the peaks remain nearly constant, while the absolute values of the intensity increase. Below 75 \,K, we observe relative increase in the intensity of (101) peak in comparison to the (011) peak, which indicates the onset of spin-reorientation (data not shown). As shown in Fig.~\ref{neutron_diff}b), at 45\,K, the  intensity of (011) peak is approximately twice in comparison to the intensity of (101) peak. Such intensity ratio suggests that a mixed magnetic structure, $G_{xz}$(${\Gamma}_{4}+{\Gamma}_{2}$), exists at 45\,K. The changes in intensity ratio of (101) and (011) peaks are much more gradual than the previously studied mixed doped orthoferrite, Nd$_{0.5}$Dy$_{0.5}$FeO$_{3}$\cite{Ankita2019b}.
\par Equal intensity ratio between (101) and (011) peaks is required for a pure ${\Gamma}_{2}$($G_{z}$) magnetic structure\cite{Epstein1969}. As evident from Fig.~\ref{neutron_diff}c), the intensity of (101) and (011) peaks is not equal, hence the spin reorientation transition does not complete even at 20\,K. In Fig.~\ref{neutron_diff}d), e), and f), the neutron diffraction patterns for 4, 2 and 1.5\,K are shown. As shown in Fig.~\ref{neutron_diff}d), e), and f), the intensity ratio of (011) and (101) peaks remain slightly higher than 1 indicating presence of a mixed structure of Fe$^{3+}$ sublattice  down to the lowest measured temperature (1.5\,K).
\begin{figure*}[tbh]
\begin{minipage}{0.5\linewidth}
\centering
\includegraphics[width=9.0cm]{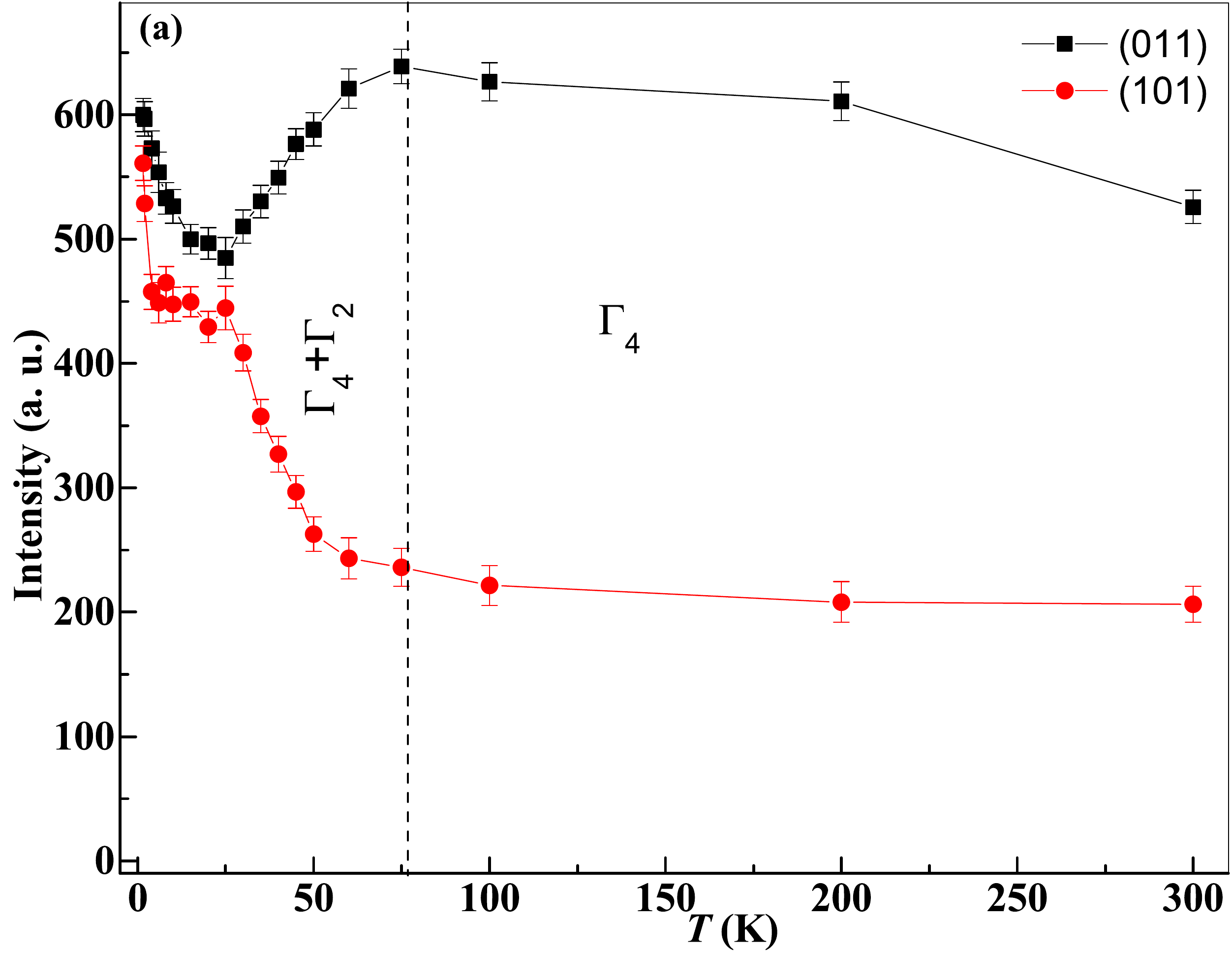}
\end{minipage}%
\begin{minipage}{0.5\linewidth}
\hspace*{-0.1cm}
\centering
\includegraphics[width=9.0cm]{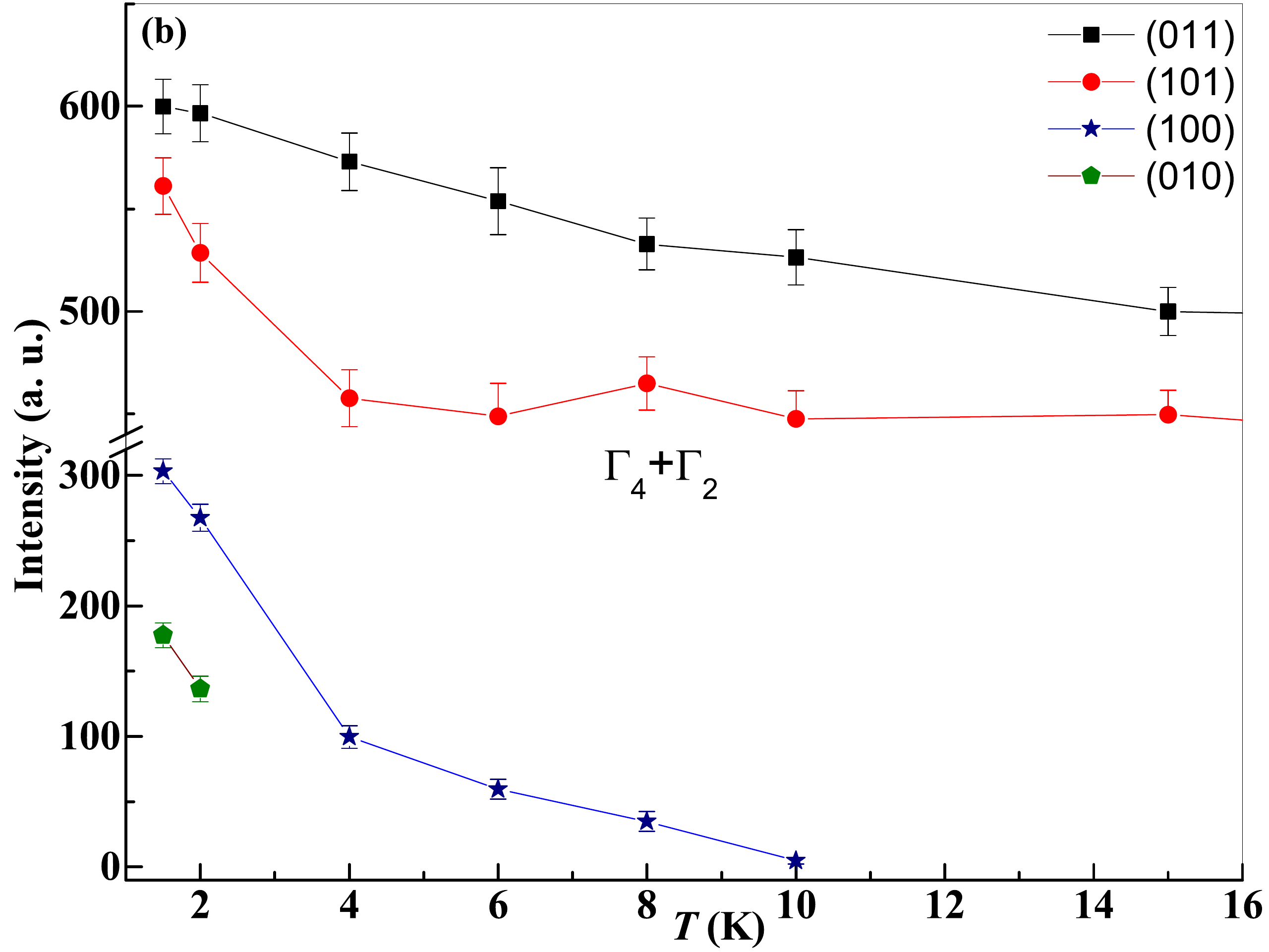}
\end{minipage}
\caption{a) Temperature variation of intensity of various magnetic peaks between 300 and 1.5\,K. b)  Temperature variation of intensity of various magnetic peaks below 15\,K. The development of peaks due to $R^{3+}$ ordering and anomalous rise in the intensity of (101) magnetic peak below 4\,K is highlighted.} 
\label{Intensity}
\end{figure*}
 \begin{figure}[b!] \center
       \begin{picture}(270,180)
        \put(-15,-10){\includegraphics[width=260\unitlength,]{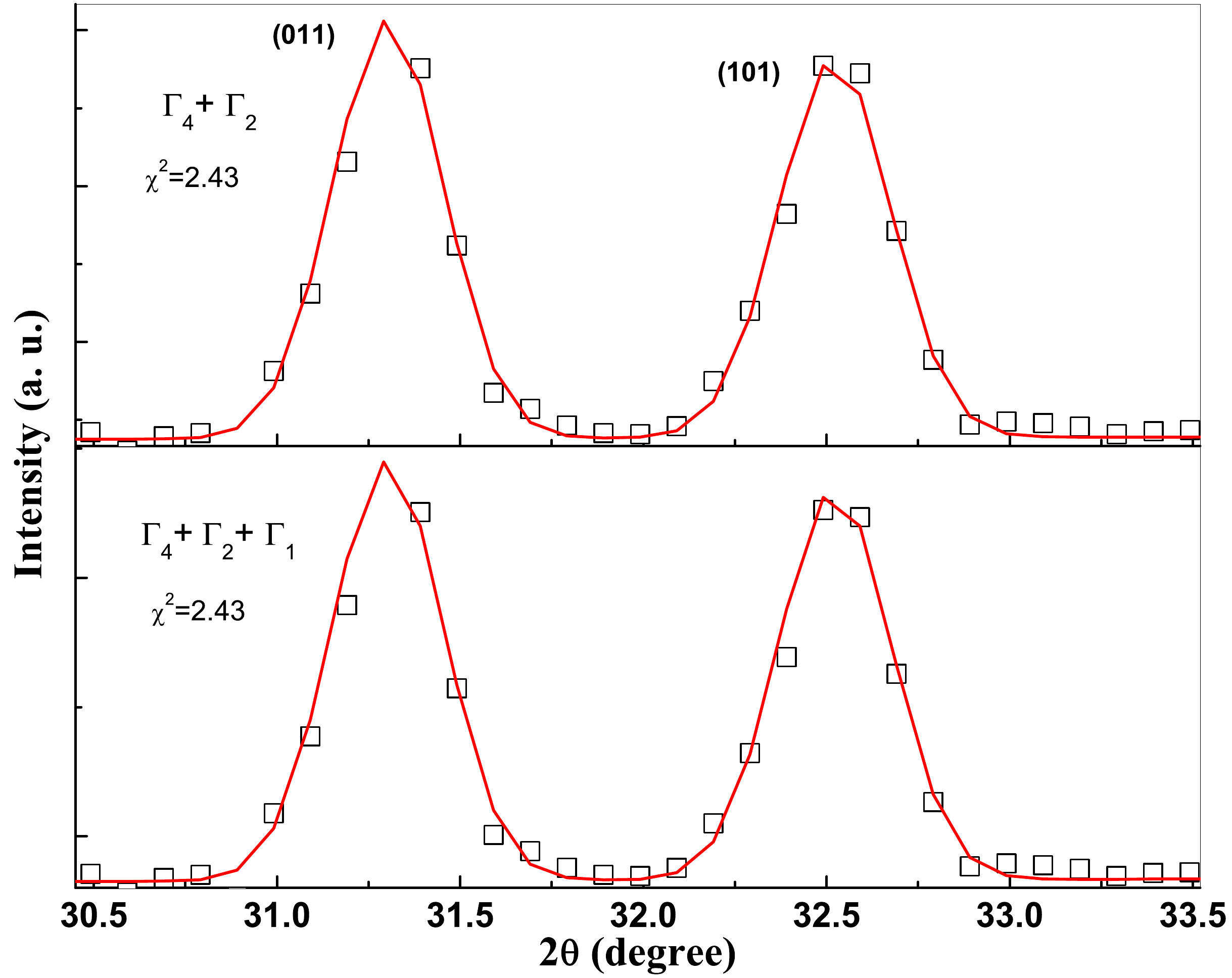}}
       \end{picture}
    \caption{(color online) (011) and (101) magnetic Bragg peaks showing results of fitting by ${\Gamma}_{4}$+${\Gamma}_{2}$(upper panel) and ${\Gamma}_{4}$+${\Gamma}_{2}$+${\Gamma}_{1}$(lower panel) magnetic structures of the Fe$^{3+}$ spins.}
        \label{Ienlarged}
      \end{figure}
  \begin{figure}[b!] \center
       \begin{picture}(260,180)
        \put(-15,-10){\includegraphics[width=260\unitlength,]{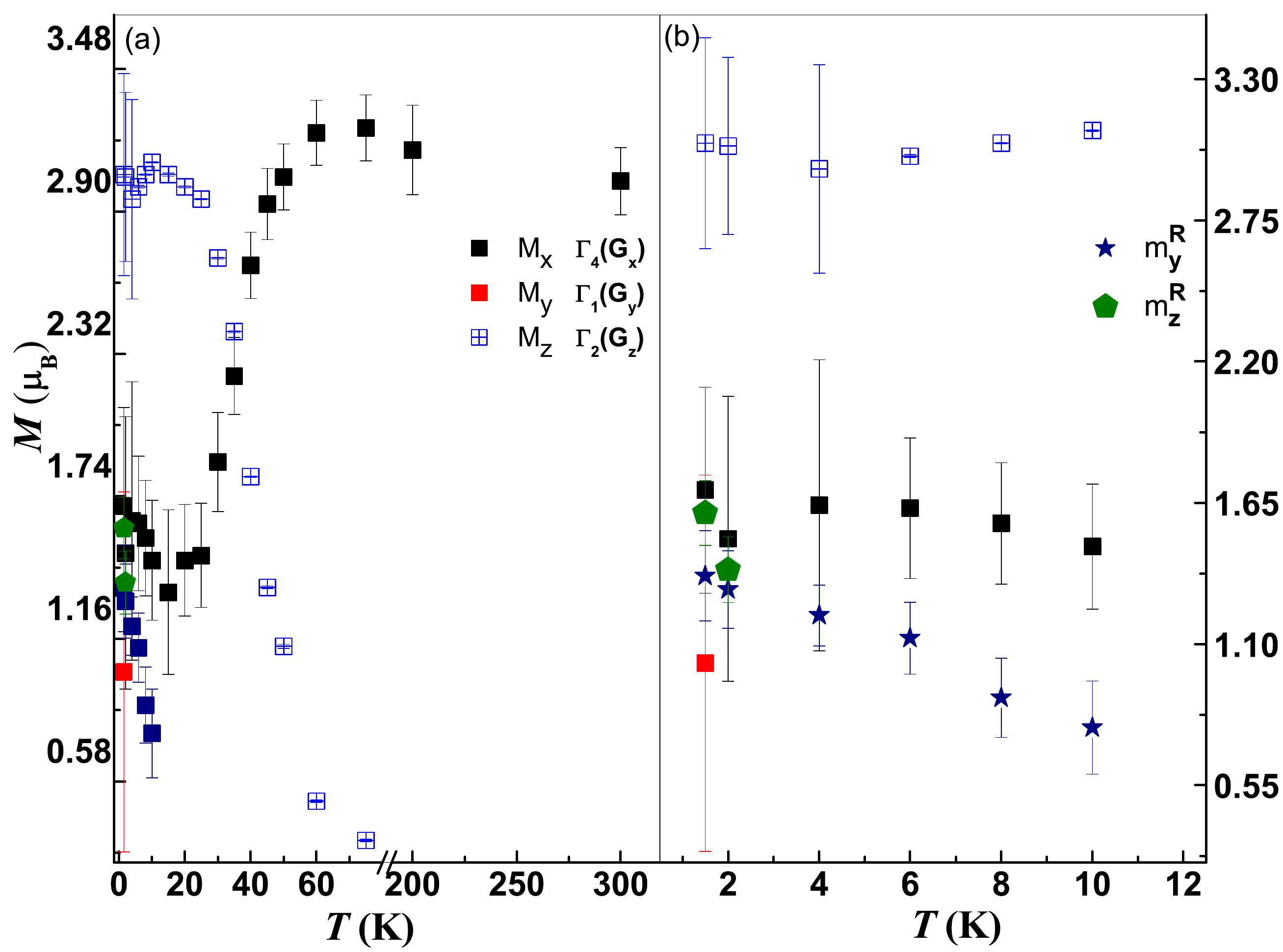}}
       \end{picture}
    \caption{(color online) a) Temperature variation of magnetic moment of Fe$^{3+}$ and Er$^{3+}$/Dy$^{3+}$ spins from 1.5\,K to 300\,K for the various magnetic structures. b) The variation of magnetic moments in the temperature range from 10 to 1.5\,K. The arrow shows the discontinuity/sudden rise of $m_{z}^R$.}
        \label{mag_moments}
      \end{figure}
%
\begin{figure*} \center
       \begin{picture}(1200,230)
       \put(0,0){\includegraphics[width=450\unitlength]{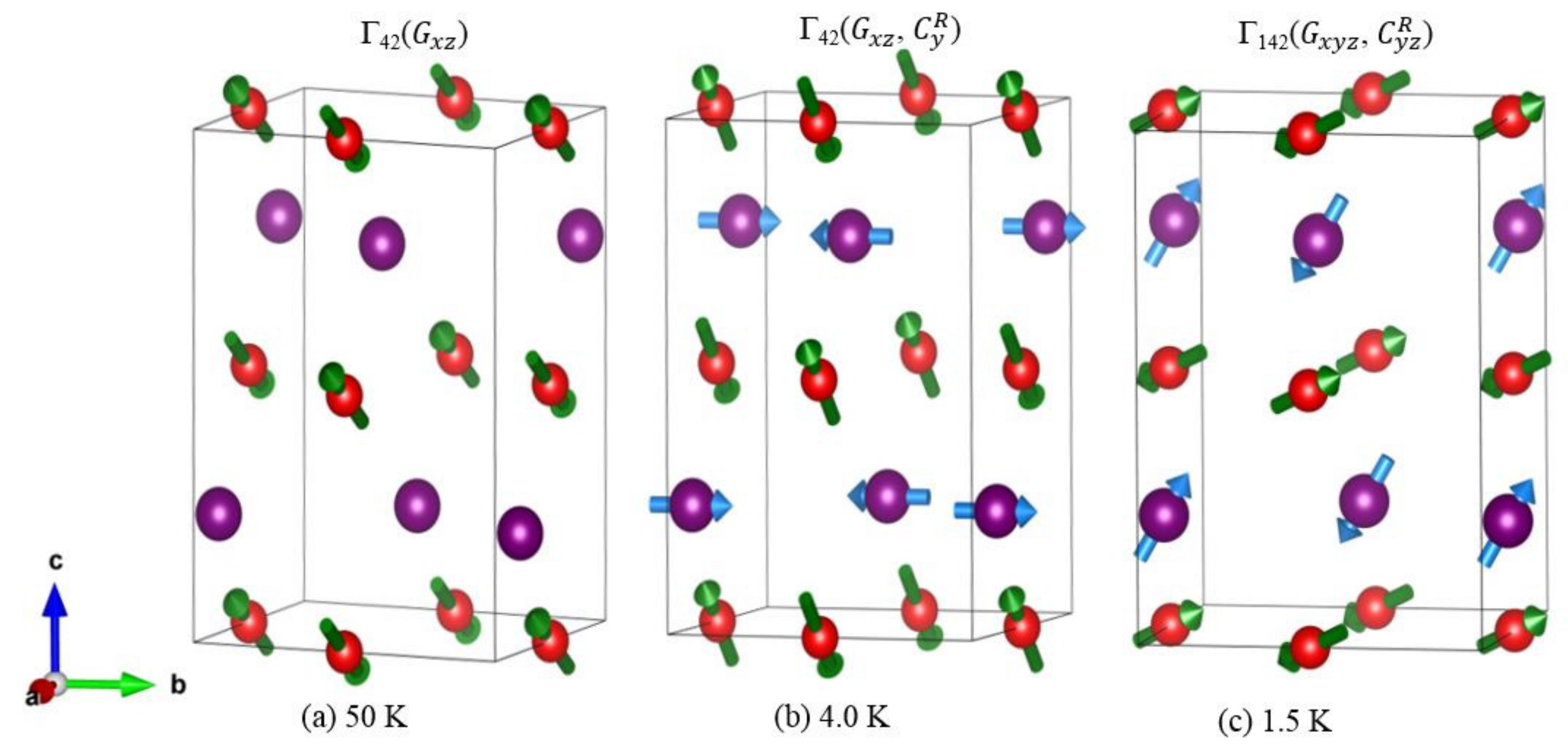}}

      \end{picture}
      \caption{(color online)  Magnetic structure of EDFO at (a) 50\,K : $G_{x}$,\,$G_{z}$ arrangement of Fe$^{3+}$ spins (b) 4\,K: ($G_{x}$,\,$G_{z}$,\,$c_{y}^{R}$) (c) 1.5\,K: ($G_{x}$,\,$G_{y}$,\,$G_{z}$,\,$c_{y}^{R}$,\,$c_{z}^{R}$). Red spheres represent Fe atoms, Violet spheres represent Er/Dy atoms}
        \label{EDFO_mag_struct}
      \end{figure*}
\par
Below 10\,K additional peaks near 26$^{\circ}$ and 45$^{\circ}$ develop. In the $Pbnm$ space group the peak at 26$^{\circ}$ can be indexed to (100), while the peaks near 45$^{\circ}$ can be indexed to (012) and (102). All the three peaks are structurally forbidden in $Pbnm$ space group. Development of additional peaks can be attributed to the onset of rare-earth ordering.
With decrease in temperature, the intensity of the three magnetic peaks increase in a systematic manner.
At 2\,K, additional magnetic peak corresponding to (010) reflection appears, along with sudden increase in intensity of (100) and (012)/(102)peaks related to rare-earth ordering and (101) peak related to Fe$^{3+}$ sub lattice ordering. 
The magnetic peaks that develop below 10\,K can also be indexed to ${\vec k}$=(0,\,0,\,0).

\par
In Fig.~\ref{Intensity}a), the temperature variation of the integrated intensities of the (011) and (101) magnetic peaks is shown for temperature range from 300 till 1.5 \,K. Below 75 \,K, the intensity of (011) peak decreases, while that of (101) peak increases in systematic manner. Such a variation of magnetic peaks' intensity confirms the initiation of the spin reorientation at 75\,K. This trend persists till 25 \,K, though they never crossover, which is a signature of complete reorientation. 
In Fig.~\ref{Intensity}b), the enlarged version of intensity variation is shown between 15 to 1.5\,K. Below 15\,K, the (011) peak shows a gradual increase till 4\,K , while that of (101) peak remains nearly constant with small fluctuations till 4\,K. The intensity of (101) peak increases considerably below 4\,K. Hence, the ratio of (011) and (101) intensity decreases below 4\,K. 
\par
Additionally, Fig.~\ref{Intensity}b) also shows the rise in intensity of the (100) and (010) peaks, associated with $R^{3+}$ ordering. Between 10 and 4\,K, the (100) peak shows a gradual rise in intensity. However,  the (100) peak, similar to (101) peak, shows "discontinuous jump" in intensity at 4\,K, which is accompanied by sudden development of (010) peak. The overall variations in intensity clearly suggests coexistence of the multiple magnetic structures due to Fe$^{3+}$ and $R^{3+}$ ordering.  
\par

The nature of the multiple phases in EDFO are understood in detail from magnetic structural refinements using representational analysis\cite{Bertaut1968}.
The Fe atom occupies the $4b$ Wycoff position, while the $R$ atoms occupy the $4c$ sites.
The Fe$^{3+}$ spins can arrange in four possible magnetic representations(Shubnikov magnetic space group) ${\Gamma}_{1}$ ($Pbnm$), ${\Gamma}_{2}$ ($Pbn'm'$), ${\Gamma}_{3}$ ($Pb'nm'$), and ${\Gamma}_{4}$ ($Pb'n'm$). 
In Bertraut's notation\cite{bertaut1963magnetism}, the three spin components for each representation(${\Gamma}_{1}$ to ${\Gamma}_{4}$) in Cartesian form are written as ($A_x$,\,$G_y$,\,$C_z$), ($F_x$,\,$C_y$,\,$G_z$), ($C_x$,\,$F_y$,\,$A_z$) and ($G_x,\,C_y,\,F_z$) respectively.\par
In EDFO,  the magnetic structure belongs to ${\Gamma}_{4}$ representation from 300\,K till 75\,K which is in agreement with the general behavior of orthoferrites. 
\begin{table*}
\label{table}
\caption{Magnetic structural details of Er$_{0.5}$Dy$_{0.5}$FeO$_{3}$ from present studies and comparison with literature for ErFeO$_{3}$ and DyFeO$_{3}$. The ``${\rightarrow}$" and ``${\sim}$" denote a complete and incomplete reorientation respectively.}
\begin{center}
\begin{tabular}{p{3.0cm}c c c c c c c c c c c c c c c c}\hline\hline 
 & \multicolumn{1}{c}{} & \multicolumn{1}{c}{} & \multicolumn{1}{c}{}& \multicolumn{2}{c}{} & \multicolumn{2}{c}{} &\multicolumn{1}{c}{} &\multicolumn{1}{c}{}\\
Compound & $T_{N}$$>$$T$$>$$T_{SR}$ & \multicolumn{1}{c}{SR} & $T_{SR}$$>$$T$$>$$T_{R}$ &  \multicolumn{2}{c}{$T_{R}$$>$$T$$>$4 \,K}  &\multicolumn{2}{c}{T=1.5 \,K}& \multicolumn{1}{c}{} \\ \hline 
 & Fe$^{3+}$  & Type &  Fe$^{3+}$ & Fe$^{3+}$ & $R^{3+}$  &  Fe$^{3+}$ & $R^{3+}$ &  Reference \\  \hline
Er$_{0.5}$Dy$_{0.5}$FeO$_{3}$ & $G_x$ $F_z$  & ${\Gamma}_{4}$${\sim}$${\Gamma}_{2}$ & $G_{xz}$  & $G_{xz}$  & $c_y^R$ & Model1: $G_{xz}$ & $c_y^R$ $c_z^R$ &This work\\
    & & & & -     & &Model2: $G_{xyz}$ & $c_y^R$ $c_z^R$ &   This work\\
    
ErFeO$_{3}$  &  $G_x$ $F_z$  & ${\Gamma}_{4}{\rightarrow}{\Gamma}_{2}$ &$F_x$ $G_z$ & $F_x$ $C_y$ $G_z$ & $c_z^R$ & &  & \cite{KOEHLER1960,Bozorth58,Grant1969, Bazaliy2004,Pinto1971,Tsymbal2007,Deng2015,Gorodetsky1973}\\
 
DyFeO$_{3}$  & $G_x$ $F_z$  & ${\Gamma}_{4}{\rightarrow}{\Gamma}_{1}$ &$G_y$  & $G_{xy}$ &-&$G_{xy}$ &$g_x^R$ $a_y^R$  & \cite{Bozorth58,Gorodetsky1968, Prelorendjo1980, Zhao2014, Wang2016,Tokunaga2008,NOWIK1966,Belov1968,White1969,Wu2017}\\ 
\hline\hline
\end{tabular}
\end{center}
\end{table*}
Based on intensity ratios of peaks, spin reorientation transition starts to occur at 75\,K similar to ErFeO$_{3}$. Below 75\,K downwards, the magnetic structure is refined as mixed structure given by  ${\Gamma}_{4}$+${\Gamma}_{2}$ representations. 
 The refinements indicate that the spin reorientation in EDFO is of ${\Gamma}_{4}{\rightarrow}{\Gamma}_{2}$ type, which is the usual second order reorientation observed in various orthoferrites. 
In the temperature range below 60\,K, we do not find signature of ${\Gamma}_{4}{\rightarrow}{\Gamma}_{1}$, abrupt transition as observed in DyFeO$_{3}$. However, the ${\Gamma}_{4}{\rightarrow}{\Gamma}_{2}$ reorientation is not complete even at 1.5 \,K, the lowest measured temperature. 
\par
The rare-earth moments, due to their lower site symmetry can arrange in eight possible representations labelled as ${\Gamma}_{1}$ to ${\Gamma}_{8}$\cite{Yamaguchi1974}. As shown in Fig.~\ref{Intensity}b), below 10\,K, (100) magnetic peak starts to originate near 26$^\circ$ value of 2$\theta$ diffraction angle. From representational analysis, the (100) peak associated with the magnetic ordering of the Er$^{3+}$/Dy$^{3+}$ moments, corresponds to the $c_y^R$ magnetic structure, which belongs the ${\Gamma}_{2}$ representation. 
The corresponding magnetic moment is denoted as $m_y^R$. 
At 2\,K, the sudden development of the (010) peak corresponds to the $c_z^R$ arrangement of the Er$^{3+}$/Dy$^{3+}$ moments, which belongs to the ${\Gamma}_{1}$ representation. The corresponding magnetic moment is denoted as $m_z^R$.
The $c_z^R$ arrangements of Er$^{3+}$ moments were also observed in ErFeO$_{3}$\cite{Gorodetsky1973, Deng2015}. In ErFeO$_{3}$, the Fe$^{3+}$ show a probable coexistence of ${\Gamma}_{1}$ and ${\Gamma}_{2}$ magnetic structures coinciding with  $c_z^R$ ordering of the rare-earth\cite{Gorodetsky1973}. Similarly,  at 1.5\,K, due to the $c_z^R$-ordering of the rare earth in EDFO, development of the ${\Gamma}_{1}$ structure for the Fe$^{3+}$ spins is also expected in addition to combined ${\Gamma}_{24}$ structure\cite{ Deng2015}. 
The signature of ${\Gamma}_{1}$ magnetic structure of Fe$^{3+}$ spins can be concurred from the increase in intensity of (101) peak below 4\,K as indicated in Fig.~\ref{Intensity}b). However to confirm this by refinement, the data at 2 and 1.5\,K are refined to combinations of ${\Gamma}_{24}$ as well as ${\Gamma}_{24}$+${\Gamma}_{1}$ structures. At 2\,K, the fitting does not converge with inclusion of three phases. However at 1.5\,K the goodness of fit is equal in both the cases. In Fig.~\ref{Ienlarged}, we show fitting in ${\Gamma}_{24}$(upper panel) and  ${\Gamma}_{24}$+${\Gamma}_{1}$ (lower panel), highlighting the (011) and (101) peak.  Considering all three representations of Fe$^{3+}$(${\Gamma}_{24}$+${\Gamma}_{1}$), the fitting quality is seems to be slightly better. Moreover, the co-existence of ${\Gamma}_{1}$ is plausible, since the $c_z^R$ arrangement of Er$^{3+}$/Dy$^{3+}$ moments are symmetry-compatible only with the $G_{y}$ arrangement of Fe$^{3+}$ spins\cite{ Deng2015}.
\par
The temperature variation of the magnetic moments for the Fe$^{3+}$ and $R^{3+}$ moments for different representations are shown in Fig.~\ref{mag_moments}a). The values of total magnetic moment of Fe$^{3+}$ is nearly 3.2\,${\mu}_\mathrm{B}$, which is lower than the theoretical expected value of 5\,${\mu}_\mathrm{B}$. Such reduction might be due to effects of covalency, disorder and also the polycrystalline nature of our samples. From 300\,K till 75\,K we observe a small increase in the magnetic moment($M_{x}$) associated with $G_{x}$ configuration. With the onset of spin reorientation, there is a decrease in $M_{x}$, while correspondingly the $M_{z}$ shows an increase. Below 20\,K both $M_{x}$ and $M_{z}$ remain nearly constant with small fluctuations as shown in Fig.~\ref{mag_moments}b). Also, as seen in Fig.~\ref{mag_moments}b), from 10\,K till 1.5\,K $m_{y}^{R}$ shows a gradual increase, while $m_{z}^{R}$ shows a sudden development at 2\,K. The $m_{y}^R$ and $m_{z}^R$ moments attain values of nearly 1.4 and 1.7\,${\mu}_\mathrm{B}$ respectively, resulting in total rare-earth sublattice moment of 2.4\,${\mu}_\mathrm{B}$ at 1.5\,K. The possible $M_{y}$ component of the Fe$^{3+}$ moments which only exists at 1.5\,K has a much larger error bar as shown in Fig.~\ref{mag_moments}b).
\par
In Table I, we list the magnetic configurations of EDFO at various temperatures. The magnetic configurations of the parent compounds of ErFeO$_{3}$ and DyFeO$_{3}$ are also listed as reference at the corresponding temperatures. The Fe$^{3+}$ and Er$^{3+}$/Dy$^{3+}$ magnetic structure of EDFO and its variation is closer to ErFeO$_{3}$ rather than DyFeO$_{3}$.
The magnetic structures of EDFO at 50 \,K, 4, K and 1.5 \,K depicting the Fe$^{3+}$ and Ed$^{3+}$/Dy$^{3+}$ spins are shown schematically in Figs.~\ref{EDFO_mag_struct}(a-c).
%
\subsubsection{Specific Heat}
\label{Specific heat}
  \begin{figure}[h!] \center
       \begin{picture}(500,170)
        \put(-5,-5){\includegraphics[width=250\unitlength]{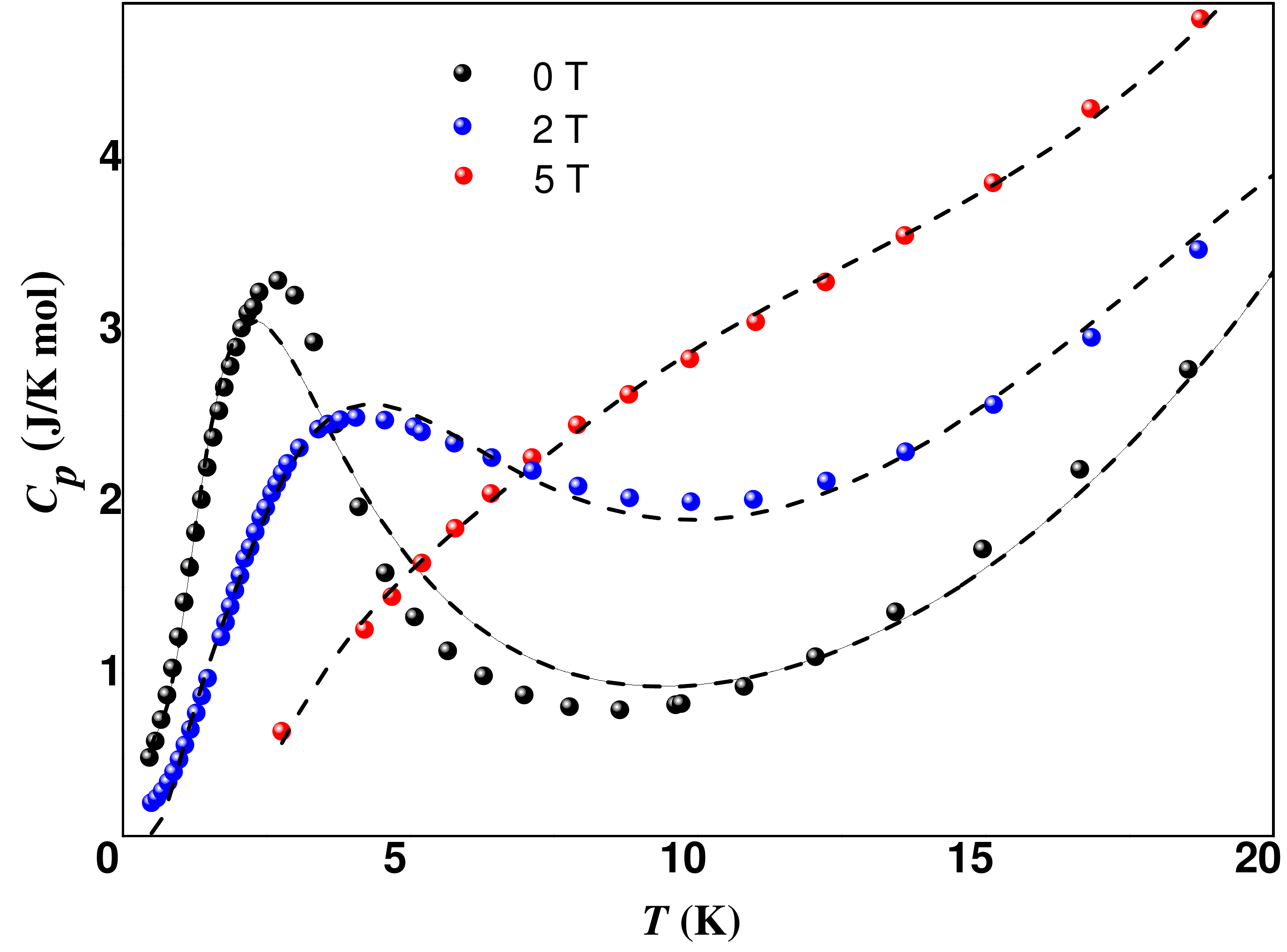}}
       \end{picture}
        \caption{The low temperature specific heat of EDFO for 0, 2 and 5 T. The dashed and solid lines shows the fitting of specific heat to Schottky and lattice terms}
        \label{Specific_heat}
      \end{figure}
The heat capacity $C_{p}$ of EDFO for 0, 2 and 5\,T are shown in Fig.~\ref{Specific_heat} from 0.4 to 20\,K. The zero field heat capacity shows a rise below 8\,K with a peak at 2.2 \,K. 
The ${\lambda}$-shaped anomaly associated with second order phase transition, seen in DyFeO$_{3}$\cite{Berton1968} at 4.2 \,K is absent in EDFO. Independent magnetic ordering of the $R^{3+}$ ions is absent till 0.4\,K which is in agreement with our neutron diffraction results. The feature observed in EDFO is similar to the peak observed in ErFeO$_{3}$ due to Schottky effect\cite{Saito2001}. 

\par
In EDFO, the Er$^{3+}$ and Dy$^{3+}$ being odd-electron systems, the ground state of each ion is a Kramer's doublet splited by molecular, exchange and dipole fields. Thus the low temperature peak in $C_{p}$ is due to the splitting of the ground state doublet of both the rare-earth ions.
To extract information about the splitting of the doublets, the specific heat is fitted in the temperature range 0.4-20 \,K as a sum of ``two-level" Schottky terms corresponding to both the rare-earth ions and the $T^{3}$ lattice term as below, 
\begin{equation}
\label{eqn3}
 C_{p}= \frac{1}{2}R\sum_{i=1}^{2}w_i(\frac{{\Delta}E_{i}}{k_\mathrm{B}T})^2\frac{exp\left[-\frac{{\Delta}E_{i}}{k_\mathrm{B}T}\right]}{(1+exp[-\frac{{\Delta}E_{i}}{k_\mathrm{B}T}])^2}+B_3T^3
\end{equation}
In Eqn.~\ref{eqn3}, ${\Delta}E_{1}$/$k_\mathrm{B}$ and ${\Delta}E_{2}$/$k_\mathrm{B}$ correspond to the doublet splitting in each $R^{3+}$ ion, while $B_{3}$ is the lattice term. A single energy splitting is insufficient to simulate the correct magnitude of the peak in $C_{p}$. 
From the fitting, the value $B_{3}$=4.04x10$^{-4}$ J/mole-$K^4$ is obtained which yields a Debye temperature of 457\,K for EDFO.
 At 0\,T, we obtain ${\Delta}E_{1}$/$k_\mathrm{B}$ = 1.5\,K and ${\Delta}E_{2}$/$k_\mathrm{B}$ = 5.6\,K.
These values are well in agreement with the optical spectroscopy studies on both the parent compounds as discussed further.
\par
Optical studies on ErFeO$_{3}$ revealed that  the splitting of the ground state doublet in the ${\Gamma}_{2}$ phase is nearly constant from 77 till 5\,K with a value of 3.17 cm$^{-1}$(0.39 meV or 4.52\,K)\cite{FaulhaberEr1967}. Similarly, above the Dy$^{3+}$ ordering temperature, the splitting of the ground state doublet in DyFeO$_{3}$ is nearly 1.5 cm$^{-1}$ (0.185\,meV or 2.14\,K) within experimental resolution\cite{FaulhaberDy1969}. The splitting in both cases is attributed to the $R^{3+}$-Fe$^{3+}$ interactions. 
Thus,  the values of ${\Delta}E_{1}$/$k_\mathrm{B}$ and ${\Delta}E_{2}$/$k_\mathrm{B}$ in EDFO can be attributed to the doublet splitting in the Dy$^{3+}$ and Er$^{3+}$ ions, respectively.  
However, due to complex temperature dependence of ${\Delta}E$, especially in the case of doublet of Er$^{3+}$ ion\cite{Hasson1975}, temperature independent ${\Delta}E$ terms cannot satisfactorily fit the Schottky peak, especially for zero field.
\par
The $C_{p}$ measured at 2\,T, fitting to Eqn.~\ref{eqn3} yields, ${\Delta}E_{1}$=5.9\,K, while ${\Delta}E_{2}$ increases to 26\,K. At 5\,T, we obtain, ${\Delta}E_{1}$=13.6\,K, and ${\Delta}E_{2}$=34\,K. Thus effect of magnetic field on the ground state of Er$^{3+}$ is more drastic, indicating greater polarizablity due to external fields. The values of ${\Delta}E$ for both ions though obtained indirectly are  close to the experimental splittings observed directly from Zeeman effect studies on both the parent compounds\cite{FaulhaberDy1969, FaulhaberEr1967, Wood1969}.

\section{DFT Calculations}
\label{theoretical}
The magnetic structure of EDFO is also explored using density functional theory. In the $Pbnm$ structure, the Er and Dy atoms occupy the $8c$ sites in a random fashion. For computational purposes, we have considered two regular arrangements of Er and Dy atoms, viz. the (111) and (001) arrangements\cite{Ankita2019b}. In the (111) arrangement, the Er and Dy atoms are placed adjacent to each other. Thus each Er atom has six Dy atoms as nearest neighbours and vice versa. In the (001) arrangement, Er and Dy atomic planes are alternately stacked along the $c$ axis\cite{Ankita2019b}. 
 \par
For both the cationic arrangements, structural relaxation of the orthorhombic unit cell have been performed using the experimental structural parameters obtained for 300\,K and 1.5\,K. The structure was relaxed considering G-type magnetic ordering of the Fe$^{3+}$ magnetic moments. 
The electronic self-consistent calculations were performed to obtain the magnetic structure and spin-resolved density of states (DOS) of EDFO, for
which the Hubbard $U$ on Fe and Er/Dy was incorporated. The Hubbard parameters $U$=8.5\,eV, $J$=0.5\,eV for Er and Dy; $U$=5.0\,eV and $J$=1\,eV for Fe are used.
The iterations were performed till an energy difference of 10$^{-6}$\,eV was achieved. Energetically, it is found that the (001) arrangement has lower energy as compared to (111) arrangement. 
\begin{figure*}[tbh]
\begin{minipage}{0.5\linewidth}
\centering
\includegraphics[width=9.0cm]{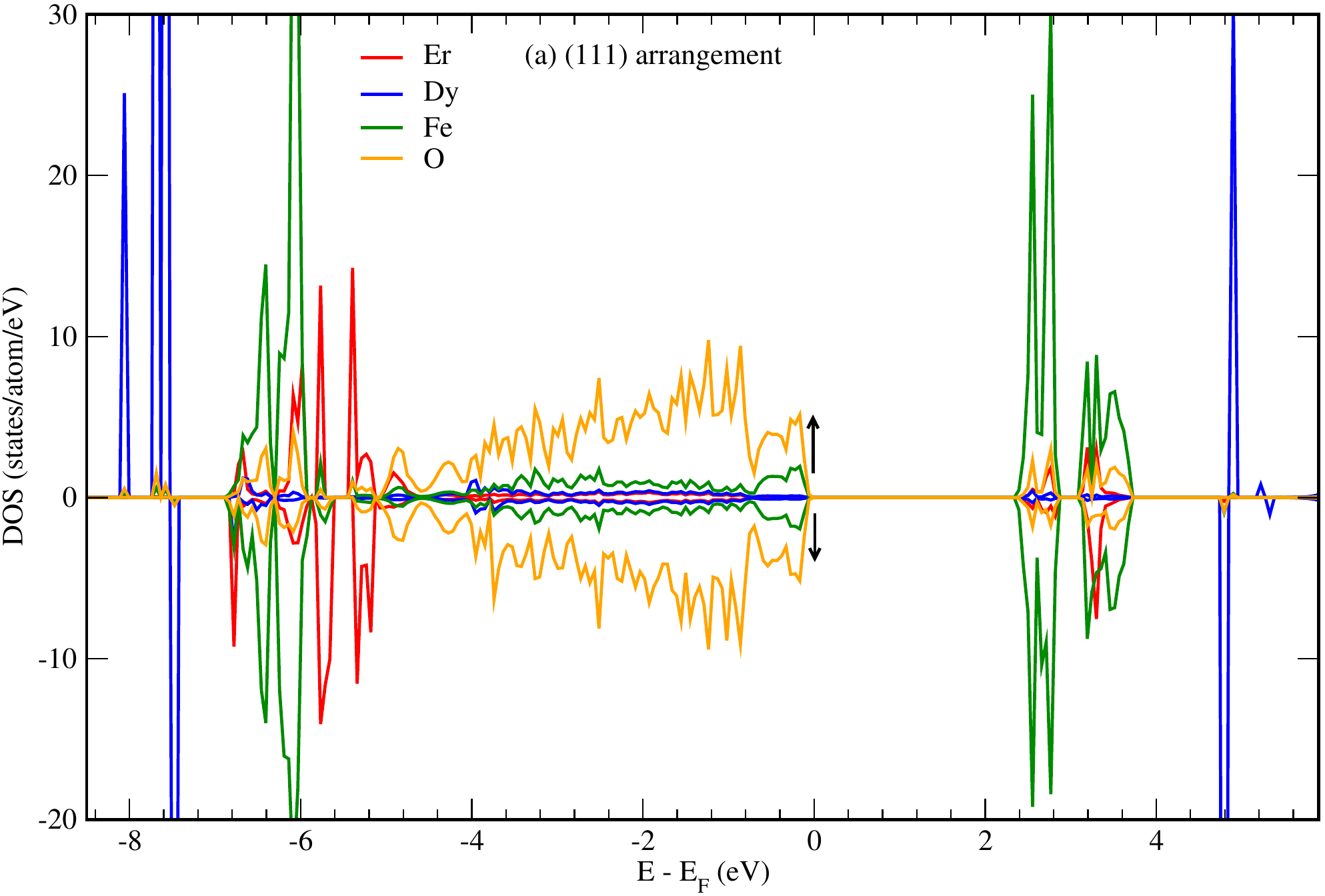}
\end{minipage}%
\begin{minipage}{0.5\linewidth}
\hspace*{0.1cm}
\centering
\includegraphics[width=9.0cm]{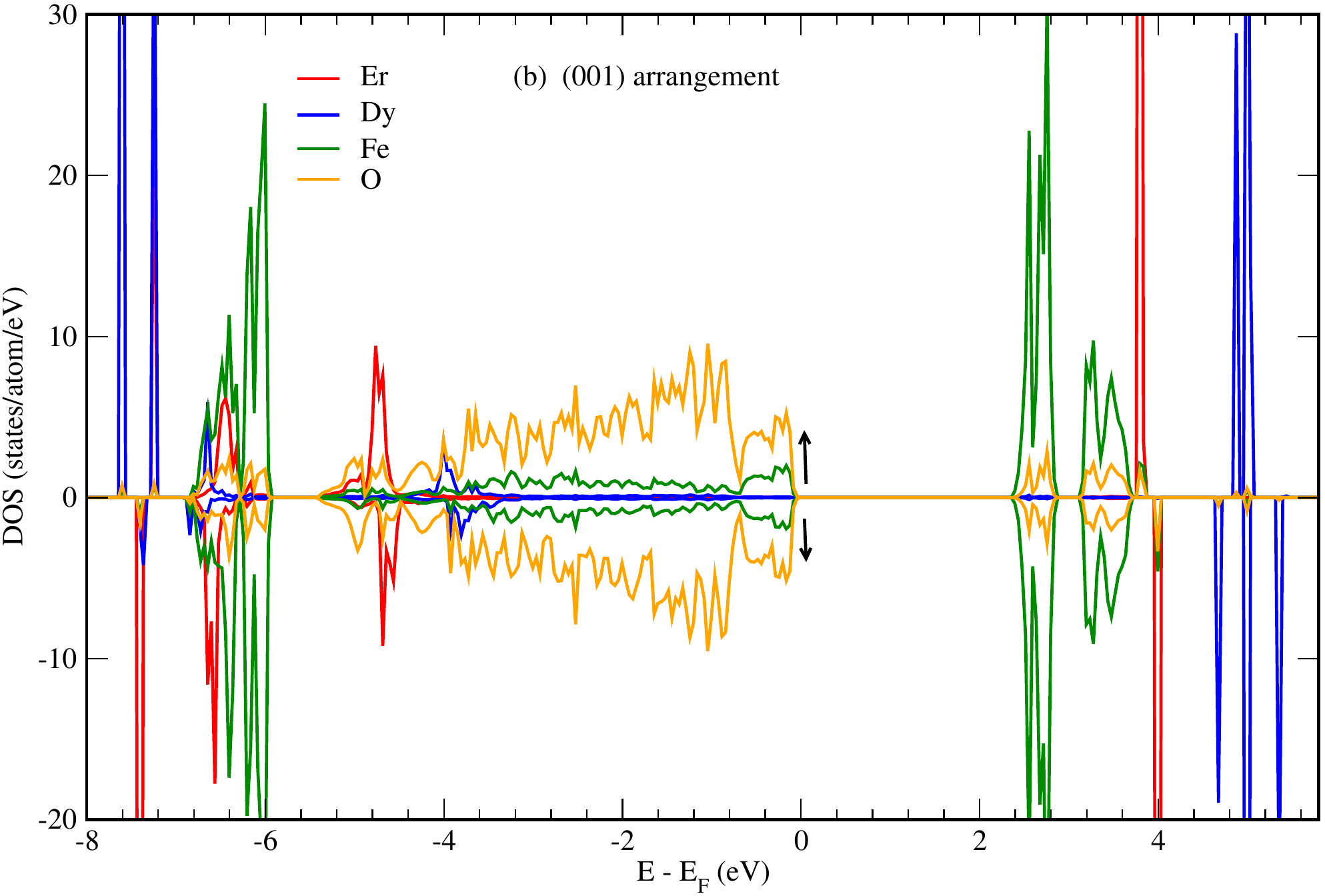}
\end{minipage}
\caption{Spin resolved density of states(DOS) of EDFO for the (a) alternate(111) (b) layered(001) arrangements of Er and Dy corresponding to the C-type ordering of rare-earth and G-type ordering of Fe moments. The ${\uparrow}$ and ${\downarrow}$ correspond to spin up and down regions respectively.} 
\label{dc}
\end{figure*}
\par
In the case of Fe$^{3+}$ sublattice, the lowest energy corresponds to G-type magnetic ordering which is common in all orthoferrites\cite{Spaldin2012}. 
In probing the rare-earth ordering, we have considered two possible arrangements of the Er$^{3+}$/Dy$^{3+}$ moments, a) C-type and b) G-type. The Fe$^{3+}$ ordering was fixed as G-type in these calculations. For both cationic arrangements of Er and Dy atoms, C-type magnetic structure emerges with lower energy, consistent with the neutron diffraction results. The relative energies are listed in Table II. The energy difference between both magnetic structures is greater for the layered arrangement as compared to the alternate arrangement of Er and Dy. The magnetic moment of Fe$^{3+}$ obtained from DFT is 4.2\,${\mu}_\mathrm{B}$, which is smaller than the free ion value of 5\,${\mu}_{B}$ and larger than the experimentally observed value. The reduction can be attributed to effects of hybridization with the O $2p$ band. Moment values of Er and Dy are 3\,${\mu}_\mathrm{B}$ and 5\,${\mu}_\mathrm{B}$ respectively, since we do not consider the effects of spin-orbit coupling in our calculations. 
\begin{table}
\caption{Relative energies (in meV) for two main antiferromagnetic orders of $R^{3+}$ ion within GGA+$U$ ($U$=8.5\,eV and J=0.5\,eV) for the two Er/Dy arrangements.  }
\begin{center}
\begin{tabular}{p{4.0cm}c c c c c}\hline\hline 
Magnetic structure & alternate (111) &  layered (001) &\\ \hline  
C-type & 0 & 0   \\ 
G-type & +9.76 & +16.46  \\ 
\hline\hline
\end{tabular}
\end{center}
\label{table2}
\end{table}
\par
 Fig.~\ref{dc}a) and b) show the spin resolved partial density of states of EDFO for (111) and (001) arrangements. In both arrangements, a band gap of ${\sim}$ 2.2\,eV is obtained.  Just below the Fermi energy, the spectral character of DOS is mainly the combination of strongly hybridized Fe $3d$ and O $2p$ states. In both the arrangements, the behavior of Fe $3d$ bands appear similar. 
 \par
The $4f$ states of Er and Dy show a different behavior in both the arrangements. 
For (111) arrangement, below the Fermi energy, Er $4f$ (${\uparrow}$ and ${\downarrow}$) DOS show a series of sharp spectral features in the range -5 to -7\,eV.  The Dy states occur much below Fermi level (-7 to -8\,eV), highly confined and does not overlap with the states of any other element.
\par
For (001) arrangement, a strong hybridization between Er $4f$ (${\uparrow}$ and ${\downarrow}$) and Fe $3d$/O $2p$ states occur around -5\,eV below which there is a gap In the range -6 to -7\,eV, the Er $4f$ ${\uparrow}$ states are strongly hybridized compared to the ${\downarrow}$ states. The Dy $4f$ states occur between -7 to -8\,eV without any signature of overlap. Above the Fermi level, $4f$ bands of both Er and Dy are more discrete and do not show any overlap with the Fe and O DOS.  
\par
The strengths of the magnetic exchange interactions between Er$^{3+}$-Er$^{3+}$, Dy$^{3+}$-Dy$^{3+}$ and Er$^{3+}$-Dy$^{3+}$ have been determined from our density functional theory calculations. Additionally, the strengths of Er$^{3+}$/Dy$^{3+}$-Fe$^{3+}$ exchange interactions are also determined. The differences in energies of ferromagnetic and anti-ferromagnetic arrangements are mapped to the Heisenberg Hamiltonian\cite{Jiachen2011, Spaldin2012}.
The calculations were performed on ``artificial unit cells using the experimental structure of 1.5\,K, in which except for the selected Fe or Er/Dy atoms, the rest of magnetic atoms are replaced by non-magnetic atoms. Thus the Fe atom is replaced by Al, while both Er and Dy are replaced by La atoms, since La$^{3+}$ ion is non-magnetic\cite{Spaldin2012}. 
\par
From the calculations, it is found that the Dy$^{3+}$-Fe$^{3+}$ interaction is the weakest, with a value of 0.019\,meV, which is smaller than the single ion anisotropy of Dy$^{3+}$ ion. The Er$^{3+}$-Fe$^{3+}$ interaction, with a much higher value of 2.88\,meV thus plays important role in the ${\Gamma}_{4}$${\rightarrow}$${\Gamma}_{2}$ reorientation.
Among the interaction strengths between the rare-earths, the Dy$^{3+}$-Dy$^{3+}$ exchange interaction is found to be 1.21\,meV. This is consistent with the fact that the interaction between the Dy$^{3+}$ ions which include the exchange and dipolar interactions result in long range ordering at 4.5\,K in DyFeO$_{3}$.
The Er$^{3+}$-Er$^{3+}$ interaction is also found to be small with a value of -0.037\,meV.
This is consistent with the fact though the Er$^{3+}$ moments can be polarized by the molecular field of Fe, suppressing the independent ordering of Er$^{3+}$ moments in ErFeO$_{3}$. 
The Er$^{3+}$-Dy$^{3+}$ exchange interactions has an appreciable value of 1.38\,meV, which helps in the establishment of the long-range ordering by polarization of the Er/Dy sublattice below 10\,K.
\section{Discussion}
\subsection{${\Gamma}_{4}$${\rightarrow}$${\Gamma}_{2}$ ``incomplete''reorientation of Fe$^{3+}$ spins}
In the ${\Gamma}_{4}$ phase below the $T_\mathrm{N1}$, due to the $F_{z}$ component of Fe$^{3+}$ spins, an effective molecular field along the $z$ axis develops. Such effective field induces a net polarization on the $R^{3+}$ moments, which by symmetry, should align as $f_z^R$. 
\par 
The Er$^{3+}$-Fe$^{3+}$ interactions are much greater than the Dy$^{3+}$-Fe$^{3+}$ interactions as seen in our first principles calculations and optical studies\cite{FaulhaberEr1967, FaulhaberDy1969}. 
The anisotropic and anti-symmetric exchange interactions part of Er$^{3+}$ and Fe$^{3+}$ exchange  interactions, cause rotation of the Fe$^{3+}$ spins below 75\,K, resulting in the ${\Gamma}_{4}$${\rightarrow}$${\Gamma}_{2}$ reorientation near 75\,K. This behavior is identical to reorientation in ErFeO$_{3}$\cite{Gorodetsky1973}. 
According to the effective field model by Yamaguchi {\it et al}\cite{Yamaguchi1973}, the anisotropic-symmetric exchange interactions between Fe$^{3+}$ and $R^{3+}$ is primarily responsible for such a gradual spin-reorientation. 
\par
During reorientation, the spontaneous weak ferromagnetic component of Fe$^{3+}$ moments changes from $F_{z}$ to $F_{x}$. 
Due to two dissimilar rare-earth ions, a local variation in exchange field develops in the system. The Dy$^{3+}$ ions do not affect the spin reorientation much, since the negligible Dy$^{3+}$-Fe$^{3+}$ interaction does not play a major role. 
Thus, instead of reorientation occurring in a short interval of 10 \,K as in ErFeO$_{3}$\cite{Gorodetsky1973}, a very gradual reorientation takes place in EDFO. 
This results in co-existence of $\Gamma_4$($G_x,\,C_y,\,F_z$) and ${\Gamma}_{2}$($F_x$,\,$C_y$,\,$G_z$)  in the entire temperature range below 75\,K, wherein $F_{z}$ and $F_{x}$ play role similar to the external applied field. 
\par
The external field along various crystal axes, cause spin reorientations in both the parent compounds. 
However, compared to ErFeO$_{3}$\cite{Zhang2019}, the effect of field is more drastic in DyFeO$_{3}$ at smaller fields.
For field along $a$ direction, above a critical field (of nearly 1-2\,T), a ${\Gamma}_{1}$${\rightarrow}$${\Gamma}_{4}$ transition is induced. A relatively small field along $c$-direction and  $b$-direction, ${\Gamma}_{1}$${\rightarrow}$${\Gamma}_{2}$ reorientation is induced in DyFeO$_{3}$\cite{Prelorendjo1980}. The effective molecular fields along the $a$,$b$ and $c$-directions eventually suppress the Morin transition in EDFO.
At 50\,K a co-existence of both the magnetic structures exists as shown in Figs.~\ref{EDFO_mag_struct}a, which persists till 1.5\,K. We also explored the possibility of an additional magnetic phase ${\Gamma}_{1}$ in addition of the ${\Gamma}_{24}$ at 1.5\,K. 
Hence, the spin reorientation in EDFO can be considered to be an ``incomplete reorientation".
\subsection{Polarization of rare-earth moments}
\subsubsection{Development of $c_{y}^R$  arrangement of $R^{3+}$  moments}
Just as the predominant Er$^{3}$-Fe$^{3}$ interactions cause the ${\Gamma}_{4}$${\rightarrow}$${\Gamma}_{2}$ reorientation, this also results in the polarization of the $R^{3}$ moments in the ($f_{x}^R$,\,$c_{y}^R$) configuration. The polarization of Er$^{3+}$ moments are lot easier as compared to the Dy$^{3+}$ moments. Though the preferred arrangement of Dy$^{3+}$ moments is ($g_{x}^R$,\,$a_{y}^R$), this is actually the spin-flopped configuration of ($f_{x}^R$,\,$c_{y}^R$) with the energy difference smaller than the single ion anisotropy of Dy$^{3+}$\cite{Yamaguchi1973}. Thus the effective field along the $x$ axis can also partially polarize the Dy$^{3+}$ moments along with the Er$^{3+}$ moments. Hence, at 4\,K, Fe$^{3+}$ spins are in  ${\Gamma}_{42}$configuration, the $R^{3+}$ moments are in $c_y^R$-configuration(Fig.~\ref{EDFO_mag_struct}b).
\subsubsection{Development of $c_{z}^R$ arrangement of $R^{3+}$ moments}
At 2\,K, a sudden appearance of the $c_{z}^{R}$ magnetic Bragg peaks occurs. As shown in  Fig.~\ref{mag_moments}b), the magnetic moment ($m_{z}^R$) shows a sudden rise.
The mechanism behind the  sudden appearance of $c_{z}^{R}$ and abrupt reorientation $G_{x}$${\rightarrow}$$G_{y}$, which is characteristic feature of ErFeO$_{3}$, is more complex to understand. In  Fig.~\ref{EDFO_mag_struct}c), we show the coexistence of $c_y^R$ and $c_z^R$ along with Fe$^{3+}$ spins having ($G_x$,\,$G_y$,\,$G_z$) at 1.5\,K.
\subsection{Analysis of the rare-earth ordering}
  The ground state of Dy$^{3+}$ and  of Er$^{3+}$ ions viz. $^6H_{15/2}$ and $^4I_{15/2}$ respectively, are split by the monoclinic crystal field($C_{s}$(m)) into eight Kramer's doublets\cite{Holmes1972,Wood1969}.
Inelastic neutron studies reveal that the splitting between the ground state and first excited doublet of Er$^{3+}$ is around 5.5\,meV, while in the case of Dy$^{3+}$ a separation of nearly 6.8\,meV is observed.
Below 50\,K, only the lowest doublet is populated in both ions. Hence, the Er$^{3+}$ and Dy$^{3+}$ ions are described as ``effective spin'' $S_R$=1/2 ($R$:rare-earth) systems which are two-fold degenerate. 
The degeneracies of ground state of both ions are Zeeman split by an energy ${\Delta}E$ due to external field as well as internal molecular fields, ${\vec H}_\mathrm{eff}$ due to $R^{3+}$-Fe$^{3+}$, $R^{3+}$-$R^{3+}$, exchange and dipolar interactions. The coupling of the Er$^{3+}$ and Dy$^{3+}$ ``spins" to the effective fields occur via the anisotropic `${\bf g}$' tensor\cite{FaulhaberEr1967, FaulhaberDy1969}. 
\par
Mean field studies pertaining to rare-earth magnetism in $R$CrO$_{3}$($R$=Nd, Er), show that temperature evolution of magnetic moment of $R^{3+}$ and ${\Delta}E$ are related in a self-consistent manner.
The sub-lattice magnetization of $R^{3+}$ is defined as $m^R$=1/2$g_y$${\mu}_\mathrm{B}$$N$$<$S$_{R}$$>$; where the thermal average of the $R^{3+}$ ``spin" is related to the doublet splitting as, $<$S$_{R}$$>$=$B_{S}$(${\Delta}E$/$k_\mathrm{B}$$T$). Here, $B_S$(x) is the Brillouin function which for $S_{R}$=1/2 becomes tanh(x)\cite{Horneich1975}.
\par
Optical spectroscopy and mean field analysis on NdCrO$_{3}$ and ErCrO$_{3}$ show that below the $T_\mathrm{N1}$, ${\Delta}E$ is proportional to temperature dependent sublattice magnetization of Cr$^{3+}$ spins $<S>$/$S$($S$=3/2 for Cr$^{3+}$). However, below the 
spin reorientation temperature($T_{SR}$), when the $R^{3+}$-$R^{3+}$ interactions become non-negligible, ${\Delta}E$ is proportional to sum of a) sublattice magnetization of Cr$^{3+}$ spins and b) tanh(${\Delta}E$($\vec{H}$,$T$)/2$k_\mathrm{B}$T) \cite{Horneich1975, Hasson1975}. 
\par
In EDFO, the situation is more complex, since the temperature dependence of ${\Delta}E$, is different for the Er$^{3+}$ and Dy$^{3+}$ ions. 
Thus in EDFO, the variation in $m_{y}^R$ can be attributed to sum of the three terms, a) Fe$^{3+}$sub-lattice magnetization, b)tanh(${\Delta}E_\mathrm{Er}$($\vec{H}$,$T$)/2$k_\mathrm{B}$T) and c)tanh(${\Delta}E_\mathrm{Dy}$($\vec{H}$,$T$)/2$k_\mathrm{B}$T). 
In EDFO, even above 75\,K, the Fe$^{3+}$ spins have attained their maximum value. In the range 1.5-10\,K the net sub-lattice magnetic moment of Fe$^{3+}$ is nearly constant. 
Thus, the temperature dependence of ${\Delta}E$ for both $R^{3+}$ ions on the Fe$^{3+}$ sub-lattice magnetization is less pronounced. 
The eventual rise of $m_{y}^R$ below 10\,K can be attributed to the interaction between the $R^{3+}$ ions. 
In view of the negligible Er$^{3+}$-Er$^{3+}$ exchange interactions, the increase in $m_{y}^R$ can be attributed to the 
the Dy$^{3+}$-Dy$^{3+}$ exchange and dipole interactions. Though optical spectroscopy studies on DyFeO$_{3}$ and DyAlO$_{3}$ suggest only a stronger dipole interaction, our first principles calculations suggest a strong exchange interaction can also exist between the Dy$^{3+}$ moments.
Additionally, the Er$^{3+}$-Dy$^{3+}$ exchange interactions which are comparable to the Dy$^{3+}$-Dy$^{3+}$ interactions, also contribute to the $c_y^R$ ordering. 
\par
At 2\,K, due to development of $c_z^R$-ordering, the total $R^{3+}$ sub-lattice magnetic moment shows sudden increase. 
Similar behavior is observed in ErFeO$_{3}$, in which below 5\,K the Er$^{3+}$ moments order as $c_z^R$ along with increase in ${\Delta}E$, which attains a maximum value of 6.5\,cm$^{-1}$(0.8\,meV or 10\,K)\cite{FaulhaberEr1967}.
In orthochromate ErCrO$_{3}$, a discontinuity in ${\Delta}E$ occurs at $T_{SR}$\cite{Hasson1975} accompanied by the $c_z^R$ ordering of Er$^{3+}$ moments. 
The Er$^{3+}$-Er$^{3+}$ dipole interaction which causes the $c_z^R$ ordering, is maximum between Er$^{3+}$ moments along the $c$ direction\cite{Wood1969,Hasson1975}.
\par
Even though the dipole interactions result in magnetic ordering of the $R^{3+}$ ions at the lowest temperature, the direction of the moments are decided by the anisotropic ${\bf g}$ tensor.
From optical spectra of ErFeO$_{3}$, it was observed that $g_{xx}$(min)$=$1.2, $g_{yy}$(max)$=$4.5 and $g_{zz}$ ($z$$||$$c$-axis)= 5.6 in $a$-$b$ plane\cite{Wood1969}. Thus, a slightly larger anisotropy along the $z$ axis facilitates the $c_{z}^R$ ordering of the Er$^{3+}$ moments below 5\,K.
In Dy$^{3+}$, $g_{xy}$(min)$=$3.2, $g_{xy}$(max)$=$18.4, $g_{zz}$$=$2.0, due to which the Dy$^{3+}$ moments have a strong Ising character with moments in direction of $g_{xy}$(max)\cite{Holmes1972} which is also clear from the long range antiferromagnetic ordering of Dy$^{3+}$ moments with ${\Gamma}_{5}$($g_{x}^{R}$,$a_{y}^{R}$) configuration in DyFeO$_{3+}$,  DyAlO$_{3+}$,  DyScO$_{3+}$. Thus the $c_z^R$ is completely opposed by the the $a$-$b$ plane anisotropy, even though optical studies on DyAlO$_{3}$ reveal that Dy$^{3+}$-Dy$^{3+}$ dipole interaction has maximum strength along the $c$ direction \cite{SchuchertDy1969}. 
Estimate of the dipolar energy in DyScO$_{3}$ for various magnetic configurations in the $a$-$b$ plane reveal that for the experimental ($g_x^R$,\,$a_y^R$) configuration, the lowest dipolar energy (-3.61\,K) is achieved which is also close to the $T_\mathrm{N2}$, while the ($f_x^R$,\,$c_y^R$) has the highest energy (+2.44\,K) at zero field\cite{Wu2017}. 
\par
In a similar manner, estimate of the ground state energy of the rare-earth in the context of dipolar energy is carried out in EDFO. The magnitude of magnetic moments are fixed according to the experimentally obtained values. The dipolar energy is calculated for Er and Dy  arranged alternately for a radii of nearly 6\,{\AA} containing eighteen atoms. 
For the four magnetic arrangements, $c_{z}^R$, $c_{y}^R$, ($c_{y}^R$,\,$c_{z}^R$) and ($g_x^R$,\,$a_y^R$), the energies obtained are -0.158\,K, +0.092\,K, -0.08\,K and -0.138\,K respectively.
Thus the $c_z^R$ ordering is most favoured by the Er$^{3+}$-Dy$^{3+}$ dipolar interaction while $c_y^R$ ordering is least favoured. 
However, the character of ${\bf g}$ tensor of Dy$^{3+}$ tends to suppress the $c_z^R$ ordering to lower temperature, due to which the peaks appear only at 2\,K with a lower magnetic moment than observed in ErFeO$_{3}$.
\subsection{Symmetry of ground state}
Finally, we discuss the symmetry aspects of the magnetic structure of EDFO at 1.5\,K.
Based on the representation analysis, the resultant magnetic structure can be written as ${\Gamma}_{4}$+${\Gamma}_{2}$+${\Gamma}_{1}$.
To simplify things, we consider the net magnetic structure at 1.5\,K as combinations of ${\Gamma}_{24}$, ${\Gamma}_{14}$ and ${\Gamma}_{12}$. The ${\Gamma}_{12}$ and ${\Gamma}_{14}$ belong to the point group $C_{2h}$($C_{2h}$)(m), while the ${\Gamma}_{24}$ belongs to the point group $C_{2h}$($C_{i}$)(m')\cite{Yamaguchi1973}. The ${\Gamma}_{14}$ and ${\Gamma}_{12}$ are invariant under the symmetry operations ($E$,${\tilde{C}}_{2x}$,$i$,  $i{\tilde{C}}_{2x}$) and ($E$,${\tilde{C}}_{2z}$,$i$, $i{\tilde{C}}_{2z}$), respectively. On the other hand, the ${\Gamma}_{24}$ is invariant under ($E$,$i$, $R{\tilde{C}}_{2y}$, $iR{\tilde{C}}_{2y}$). Here $i$ corresponds to the inversion symmetry operation and $R$ corresponds to time reversal symmetry operator.
In the case of ${\Gamma}_{24}$, there occurs an effective field, which is absent in the case of ${\Gamma}_{1}$. 
Considering the coexistence of three phases, the only symmetry elements remaining are ($E$,$i$). The presence of inversion symmetry rules out possibility of a spontaneous ferroelectric polarization. Moreover, reduction in the structural symmetry due to the simultaneous presence of $c_{y}^R$ and $c_{z}^R$ does not occur, unlike speculated by Deng {\it et al.}\cite{Deng2015}. The absence of any ferroelectric polarization or magnetodielectric effect is also confirmed by  temperature dependent dielectric studies in presence of the magnetic field (data is not shown).
\par
However the coexistence $c_y^R$ and $c_z^R$ structures coincides with a negative volume expansion occurs below 10\,K as shown in Fig.~\ref{latticeparams}(e). Though this suggests a magneto-volume effect, the detailed analyses are beyond the scope of this paper.
\subsection{Conclusion}
In conclusion, we have investigated magnetic behavior of polycrystalline EDFO using various experimental techniques like bulk magnetization, neutron powder diffraction and specific heat, while the observed properties were correlated with theoretical estimations from density functional theory. 
At 300 K, magnetic structure belongs to ${\Gamma_4}$ configuration.
The 50${\%}$ substitution of Er$^{3+}$ and Dy$^{3+}$ results in a complex spin reorientation of EDFO. The gradual $\Gamma_4$${\rightarrow}$${\Gamma}_{2}$ reorientation of the Fe$^{3+}$ spins begin below 75\,K. 
However, the reorientation remains incomplete even at 10\,K and continue to coexist at lower temperatures. The ordering of rare-earth due to its polarization starts below 10\,K resulting in $c_{y}^R$ peak in the ${\Gamma}_{2}$ representation. 
At 2\,K, the sudden development of the $c_{z}^R$ magnetic peak occurs, which can initiate the sudden rotation of the Fe$^{3+}$ spins in the $y$ direction. 
The specific heat shows a Schottky peak of relatively larger intensity, indicating the absence of second order phase transition.
The strengths of exchange interactions estimated from density functional theory calculations suggest that the Er$^{3+}$-Fe$^{3+}$, Er$^{3+}$-Dy$^{3+}$ and Dy$^{3+}$-Dy$^{3+}$ exchange interactions are of comparable strength, while the Er$^{3+}$-Er$^{3+}$ and Dy$^{3+}$-Fe$^{3+}$ interactions are the weakest. The C-type ordering of rare-earth magnetic moments is consistent with experimental data.
The development of $c_z^R$ can be attributed to the Er$^{3+}$-Er $^{3+}$dipole interactions, while the strong anisotropy of Dy$^{3+}$ ions tend to suppress this transition to lower temperature.
\subsection{Acknowledgment}
This work was supported by the UGC-DAE Consortium for Scientific Research (CSR) and Science and Engineering Research Board (SERB) through CRS-M-228, ECR/2015/000136, respectively. We acknowledge the support from IIT Roorkee through SMILE-13 grant. AS and SR acknowledge MHRD for research fellowships. CMNK and WT acknowledge support from the Polish National Agency for Academic Exchange under the ÔPolish Returns 2019Õ programme, grant PPN/PPO/2019/1/00014"
\bibliography{EDFO_Neutron}

\begin{thebibliography}{85}%
\makeatletter
\providecommand \@ifxundefined [1]{%
 \@ifx{#1\undefined}
}%
\providecommand \@ifnum [1]{%
 \ifnum #1\expandafter \@firstoftwo
 \else \expandafter \@secondoftwo
 \fi
}%
\providecommand \@ifx [1]{%
 \ifx #1\expandafter \@firstoftwo
 \else \expandafter \@secondoftwo
 \fi
}%
\providecommand \natexlab [1]{#1}%
\providecommand \enquote  [1]{``#1''}%
\providecommand \bibnamefont  [1]{#1}%
\providecommand \bibfnamefont [1]{#1}%
\providecommand \citenamefont [1]{#1}%
\providecommand \href@noop [0]{\@secondoftwo}%
\providecommand \href [0]{\begingroup \@sanitize@url \@href}%
\providecommand \@href[1]{\@@startlink{#1}\@@href}%
\providecommand \@@href[1]{\endgroup#1\@@endlink}%
\providecommand \@sanitize@url [0]{\catcode `\\12\catcode `\$12\catcode
  `\&12\catcode `\#12\catcode `\^12\catcode `\_12\catcode `\%12\relax}%
\providecommand \@@startlink[1]{}%
\providecommand \@@endlink[0]{}%
\providecommand \url  [0]{\begingroup\@sanitize@url \@url }%
\providecommand \@url [1]{\endgroup\@href {#1}{\urlprefix }}%
\providecommand \urlprefix  [0]{URL }%
\providecommand \Eprint [0]{\href }%
\providecommand \doibase [0]{http://dx.doi.org/}%
\providecommand \selectlanguage [0]{\@gobble}%
\providecommand \bibinfo  [0]{\@secondoftwo}%
\providecommand \bibfield  [0]{\@secondoftwo}%
\providecommand \translation [1]{[#1]}%
\providecommand \BibitemOpen [0]{}%
\providecommand \bibitemStop [0]{}%
\providecommand \bibitemNoStop [0]{.\EOS\space}%
\providecommand \EOS [0]{\spacefactor3000\relax}%
\providecommand \BibitemShut  [1]{\csname bibitem#1\endcsname}%
\let\auto@bib@innerbib\@empty
\bibitem [{\citenamefont {Guo}\ \emph {et~al.}(2020)\citenamefont {Guo},
  \citenamefont {Cheng}, \citenamefont {Ren}, \citenamefont {Zhang},
  \citenamefont {Lin}, \citenamefont {Jin}, \citenamefont {Cao}, \citenamefont
  {Sheng},\ and\ \citenamefont {Ma}}]{Guo2020}%
  \BibitemOpen
  \bibfield  {author} {\bibinfo {author} {\bibfnamefont {J.}~\bibnamefont
  {Guo}}, \bibinfo {author} {\bibfnamefont {L.}~\bibnamefont {Cheng}}, \bibinfo
  {author} {\bibfnamefont {Z.}~\bibnamefont {Ren}}, \bibinfo {author}
  {\bibfnamefont {W.}~\bibnamefont {Zhang}}, \bibinfo {author} {\bibfnamefont
  {X.}~\bibnamefont {Lin}}, \bibinfo {author} {\bibfnamefont {Z.}~\bibnamefont
  {Jin}}, \bibinfo {author} {\bibfnamefont {S.}~\bibnamefont {Cao}}, \bibinfo
  {author} {\bibfnamefont {Z.}~\bibnamefont {Sheng}}, \ and\ \bibinfo {author}
  {\bibfnamefont {G.}~\bibnamefont {Ma}},\ }\href@noop {} {\bibfield  {journal}
  {\bibinfo  {journal} {Journal of Physics: Condensed Matter}\ }\textbf
  {\bibinfo {volume} {32}},\ \bibinfo {pages} {185401} (\bibinfo {year}
  {2020})}\BibitemShut {NoStop}%
\bibitem [{\citenamefont {Mikhaylovskiy}\ \emph
  {et~al.}(2015{\natexlab{a}})\citenamefont {Mikhaylovskiy}, \citenamefont
  {Huisman}, \citenamefont {Popov}, \citenamefont {Zvezdin}, \citenamefont
  {Rasing}, \citenamefont {Pisarev},\ and\ \citenamefont
  {Kimel}}]{Mikhaylovskiy2015}%
  \BibitemOpen
  \bibfield  {author} {\bibinfo {author} {\bibfnamefont {R.~V.}\ \bibnamefont
  {Mikhaylovskiy}}, \bibinfo {author} {\bibfnamefont {T.~J.}\ \bibnamefont
  {Huisman}}, \bibinfo {author} {\bibfnamefont {A.~I.}\ \bibnamefont {Popov}},
  \bibinfo {author} {\bibfnamefont {A.~K.}\ \bibnamefont {Zvezdin}}, \bibinfo
  {author} {\bibfnamefont {T.}~\bibnamefont {Rasing}}, \bibinfo {author}
  {\bibfnamefont {R.~V.}\ \bibnamefont {Pisarev}}, \ and\ \bibinfo {author}
  {\bibfnamefont {A.~V.}\ \bibnamefont {Kimel}},\ }\href@noop {} {\bibfield
  {journal} {\bibinfo  {journal} {Phys. Rev. B}\ }\textbf {\bibinfo {volume}
  {92}},\ \bibinfo {pages} {094437} (\bibinfo {year}
  {2015}{\natexlab{a}})}\BibitemShut {NoStop}%
\bibitem [{\citenamefont {Mikhaylovskiy}\ \emph
  {et~al.}(2015{\natexlab{b}})\citenamefont {Mikhaylovskiy}, \citenamefont
  {Hendry}, \citenamefont {Secchi}, \citenamefont {Mentink}, \citenamefont
  {Eckstein}, \citenamefont {Wu}, \citenamefont {Pisarev}, \citenamefont
  {Kruglyak}, \citenamefont {Katsnelson}, \citenamefont {Rasing},\ and\
  \citenamefont {Kimel}}]{Mikhaylovskiy2015b}%
  \BibitemOpen
  \bibfield  {author} {\bibinfo {author} {\bibfnamefont {R.~V.}\ \bibnamefont
  {Mikhaylovskiy}}, \bibinfo {author} {\bibfnamefont {E.}~\bibnamefont
  {Hendry}}, \bibinfo {author} {\bibfnamefont {A.}~\bibnamefont {Secchi}},
  \bibinfo {author} {\bibfnamefont {J.~H.}\ \bibnamefont {Mentink}}, \bibinfo
  {author} {\bibfnamefont {M.}~\bibnamefont {Eckstein}}, \bibinfo {author}
  {\bibfnamefont {A.}~\bibnamefont {Wu}}, \bibinfo {author} {\bibfnamefont
  {R.~V.}\ \bibnamefont {Pisarev}}, \bibinfo {author} {\bibfnamefont {V.~V.}\
  \bibnamefont {Kruglyak}}, \bibinfo {author} {\bibfnamefont {M.~I.}\
  \bibnamefont {Katsnelson}}, \bibinfo {author} {\bibfnamefont
  {T.}~\bibnamefont {Rasing}}, \ and\ \bibinfo {author} {\bibfnamefont {A.~V.}\
  \bibnamefont {Kimel}},\ }\href@noop {} {\bibfield  {journal} {\bibinfo
  {journal} {Nature Communications}\ }\textbf {\bibinfo {volume} {6}},\
  \bibinfo {pages} {8190} (\bibinfo {year} {2015}{\natexlab{b}})}\BibitemShut
  {NoStop}%
\bibitem [{\citenamefont {Mikhaylovskiy}\ \emph {et~al.}(2014)\citenamefont
  {Mikhaylovskiy}, \citenamefont {Hendry}, \citenamefont {Kruglyak},
  \citenamefont {Pisarev}, \citenamefont {Rasing},\ and\ \citenamefont
  {Kimel}}]{Mikhaylovskiy2014}%
  \BibitemOpen
  \bibfield  {author} {\bibinfo {author} {\bibfnamefont {R.~V.}\ \bibnamefont
  {Mikhaylovskiy}}, \bibinfo {author} {\bibfnamefont {E.}~\bibnamefont
  {Hendry}}, \bibinfo {author} {\bibfnamefont {V.~V.}\ \bibnamefont
  {Kruglyak}}, \bibinfo {author} {\bibfnamefont {R.~V.}\ \bibnamefont
  {Pisarev}}, \bibinfo {author} {\bibfnamefont {T.}~\bibnamefont {Rasing}}, \
  and\ \bibinfo {author} {\bibfnamefont {A.~V.}\ \bibnamefont {Kimel}},\
  }\href@noop {} {\bibfield  {journal} {\bibinfo  {journal} {Phys. Rev. B}\
  }\textbf {\bibinfo {volume} {90}},\ \bibinfo {pages} {184405} (\bibinfo
  {year} {2014})}\BibitemShut {NoStop}%
\bibitem [{\citenamefont {Yamaguchi}\ \emph {et~al.}(2013)\citenamefont
  {Yamaguchi}, \citenamefont {Kurihara}, \citenamefont {Minami}, \citenamefont
  {Nakajima},\ and\ \citenamefont {Suemoto}}]{Yamaguchi2013}%
  \BibitemOpen
  \bibfield  {author} {\bibinfo {author} {\bibfnamefont {K.}~\bibnamefont
  {Yamaguchi}}, \bibinfo {author} {\bibfnamefont {T.}~\bibnamefont {Kurihara}},
  \bibinfo {author} {\bibfnamefont {Y.}~\bibnamefont {Minami}}, \bibinfo
  {author} {\bibfnamefont {M.}~\bibnamefont {Nakajima}}, \ and\ \bibinfo
  {author} {\bibfnamefont {T.}~\bibnamefont {Suemoto}},\ }\href@noop {}
  {\bibfield  {journal} {\bibinfo  {journal} {Phys. Rev. Lett.}\ }\textbf
  {\bibinfo {volume} {110}},\ \bibinfo {pages} {137204} (\bibinfo {year}
  {2013})}\BibitemShut {NoStop}%
\bibitem [{\citenamefont {Ding}\ \emph {et~al.}(2019)\citenamefont {Ding},
  \citenamefont {Xue}, \citenamefont {Liang}, \citenamefont {Liu},
  \citenamefont {Li}, \citenamefont {Cao}, \citenamefont {Sun}, \citenamefont
  {Zhao}, \citenamefont {Yang},\ and\ \citenamefont {Yang}}]{Ding2019}%
  \BibitemOpen
  \bibfield  {author} {\bibinfo {author} {\bibfnamefont {S.}~\bibnamefont
  {Ding}}, \bibinfo {author} {\bibfnamefont {M.}~\bibnamefont {Xue}}, \bibinfo
  {author} {\bibfnamefont {Z.}~\bibnamefont {Liang}}, \bibinfo {author}
  {\bibfnamefont {Z.}~\bibnamefont {Liu}}, \bibinfo {author} {\bibfnamefont
  {R.}~\bibnamefont {Li}}, \bibinfo {author} {\bibfnamefont {S.}~\bibnamefont
  {Cao}}, \bibinfo {author} {\bibfnamefont {Y.}~\bibnamefont {Sun}}, \bibinfo
  {author} {\bibfnamefont {J.}~\bibnamefont {Zhao}}, \bibinfo {author}
  {\bibfnamefont {W.}~\bibnamefont {Yang}}, \ and\ \bibinfo {author}
  {\bibfnamefont {J.}~\bibnamefont {Yang}},\ }\href@noop {} {\bibfield
  {journal} {\bibinfo  {journal} {Journal of Physics: Condensed Matter}\
  }\textbf {\bibinfo {volume} {31}},\ \bibinfo {pages} {435801} (\bibinfo
  {year} {2019})}\BibitemShut {NoStop}%
\bibitem [{\citenamefont {Kimel}\ \emph {et~al.}(2004)\citenamefont {Kimel},
  \citenamefont {Kirilyuk}, \citenamefont {Tsvetkov}, \citenamefont {Pisarev},\
  and\ \citenamefont {Rasing}}]{Kimel2004}%
  \BibitemOpen
  \bibfield  {author} {\bibinfo {author} {\bibfnamefont {A.~V.}\ \bibnamefont
  {Kimel}}, \bibinfo {author} {\bibfnamefont {A.}~\bibnamefont {Kirilyuk}},
  \bibinfo {author} {\bibfnamefont {A.}~\bibnamefont {Tsvetkov}}, \bibinfo
  {author} {\bibfnamefont {R.~V.}\ \bibnamefont {Pisarev}}, \ and\ \bibinfo
  {author} {\bibfnamefont {T.}~\bibnamefont {Rasing}},\ }\href@noop {}
  {\bibfield  {journal} {\bibinfo  {journal} {Nature}\ }\textbf {\bibinfo
  {volume} {429}},\ \bibinfo {pages} {850} (\bibinfo {year}
  {2004})}\BibitemShut {NoStop}%
\bibitem [{\citenamefont {Deng}\ \emph {et~al.}(2015)\citenamefont {Deng},
  \citenamefont {Guo}, \citenamefont {Ren}, \citenamefont {Cao}, \citenamefont
  {Maynard-Casely}, \citenamefont {Avdeev},\ and\ \citenamefont
  {Mclntyre}}]{Deng2015}%
  \BibitemOpen
  \bibfield  {author} {\bibinfo {author} {\bibfnamefont {G.}~\bibnamefont
  {Deng}}, \bibinfo {author} {\bibfnamefont {P.}~\bibnamefont {Guo}}, \bibinfo
  {author} {\bibfnamefont {W.}~\bibnamefont {Ren}}, \bibinfo {author}
  {\bibfnamefont {S.}~\bibnamefont {Cao}}, \bibinfo {author} {\bibfnamefont
  {H.~E.}\ \bibnamefont {Maynard-Casely}}, \bibinfo {author} {\bibfnamefont
  {M.}~\bibnamefont {Avdeev}}, \ and\ \bibinfo {author} {\bibfnamefont {G.~J.}\
  \bibnamefont {Mclntyre}},\ }\href@noop {} {\bibfield  {journal} {\bibinfo
  {journal} {Journal of Applied Physics}\ }\textbf {\bibinfo {volume} {117}},\
  \bibinfo {pages} {164105} (\bibinfo {year} {2015})}\BibitemShut {NoStop}%
\bibitem [{\citenamefont {Du}\ \emph {et~al.}(2010)\citenamefont {Du},
  \citenamefont {Cheng}, \citenamefont {Wang},\ and\ \citenamefont
  {Dou}}]{Du2010}%
  \BibitemOpen
  \bibfield  {author} {\bibinfo {author} {\bibfnamefont {Y.}~\bibnamefont
  {Du}}, \bibinfo {author} {\bibfnamefont {Z.~X.}\ \bibnamefont {Cheng}},
  \bibinfo {author} {\bibfnamefont {X.~L.}\ \bibnamefont {Wang}}, \ and\
  \bibinfo {author} {\bibfnamefont {S.~X.}\ \bibnamefont {Dou}},\ }\href@noop
  {} {\bibfield  {journal} {\bibinfo  {journal} {Journal of Applied Physics}\
  }\textbf {\bibinfo {volume} {107}},\ \bibinfo {pages} {09D908} (\bibinfo
  {year} {2010})}\BibitemShut {NoStop}%
\bibitem [{\citenamefont {Tokunaga}\ \emph {et~al.}(2008)\citenamefont
  {Tokunaga}, \citenamefont {Iguchi}, \citenamefont {Arima},\ and\
  \citenamefont {Tokura}}]{Tokunaga2008}%
  \BibitemOpen
  \bibfield  {author} {\bibinfo {author} {\bibfnamefont {Y.}~\bibnamefont
  {Tokunaga}}, \bibinfo {author} {\bibfnamefont {S.}~\bibnamefont {Iguchi}},
  \bibinfo {author} {\bibfnamefont {T.}~\bibnamefont {Arima}}, \ and\ \bibinfo
  {author} {\bibfnamefont {Y.}~\bibnamefont {Tokura}},\ }\href@noop {}
  {\bibfield  {journal} {\bibinfo  {journal} {Physical Review Letters}\
  }\textbf {\bibinfo {volume} {101}},\ \bibinfo {pages} {097205} (\bibinfo
  {year} {2008})}\BibitemShut {NoStop}%
\bibitem [{\citenamefont {Yokota}\ \emph {et~al.}(2015)\citenamefont {Yokota},
  \citenamefont {Nozue}, \citenamefont {Nakamura}, \citenamefont {Hojo},
  \citenamefont {Fukunaga}, \citenamefont {Janolin}, \citenamefont {Kiat},\
  and\ \citenamefont {Fuwa}}]{Yokota2015}%
  \BibitemOpen
  \bibfield  {author} {\bibinfo {author} {\bibfnamefont {H.}~\bibnamefont
  {Yokota}}, \bibinfo {author} {\bibfnamefont {T.}~\bibnamefont {Nozue}},
  \bibinfo {author} {\bibfnamefont {S.}~\bibnamefont {Nakamura}}, \bibinfo
  {author} {\bibfnamefont {H.}~\bibnamefont {Hojo}}, \bibinfo {author}
  {\bibfnamefont {M.}~\bibnamefont {Fukunaga}}, \bibinfo {author}
  {\bibfnamefont {P.-E.}\ \bibnamefont {Janolin}}, \bibinfo {author}
  {\bibfnamefont {J.-M.}\ \bibnamefont {Kiat}}, \ and\ \bibinfo {author}
  {\bibfnamefont {A.}~\bibnamefont {Fuwa}},\ }\href@noop {} {\bibfield
  {journal} {\bibinfo  {journal} {Phys. Rev. B}\ }\textbf {\bibinfo {volume}
  {92}},\ \bibinfo {pages} {054101} (\bibinfo {year} {2015})}\BibitemShut
  {NoStop}%
\bibitem [{\citenamefont {Singh}\ \emph {et~al.}(2019)\citenamefont {Singh},
  \citenamefont {Rajput}, \citenamefont {Padmanabhan}, \citenamefont {Kedarsh},
  \citenamefont {Anas}, \citenamefont {Maitra},\ and\ \citenamefont
  {Malik}}]{Ankita2019}%
  \BibitemOpen
  \bibfield  {author} {\bibinfo {author} {\bibfnamefont {A.}~\bibnamefont
  {Singh}}, \bibinfo {author} {\bibfnamefont {S.}~\bibnamefont {Rajput}},
  \bibinfo {author} {\bibfnamefont {B.}~\bibnamefont {Padmanabhan}}, \bibinfo
  {author} {\bibfnamefont {K.}~\bibnamefont {Kedarsh}}, \bibinfo {author}
  {\bibfnamefont {M.}~\bibnamefont {Anas}}, \bibinfo {author} {\bibfnamefont
  {T.}~\bibnamefont {Maitra}}, \ and\ \bibinfo {author} {\bibfnamefont {V.~K.}\
  \bibnamefont {Malik}},\ }\href@noop {} {\bibfield  {journal} {\bibinfo
  {journal} {Journal of Physics:Condensed Matter}\ }\textbf {\bibinfo {volume}
  {31}},\ \bibinfo {pages} {355802} (\bibinfo {year} {2019})}\BibitemShut
  {NoStop}%
\bibitem [{\citenamefont {Ke}\ \emph {et~al.}(2016)\citenamefont {Ke},
  \citenamefont {Zhang}, \citenamefont {Ma},\ and\ \citenamefont
  {Cheng}}]{Ke2016}%
  \BibitemOpen
  \bibfield  {author} {\bibinfo {author} {\bibfnamefont {Y.~J.}\ \bibnamefont
  {Ke}}, \bibinfo {author} {\bibfnamefont {X.~Q.}\ \bibnamefont {Zhang}},
  \bibinfo {author} {\bibfnamefont {Y.}~\bibnamefont {Ma}}, \ and\ \bibinfo
  {author} {\bibfnamefont {Z.~H.}\ \bibnamefont {Cheng}},\ }\href@noop {}
  {\bibfield  {journal} {\bibinfo  {journal} {Scientific Reports}\ }\textbf
  {\bibinfo {volume} {6}},\ \bibinfo {pages} {19775} (\bibinfo {year}
  {2016})}\BibitemShut {NoStop}%
\bibitem [{\citenamefont {Chakraborty}\ \emph
  {et~al.}(2016{\natexlab{a}})\citenamefont {Chakraborty}, \citenamefont
  {Yadav}, \citenamefont {Elizabeth},\ and\ \citenamefont
  {Bhat}}]{Chakraborty2016}%
  \BibitemOpen
  \bibfield  {author} {\bibinfo {author} {\bibfnamefont {T.}~\bibnamefont
  {Chakraborty}}, \bibinfo {author} {\bibfnamefont {R.}~\bibnamefont {Yadav}},
  \bibinfo {author} {\bibfnamefont {S.}~\bibnamefont {Elizabeth}}, \ and\
  \bibinfo {author} {\bibfnamefont {H.~L.}\ \bibnamefont {Bhat}},\ }\href@noop
  {} {\bibfield  {journal} {\bibinfo  {journal} {Physical Chemistry Chemical
  Physics}\ }\textbf {\bibinfo {volume} {18}},\ \bibinfo {pages} {5316}
  (\bibinfo {year} {2016}{\natexlab{a}})}\BibitemShut {NoStop}%
\bibitem [{\citenamefont {Kn\'{i}\v{z}ek}\ \emph {et~al.}(2009)\citenamefont
  {Kn\'{i}\v{z}ek}, \citenamefont {Hejtm\'{a}nek}, \citenamefont {Jir\'{a}k},
  \citenamefont {Tome\v{s}}, \citenamefont {Henry},\ and\ \citenamefont
  {Gilles}}]{Karel2009}%
  \BibitemOpen
  \bibfield  {author} {\bibinfo {author} {\bibfnamefont {K.}~\bibnamefont
  {Kn\'{i}\v{z}ek}}, \bibinfo {author} {\bibfnamefont {J.}~\bibnamefont
  {Hejtm\'{a}nek}}, \bibinfo {author} {\bibfnamefont {Z.}~\bibnamefont
  {Jir\'{a}k}}, \bibinfo {author} {\bibfnamefont {P.}~\bibnamefont
  {Tome\v{s}}}, \bibinfo {author} {\bibfnamefont {P.}~\bibnamefont {Henry}}, \
  and\ \bibinfo {author} {\bibfnamefont {A.}~\bibnamefont {Gilles}},\
  }\href@noop {} {\bibfield  {journal} {\bibinfo  {journal} {Physical Review
  B}\ }\textbf {\bibinfo {volume} {79}},\ \bibinfo {pages} {134103} (\bibinfo
  {year} {2009})}\BibitemShut {NoStop}%
\bibitem [{\citenamefont {S{\l}awi{\'n}ski}\ \emph {et~al.}(2005)\citenamefont
  {S{\l}awi{\'n}ski}, \citenamefont {Przenios{\l}o}, \citenamefont
  {Sosnowska},\ and\ \citenamefont {Suard}}]{Slawinski2005}%
  \BibitemOpen
  \bibfield  {author} {\bibinfo {author} {\bibfnamefont {W.}~\bibnamefont
  {S{\l}awi{\'n}ski}}, \bibinfo {author} {\bibfnamefont {R.}~\bibnamefont
  {Przenios{\l}o}}, \bibinfo {author} {\bibfnamefont {I.}~\bibnamefont
  {Sosnowska}}, \ and\ \bibinfo {author} {\bibfnamefont {E.}~\bibnamefont
  {Suard}},\ }\href@noop {} {\bibfield  {journal} {\bibinfo  {journal} {Journal
  of Physics: Condensed Matter}\ }\textbf {\bibinfo {volume} {17}},\ \bibinfo
  {pages} {4605} (\bibinfo {year} {2005})}\BibitemShut {NoStop}%
\bibitem [{\citenamefont {Ross}\ \emph {et~al.}(2004)\citenamefont {Ross},
  \citenamefont {Zhao}, \citenamefont {Burt},\ and\ \citenamefont
  {Chaplin}}]{Ross2004}%
  \BibitemOpen
  \bibfield  {author} {\bibinfo {author} {\bibfnamefont {N.~L.}\ \bibnamefont
  {Ross}}, \bibinfo {author} {\bibfnamefont {J.}~\bibnamefont {Zhao}}, \bibinfo
  {author} {\bibfnamefont {J.~B.}\ \bibnamefont {Burt}}, \ and\ \bibinfo
  {author} {\bibfnamefont {T.~D.}\ \bibnamefont {Chaplin}},\ }\href@noop {}
  {\bibfield  {journal} {\bibinfo  {journal} {Journal of Physics: Condensed
  Matter}\ }\textbf {\bibinfo {volume} {16}},\ \bibinfo {pages} {5721}
  (\bibinfo {year} {2004})}\BibitemShut {NoStop}%
\bibitem [{\citenamefont {Glazer}(1972)}]{Glazer1972}%
  \BibitemOpen
  \bibfield  {author} {\bibinfo {author} {\bibfnamefont {A.~M.}\ \bibnamefont
  {Glazer}},\ }\href@noop {} {\bibfield  {journal} {\bibinfo  {journal} {Acta
  Crystallographica Section B}\ }\textbf {\bibinfo {volume} {28}},\ \bibinfo
  {pages} {3384} (\bibinfo {year} {1972})}\BibitemShut {NoStop}%
\bibitem [{\citenamefont {Woodward}(1997{\natexlab{a}})}]{Woodward1997a}%
  \BibitemOpen
  \bibfield  {author} {\bibinfo {author} {\bibfnamefont {P.~M.}\ \bibnamefont
  {Woodward}},\ }\href@noop {} {\bibfield  {journal} {\bibinfo  {journal} {Acta
  Crystallographica Section B}\ }\textbf {\bibinfo {volume} {53}},\ \bibinfo
  {pages} {32} (\bibinfo {year} {1997}{\natexlab{a}})}\BibitemShut {NoStop}%
\bibitem [{\citenamefont {Woodward}(1997{\natexlab{b}})}]{Woodward1997b}%
  \BibitemOpen
  \bibfield  {author} {\bibinfo {author} {\bibfnamefont {P.~M.}\ \bibnamefont
  {Woodward}},\ }\href@noop {} {\bibfield  {journal} {\bibinfo  {journal} {Acta
  Crystallographica Section B}\ }\textbf {\bibinfo {volume} {53}},\ \bibinfo
  {pages} {44} (\bibinfo {year} {1997}{\natexlab{b}})}\BibitemShut {NoStop}%
\bibitem [{\citenamefont {Marezio~M.}\ and\ \citenamefont
  {D.}(1970)}]{Marezio70}%
  \BibitemOpen
  \bibfield  {author} {\bibinfo {author} {\bibfnamefont {R.~J.~P.}\
  \bibnamefont {Marezio~M.}}\ and\ \bibinfo {author} {\bibfnamefont {D.~P.}\
  \bibnamefont {D.}},\ }\href@noop {} {\bibfield  {journal} {\bibinfo
  {journal} {Acta Crystallogr. B}\ }\textbf {\bibinfo {volume} {26}},\ \bibinfo
  {pages} {2008} (\bibinfo {year} {1970})}\BibitemShut {NoStop}%
\bibitem [{\citenamefont {Koehler}\ \emph {et~al.}(1960)\citenamefont
  {Koehler}, \citenamefont {Wollan},\ and\ \citenamefont
  {Wilkinson}}]{KOEHLER1960}%
  \BibitemOpen
  \bibfield  {author} {\bibinfo {author} {\bibfnamefont {W.~C.}\ \bibnamefont
  {Koehler}}, \bibinfo {author} {\bibfnamefont {E.~O.}\ \bibnamefont {Wollan}},
  \ and\ \bibinfo {author} {\bibfnamefont {M.~K.}\ \bibnamefont {Wilkinson}},\
  }\href@noop {} {\bibfield  {journal} {\bibinfo  {journal} {Phys. Rev.}\
  }\textbf {\bibinfo {volume} {118}},\ \bibinfo {pages} {58} (\bibinfo {year}
  {1960})}\BibitemShut {NoStop}%
\bibitem [{\citenamefont {White}(1969)}]{White1969}%
  \BibitemOpen
  \bibfield  {author} {\bibinfo {author} {\bibfnamefont {R.~L.}\ \bibnamefont
  {White}},\ }\href@noop {} {\bibfield  {journal} {\bibinfo  {journal} {Journal
  of Applied Physics}\ }\textbf {\bibinfo {volume} {40}},\ \bibinfo {pages}
  {1061} (\bibinfo {year} {1969})}\BibitemShut {NoStop}%
\bibitem [{\citenamefont {Bertaut}(1963)}]{bertaut1963magnetism}%
  \BibitemOpen
  \bibfield  {author} {\bibinfo {author} {\bibfnamefont {E.~F.}\ \bibnamefont
  {Bertaut}},\ }\href@noop {} {\emph {\bibinfo {title} {Magnetism III}}},\
  edited by\ \bibinfo {editor} {\bibfnamefont {G.~T.}\ \bibnamefont {Rado}}\
  and\ \bibinfo {editor} {\bibfnamefont {H.}~\bibnamefont {Suhl}}\ (\bibinfo
  {publisher} {Academic Press, New York},\ \bibinfo {year} {1963})\BibitemShut
  {NoStop}%
\bibitem [{\citenamefont {Yamaguchi}(1974)}]{Yamaguchi1974}%
  \BibitemOpen
  \bibfield  {author} {\bibinfo {author} {\bibfnamefont {T.}~\bibnamefont
  {Yamaguchi}},\ }\href@noop {} {\bibfield  {journal} {\bibinfo  {journal} {J.
  Phys. Chem. Solids.}\ }\textbf {\bibinfo {volume} {35}},\ \bibinfo {pages}
  {479} (\bibinfo {year} {1974})}\BibitemShut {NoStop}%
\bibitem [{\citenamefont {Bartolom{\'e}}\ \emph {et~al.}(1997)\citenamefont
  {Bartolom{\'e}}, \citenamefont {Palacios}, \citenamefont {Kuz'min},
  \citenamefont {Bartolom{\'e}}, \citenamefont {Sosnowska}, \citenamefont
  {Przenios{\l}o}, \citenamefont {Sonntag},\ and\ \citenamefont
  {Lukina}}]{bartolome1997single}%
  \BibitemOpen
  \bibfield  {author} {\bibinfo {author} {\bibfnamefont {J.}~\bibnamefont
  {Bartolom{\'e}}}, \bibinfo {author} {\bibfnamefont {E.}~\bibnamefont
  {Palacios}}, \bibinfo {author} {\bibfnamefont {M.~D.}\ \bibnamefont
  {Kuz'min}}, \bibinfo {author} {\bibfnamefont {F.}~\bibnamefont
  {Bartolom{\'e}}}, \bibinfo {author} {\bibfnamefont {I.}~\bibnamefont
  {Sosnowska}}, \bibinfo {author} {\bibfnamefont {R.}~\bibnamefont
  {Przenios{\l}o}}, \bibinfo {author} {\bibfnamefont {R.}~\bibnamefont
  {Sonntag}}, \ and\ \bibinfo {author} {\bibfnamefont {M.~M.}\ \bibnamefont
  {Lukina}},\ }\href@noop {} {\bibfield  {journal} {\bibinfo  {journal}
  {Physical Review B}\ }\textbf {\bibinfo {volume} {55}},\ \bibinfo {pages}
  {11432} (\bibinfo {year} {1997})}\BibitemShut {NoStop}%
\bibitem [{\citenamefont {Pinto}\ and\ \citenamefont
  {Shaked}(1972)}]{Pinto1972}%
  \BibitemOpen
  \bibfield  {author} {\bibinfo {author} {\bibfnamefont {H.}~\bibnamefont
  {Pinto}}\ and\ \bibinfo {author} {\bibfnamefont {H.}~\bibnamefont {Shaked}},\
  }\href@noop {} {\bibfield  {journal} {\bibinfo  {journal} {Solid State
  Communication}\ }\textbf {\bibinfo {volume} {10}},\ \bibinfo {pages} {663}
  (\bibinfo {year} {1972})}\BibitemShut {NoStop}%
\bibitem [{\citenamefont {Sosnowsk}\ \emph {et~al.}(1986)\citenamefont
  {Sosnowsk}, \citenamefont {Steichelea},\ and\ \citenamefont
  {Hewatc}}]{SOSNOWSK1986}%
  \BibitemOpen
  \bibfield  {author} {\bibinfo {author} {\bibfnamefont {I.}~\bibnamefont
  {Sosnowsk}}, \bibinfo {author} {\bibfnamefont {E.}~\bibnamefont
  {Steichelea}}, \ and\ \bibinfo {author} {\bibfnamefont {A.}~\bibnamefont
  {Hewatc}},\ }\href@noop {} {\bibfield  {journal} {\bibinfo  {journal}
  {Physica B+C}\ }\textbf {\bibinfo {volume} {136}},\ \bibinfo {pages} {394 }
  (\bibinfo {year} {1986})}\BibitemShut {NoStop}%
\bibitem [{\citenamefont {Przenios{\l}o}\ \emph {et~al.}(1996)\citenamefont
  {Przenios{\l}o}, \citenamefont {Sosnowska}, \citenamefont {Fischer},
  \citenamefont {Marti}, \citenamefont {Bartolom{\'e}}, \citenamefont
  {Bartolom{\'e}}, \citenamefont {Palacios},\ and\ \citenamefont
  {Sonntag}}]{Przeniosto1996}%
  \BibitemOpen
  \bibfield  {author} {\bibinfo {author} {\bibfnamefont {R.}~\bibnamefont
  {Przenios{\l}o}}, \bibinfo {author} {\bibfnamefont {I.}~\bibnamefont
  {Sosnowska}}, \bibinfo {author} {\bibfnamefont {P.}~\bibnamefont {Fischer}},
  \bibinfo {author} {\bibfnamefont {W.}~\bibnamefont {Marti}}, \bibinfo
  {author} {\bibfnamefont {F.}~\bibnamefont {Bartolom{\'e}}}, \bibinfo {author}
  {\bibfnamefont {J.}~\bibnamefont {Bartolom{\'e}}}, \bibinfo {author}
  {\bibfnamefont {E.}~\bibnamefont {Palacios}}, \ and\ \bibinfo {author}
  {\bibfnamefont {R.}~\bibnamefont {Sonntag}},\ }\href@noop {} {\bibfield
  {journal} {\bibinfo  {journal} {Journal of Magnetism and Magnetic Materials}\
  }\textbf {\bibinfo {volume} {160}},\ \bibinfo {pages} {370} (\bibinfo {year}
  {1996})}\BibitemShut {NoStop}%
\bibitem [{\citenamefont {Yuan}\ \emph {et~al.}(2011)\citenamefont {Yuan},
  \citenamefont {Wang}, \citenamefont {Shao}, \citenamefont {Chang},
  \citenamefont {Kang}, \citenamefont {Isikawa},\ and\ \citenamefont
  {Cao}}]{Yuan2011}%
  \BibitemOpen
  \bibfield  {author} {\bibinfo {author} {\bibfnamefont {S.}~\bibnamefont
  {Yuan}}, \bibinfo {author} {\bibfnamefont {Y.}~\bibnamefont {Wang}}, \bibinfo
  {author} {\bibfnamefont {M.}~\bibnamefont {Shao}}, \bibinfo {author}
  {\bibfnamefont {F.}~\bibnamefont {Chang}}, \bibinfo {author} {\bibfnamefont
  {B.}~\bibnamefont {Kang}}, \bibinfo {author} {\bibfnamefont {Y.}~\bibnamefont
  {Isikawa}}, \ and\ \bibinfo {author} {\bibfnamefont {S.}~\bibnamefont
  {Cao}},\ }\href@noop {} {\bibfield  {journal} {\bibinfo  {journal} {Journal
  of applied physics}\ }\textbf {\bibinfo {volume} {109}},\ \bibinfo {pages}
  {07E141} (\bibinfo {year} {2011})}\BibitemShut {NoStop}%
\bibitem [{\citenamefont {Song}\ \emph {et~al.}(2015)\citenamefont {Song},
  \citenamefont {Jiang}, \citenamefont {Kang}, \citenamefont {Zhang},
  \citenamefont {Cheng}, \citenamefont {Ma},\ and\ \citenamefont
  {Cao}}]{SONG2015}%
  \BibitemOpen
  \bibfield  {author} {\bibinfo {author} {\bibfnamefont {G.}~\bibnamefont
  {Song}}, \bibinfo {author} {\bibfnamefont {J.}~\bibnamefont {Jiang}},
  \bibinfo {author} {\bibfnamefont {B.}~\bibnamefont {Kang}}, \bibinfo {author}
  {\bibfnamefont {J.}~\bibnamefont {Zhang}}, \bibinfo {author} {\bibfnamefont
  {Z.}~\bibnamefont {Cheng}}, \bibinfo {author} {\bibfnamefont
  {G.}~\bibnamefont {Ma}}, \ and\ \bibinfo {author} {\bibfnamefont
  {S.}~\bibnamefont {Cao}},\ }\href@noop {} {\bibfield  {journal} {\bibinfo
  {journal} {Solid State Communications}\ }\textbf {\bibinfo {volume} {211}},\
  \bibinfo {pages} {47} (\bibinfo {year} {2015})}\BibitemShut {NoStop}%
\bibitem [{\citenamefont {Jiang}\ \emph {et~al.}(2016)\citenamefont {Jiang},
  \citenamefont {Song}, \citenamefont {Wang}, \citenamefont {Jin},
  \citenamefont {Tian}, \citenamefont {Lin}, \citenamefont {Han}, \citenamefont
  {Ma}, \citenamefont {Cao},\ and\ \citenamefont {Cheng}}]{Jiang2016}%
  \BibitemOpen
  \bibfield  {author} {\bibinfo {author} {\bibfnamefont {J.}~\bibnamefont
  {Jiang}}, \bibinfo {author} {\bibfnamefont {G.}~\bibnamefont {Song}},
  \bibinfo {author} {\bibfnamefont {D.}~\bibnamefont {Wang}}, \bibinfo {author}
  {\bibfnamefont {Z.}~\bibnamefont {Jin}}, \bibinfo {author} {\bibfnamefont
  {Z.}~\bibnamefont {Tian}}, \bibinfo {author} {\bibfnamefont {X.}~\bibnamefont
  {Lin}}, \bibinfo {author} {\bibfnamefont {J.}~\bibnamefont {Han}}, \bibinfo
  {author} {\bibfnamefont {G.}~\bibnamefont {Ma}}, \bibinfo {author}
  {\bibfnamefont {S.}~\bibnamefont {Cao}}, \ and\ \bibinfo {author}
  {\bibfnamefont {Z.}~\bibnamefont {Cheng}},\ }\href@noop {} {\bibfield
  {journal} {\bibinfo  {journal} {Journal of Physics: Condensed Matter}\
  }\textbf {\bibinfo {volume} {28}},\ \bibinfo {pages} {116002} (\bibinfo
  {year} {2016})}\BibitemShut {NoStop}%
\bibitem [{\citenamefont {Chen}\ \emph {et~al.}(2012)\citenamefont {Chen},
  \citenamefont {Li}, \citenamefont {Cao}, \citenamefont {Yuan}, \citenamefont
  {Hong},\ and\ \citenamefont {Zhang}}]{Chen2012}%
  \BibitemOpen
  \bibfield  {author} {\bibinfo {author} {\bibfnamefont {L.}~\bibnamefont
  {Chen}}, \bibinfo {author} {\bibfnamefont {T.}~\bibnamefont {Li}}, \bibinfo
  {author} {\bibfnamefont {S.}~\bibnamefont {Cao}}, \bibinfo {author}
  {\bibfnamefont {S.}~\bibnamefont {Yuan}}, \bibinfo {author} {\bibfnamefont
  {F.}~\bibnamefont {Hong}}, \ and\ \bibinfo {author} {\bibfnamefont
  {J.}~\bibnamefont {Zhang}},\ }\href@noop {} {\bibfield  {journal} {\bibinfo
  {journal} {Journal of applied physics}\ }\textbf {\bibinfo {volume} {111}},\
  \bibinfo {pages} {103905} (\bibinfo {year} {2012})}\BibitemShut {NoStop}%
\bibitem [{\citenamefont {Przenios{\l}o}\ \emph {et~al.}(1995)\citenamefont
  {Przenios{\l}o}, \citenamefont {Sosnowska},\ and\ \citenamefont
  {Fischer}}]{PRZENIOSLO1995}%
  \BibitemOpen
  \bibfield  {author} {\bibinfo {author} {\bibfnamefont {R.}~\bibnamefont
  {Przenios{\l}o}}, \bibinfo {author} {\bibfnamefont {I.}~\bibnamefont
  {Sosnowska}}, \ and\ \bibinfo {author} {\bibfnamefont {P.}~\bibnamefont
  {Fischer}},\ }\href@noop {} {\bibfield  {journal} {\bibinfo  {journal}
  {Journal of Magnetism and Magnetic Materials}\ }\textbf {\bibinfo {volume}
  {140-144}},\ \bibinfo {pages} {2153} (\bibinfo {year} {1995})}\BibitemShut
  {NoStop}%
\bibitem [{\citenamefont {F.~Bartolom{\'e}}\ \emph {et~al.}(1994)\citenamefont
  {F.~Bartolom{\'e}}, \citenamefont {Bartolom{\'e}}, \citenamefont {Blasco},
  \citenamefont {Garcia},\ and\ \citenamefont {Sapifia}}]{Bartolome1994}%
  \BibitemOpen
  \bibfield  {author} {\bibinfo {author} {\bibfnamefont {M.~D.}\ \bibnamefont
  {F.~Bartolom{\'e}}, \bibfnamefont {M.D.~Kuz'min}}, \bibinfo {author}
  {\bibfnamefont {J.}~\bibnamefont {Bartolom{\'e}}}, \bibinfo {author}
  {\bibfnamefont {J.}~\bibnamefont {Blasco}}, \bibinfo {author} {\bibfnamefont
  {J.}~\bibnamefont {Garcia}}, \ and\ \bibinfo {author} {\bibfnamefont
  {F.}~\bibnamefont {Sapifia}},\ }\href@noop {} {\bibfield  {journal} {\bibinfo
   {journal} {Solid State Communications}\ }\textbf {\bibinfo {volume} {91}},\
  \bibinfo {pages} {177} (\bibinfo {year} {1994})}\BibitemShut {NoStop}%
\bibitem [{\citenamefont {Bozorth}\ \emph {et~al.}(1958)\citenamefont
  {Bozorth}, \citenamefont {Kramer},\ and\ \citenamefont
  {Remeika}}]{Bozorth58}%
  \BibitemOpen
  \bibfield  {author} {\bibinfo {author} {\bibfnamefont {R.~M.}\ \bibnamefont
  {Bozorth}}, \bibinfo {author} {\bibfnamefont {V.}~\bibnamefont {Kramer}}, \
  and\ \bibinfo {author} {\bibfnamefont {J.~P.}\ \bibnamefont {Remeika}},\
  }\href@noop {} {\bibfield  {journal} {\bibinfo  {journal} {Phys. Rev. Lett.}\
  }\textbf {\bibinfo {volume} {1}},\ \bibinfo {pages} {3} (\bibinfo {year}
  {1958})}\BibitemShut {NoStop}%
\bibitem [{\citenamefont {Grant}\ and\ \citenamefont
  {Geller}(1969)}]{Grant1969}%
  \BibitemOpen
  \bibfield  {author} {\bibinfo {author} {\bibfnamefont {R.~W.}\ \bibnamefont
  {Grant}}\ and\ \bibinfo {author} {\bibfnamefont {S.}~\bibnamefont {Geller}},\
  }\href@noop {} {\bibfield  {journal} {\bibinfo  {journal} {Solid State
  Communications}\ }\textbf {\bibinfo {volume} {7}},\ \bibinfo {pages} {1291}
  (\bibinfo {year} {1969})}\BibitemShut {NoStop}%
\bibitem [{\citenamefont {Bazaliy}\ \emph {et~al.}(2004)\citenamefont
  {Bazaliy}, \citenamefont {Tsymba}, \citenamefont {Kakazei}, \citenamefont
  {Izotov},\ and\ \citenamefont {Wige}}]{Bazaliy2004}%
  \BibitemOpen
  \bibfield  {author} {\bibinfo {author} {\bibfnamefont {Y.~B.}\ \bibnamefont
  {Bazaliy}}, \bibinfo {author} {\bibfnamefont {L.~T.}\ \bibnamefont {Tsymba}},
  \bibinfo {author} {\bibfnamefont {G.~N.}\ \bibnamefont {Kakazei}}, \bibinfo
  {author} {\bibfnamefont {A.~I.}\ \bibnamefont {Izotov}}, \ and\ \bibinfo
  {author} {\bibfnamefont {P.~E.}\ \bibnamefont {Wige}},\ }\href@noop {}
  {\bibfield  {journal} {\bibinfo  {journal} {Physical Review B}\ }\textbf
  {\bibinfo {volume} {69}},\ \bibinfo {pages} {104429} (\bibinfo {year}
  {2004})}\BibitemShut {NoStop}%
\bibitem [{\citenamefont {Pinto}\ \emph {et~al.}(1971)\citenamefont {Pinto},
  \citenamefont {Shachar}, \citenamefont {Shaked},\ and\ \citenamefont
  {Shtrikman}}]{Pinto1971}%
  \BibitemOpen
  \bibfield  {author} {\bibinfo {author} {\bibfnamefont {H.}~\bibnamefont
  {Pinto}}, \bibinfo {author} {\bibfnamefont {G.}~\bibnamefont {Shachar}},
  \bibinfo {author} {\bibfnamefont {H.}~\bibnamefont {Shaked}}, \ and\ \bibinfo
  {author} {\bibfnamefont {S.}~\bibnamefont {Shtrikman}},\ }\href@noop {}
  {\bibfield  {journal} {\bibinfo  {journal} {Phys. Rev. B}\ }\textbf {\bibinfo
  {volume} {3}},\ \bibinfo {pages} {3861} (\bibinfo {year} {1971})}\BibitemShut
  {NoStop}%
\bibitem [{\citenamefont {Tsymbal}\ \emph {et~al.}(2007)\citenamefont
  {Tsymbal}, \citenamefont {Bazaliy}, \citenamefont {Derkachenko},
  \citenamefont {Kamenevand}, \citenamefont {Kakazei}, \citenamefont
  {Palomares},\ and\ \citenamefont {Wigen}}]{Tsymbal2007}%
  \BibitemOpen
  \bibfield  {author} {\bibinfo {author} {\bibfnamefont {L.~T.}\ \bibnamefont
  {Tsymbal}}, \bibinfo {author} {\bibfnamefont {Y.~B.}\ \bibnamefont
  {Bazaliy}}, \bibinfo {author} {\bibfnamefont {V.~N.}\ \bibnamefont
  {Derkachenko}}, \bibinfo {author} {\bibfnamefont {V.~I.}\ \bibnamefont
  {Kamenevand}}, \bibinfo {author} {\bibfnamefont {G.~N.}\ \bibnamefont
  {Kakazei}}, \bibinfo {author} {\bibfnamefont {F.~J.}\ \bibnamefont
  {Palomares}}, \ and\ \bibinfo {author} {\bibfnamefont {P.~E.}\ \bibnamefont
  {Wigen}},\ }\href@noop {} {\bibfield  {journal} {\bibinfo  {journal} {Journal
  of applied physics}\ }\textbf {\bibinfo {volume} {101}},\ \bibinfo {pages}
  {123919} (\bibinfo {year} {2007})}\BibitemShut {NoStop}%
\bibitem [{\citenamefont {Gorodetsky}\ \emph {et~al.}(1973)\citenamefont
  {Gorodetsky}, \citenamefont {Hornreich}, \citenamefont {Yaeger},
  \citenamefont {Pinto}, \citenamefont {Shachar},\ and\ \citenamefont
  {Shaked}}]{Gorodetsky1973}%
  \BibitemOpen
  \bibfield  {author} {\bibinfo {author} {\bibfnamefont {G.}~\bibnamefont
  {Gorodetsky}}, \bibinfo {author} {\bibfnamefont {R.~M.}\ \bibnamefont
  {Hornreich}}, \bibinfo {author} {\bibfnamefont {I.}~\bibnamefont {Yaeger}},
  \bibinfo {author} {\bibfnamefont {H.}~\bibnamefont {Pinto}}, \bibinfo
  {author} {\bibfnamefont {G.}~\bibnamefont {Shachar}}, \ and\ \bibinfo
  {author} {\bibfnamefont {H.}~\bibnamefont {Shaked}},\ }\href@noop {}
  {\bibfield  {journal} {\bibinfo  {journal} {Phys. Rev. B}\ }\textbf {\bibinfo
  {volume} {8}},\ \bibinfo {pages} {3398} (\bibinfo {year} {1973})}\BibitemShut
  {NoStop}%
\bibitem [{\citenamefont {Gorodetsky}\ \emph {et~al.}(1968)\citenamefont
  {Gorodetsky}, \citenamefont {Sharon},\ and\ \citenamefont
  {Shtrikman}}]{Gorodetsky1968}%
  \BibitemOpen
  \bibfield  {author} {\bibinfo {author} {\bibfnamefont {G.}~\bibnamefont
  {Gorodetsky}}, \bibinfo {author} {\bibfnamefont {B.}~\bibnamefont {Sharon}},
  \ and\ \bibinfo {author} {\bibfnamefont {S.}~\bibnamefont {Shtrikman}},\
  }\href@noop {} {\bibfield  {journal} {\bibinfo  {journal} {Journal of Applied
  Physics}\ }\textbf {\bibinfo {volume} {39}},\ \bibinfo {pages} {1371}
  (\bibinfo {year} {1968})}\BibitemShut {NoStop}%
\bibitem [{\citenamefont {Prelorendjo}\ \emph {et~al.}(1980)\citenamefont
  {Prelorendjo}, \citenamefont {Johnson}, \citenamefont {Thomas},\ and\
  \citenamefont {Wanklyn}}]{Prelorendjo1980}%
  \BibitemOpen
  \bibfield  {author} {\bibinfo {author} {\bibfnamefont {L.~A.}\ \bibnamefont
  {Prelorendjo}}, \bibinfo {author} {\bibfnamefont {C.~E.}\ \bibnamefont
  {Johnson}}, \bibinfo {author} {\bibfnamefont {M.~F.}\ \bibnamefont {Thomas}},
  \ and\ \bibinfo {author} {\bibfnamefont {B.~M.}\ \bibnamefont {Wanklyn}},\
  }\href@noop {} {\bibfield  {journal} {\bibinfo  {journal} {Journal of Physics
  C: Solid State Physics}\ }\textbf {\bibinfo {volume} {13}},\ \bibinfo {pages}
  {2567} (\bibinfo {year} {1980})}\BibitemShut {NoStop}%
\bibitem [{\citenamefont {Zhao}\ \emph {et~al.}(2014)\citenamefont {Zhao},
  \citenamefont {Zhao}, \citenamefont {Zhou}, \citenamefont {Zhang},
  \citenamefont {Li}, \citenamefont {Fan}, \citenamefont {Sun},\ and\
  \citenamefont {Li}}]{Zhao2014}%
  \BibitemOpen
  \bibfield  {author} {\bibinfo {author} {\bibfnamefont {Z.~Y.}\ \bibnamefont
  {Zhao}}, \bibinfo {author} {\bibfnamefont {X.}~\bibnamefont {Zhao}}, \bibinfo
  {author} {\bibfnamefont {H.~D.}\ \bibnamefont {Zhou}}, \bibinfo {author}
  {\bibfnamefont {F.~B.}\ \bibnamefont {Zhang}}, \bibinfo {author}
  {\bibfnamefont {Q.~J.}\ \bibnamefont {Li}}, \bibinfo {author} {\bibfnamefont
  {C.}~\bibnamefont {Fan}}, \bibinfo {author} {\bibfnamefont {X.~F.}\
  \bibnamefont {Sun}}, \ and\ \bibinfo {author} {\bibfnamefont {X.~G.}\
  \bibnamefont {Li}},\ }\href@noop {} {\bibfield  {journal} {\bibinfo
  {journal} {Physical Review B}\ }\textbf {\bibinfo {volume} {89}},\ \bibinfo
  {pages} {224205} (\bibinfo {year} {2014})}\BibitemShut {NoStop}%
\bibitem [{\citenamefont {Wang}\ \emph {et~al.}(2016)\citenamefont {Wang},
  \citenamefont {Liu}, \citenamefont {Sheng}, \citenamefont {Luo},
  \citenamefont {Ye}, \citenamefont {Zhao}, \citenamefont {Sun}, \citenamefont
  {Danilkin}, \citenamefont {Deng},\ and\ \citenamefont {Bao}}]{Wang2016}%
  \BibitemOpen
  \bibfield  {author} {\bibinfo {author} {\bibfnamefont {J.}~\bibnamefont
  {Wang}}, \bibinfo {author} {\bibfnamefont {J.}~\bibnamefont {Liu}}, \bibinfo
  {author} {\bibfnamefont {J.}~\bibnamefont {Sheng}}, \bibinfo {author}
  {\bibfnamefont {W.}~\bibnamefont {Luo}}, \bibinfo {author} {\bibfnamefont
  {F.}~\bibnamefont {Ye}}, \bibinfo {author} {\bibfnamefont {Z.}~\bibnamefont
  {Zhao}}, \bibinfo {author} {\bibfnamefont {X.}~\bibnamefont {Sun}}, \bibinfo
  {author} {\bibfnamefont {S.~A.}\ \bibnamefont {Danilkin}}, \bibinfo {author}
  {\bibfnamefont {G.}~\bibnamefont {Deng}}, \ and\ \bibinfo {author}
  {\bibfnamefont {W.}~\bibnamefont {Bao}},\ }\href@noop {} {\bibfield
  {journal} {\bibinfo  {journal} {Physical Review B}\ }\textbf {\bibinfo
  {volume} {93}},\ \bibinfo {pages} {140403} (\bibinfo {year}
  {2016})}\BibitemShut {NoStop}%
\bibitem [{\citenamefont {Morin}(1950)}]{Morin50}%
  \BibitemOpen
  \bibfield  {author} {\bibinfo {author} {\bibfnamefont {F.~J.}\ \bibnamefont
  {Morin}},\ }\href@noop {} {\bibfield  {journal} {\bibinfo  {journal} {Phys.
  Rev.}\ }\textbf {\bibinfo {volume} {78}},\ \bibinfo {pages} {819} (\bibinfo
  {year} {1950})}\BibitemShut {NoStop}%
\bibitem [{\citenamefont {Yamaguchi}(1973)}]{Yamaguchi1973}%
  \BibitemOpen
  \bibfield  {author} {\bibinfo {author} {\bibfnamefont {T.}~\bibnamefont
  {Yamaguchi}},\ }\href@noop {} {\bibfield  {journal} {\bibinfo  {journal}
  {Physical review B}\ }\textbf {\bibinfo {volume} {8}},\ \bibinfo {pages}
  {5187} (\bibinfo {year} {1973})}\BibitemShut {NoStop}%
\bibitem [{\citenamefont {Berton}\ and\ \citenamefont
  {Sharon}(1968)}]{Berton1968}%
  \BibitemOpen
  \bibfield  {author} {\bibinfo {author} {\bibfnamefont {A.}~\bibnamefont
  {Berton}}\ and\ \bibinfo {author} {\bibfnamefont {B.}~\bibnamefont
  {Sharon}},\ }\href@noop {} {\bibfield  {journal} {\bibinfo  {journal}
  {Journal of Applied Physics}\ }\textbf {\bibinfo {volume} {39}},\ \bibinfo
  {pages} {1367} (\bibinfo {year} {1968})}\BibitemShut {NoStop}%
\bibitem [{\citenamefont {Nowik}\ and\ \citenamefont
  {Williams}(1966)}]{NOWIK1966}%
  \BibitemOpen
  \bibfield  {author} {\bibinfo {author} {\bibfnamefont {I.}~\bibnamefont
  {Nowik}}\ and\ \bibinfo {author} {\bibfnamefont {H.}~\bibnamefont
  {Williams}},\ }\href@noop {} {\bibfield  {journal} {\bibinfo  {journal}
  {Physics Letters}\ }\textbf {\bibinfo {volume} {20}},\ \bibinfo {pages} {154}
  (\bibinfo {year} {1966})}\BibitemShut {NoStop}%
\bibitem [{\citenamefont {Belov}\ \emph {et~al.}(1968)\citenamefont {Belov},
  \citenamefont {Kadomtseva}, \citenamefont {Ledneva}, \citenamefont
  {Ovchinnikova}, \citenamefont {Panomarev},\ and\ \citenamefont
  {Timofeeva}}]{Belov1968}%
  \BibitemOpen
  \bibfield  {author} {\bibinfo {author} {\bibfnamefont {K.~P.}\ \bibnamefont
  {Belov}}, \bibinfo {author} {\bibfnamefont {A.~M.}\ \bibnamefont
  {Kadomtseva}}, \bibinfo {author} {\bibfnamefont {L.~M.}\ \bibnamefont
  {Ledneva}}, \bibinfo {author} {\bibfnamefont {T.~L.}\ \bibnamefont
  {Ovchinnikova}}, \bibinfo {author} {\bibfnamefont {Y.~G.}\ \bibnamefont
  {Panomarev}}, \ and\ \bibinfo {author} {\bibfnamefont {V.~A.}\ \bibnamefont
  {Timofeeva}},\ }\href@noop {} {\bibfield  {journal} {\bibinfo  {journal}
  {Soviet Physics-Solid State}\ }\textbf {\bibinfo {volume} {9}},\ \bibinfo
  {pages} {2190} (\bibinfo {year} {1968})}\BibitemShut {NoStop}%
\bibitem [{\citenamefont {Holmes}\ \emph {et~al.}(1972)\citenamefont {Holmes},
  \citenamefont {Uitert}, \citenamefont {Hecker},\ and\ \citenamefont
  {Hull}}]{Holmes1972}%
  \BibitemOpen
  \bibfield  {author} {\bibinfo {author} {\bibfnamefont {L.~M.}\ \bibnamefont
  {Holmes}}, \bibinfo {author} {\bibfnamefont {L.~G.~V.}\ \bibnamefont
  {Uitert}}, \bibinfo {author} {\bibfnamefont {R.~R.}\ \bibnamefont {Hecker}},
  \ and\ \bibinfo {author} {\bibfnamefont {G.~W.}\ \bibnamefont {Hull}},\
  }\href@noop {} {\bibfield  {journal} {\bibinfo  {journal} {Physical Review
  B}\ }\textbf {\bibinfo {volume} {5}},\ \bibinfo {pages} {138} (\bibinfo
  {year} {1972})}\BibitemShut {NoStop}%
\bibitem [{\citenamefont {Krynetskii}\ and\ \citenamefont
  {Matveev}(1997)}]{Krynetskii1997}%
  \BibitemOpen
  \bibfield  {author} {\bibinfo {author} {\bibfnamefont {I.~B.}\ \bibnamefont
  {Krynetskii}}\ and\ \bibinfo {author} {\bibfnamefont {V.~M.}\ \bibnamefont
  {Matveev}},\ }\href@noop {} {\bibfield  {journal} {\bibinfo  {journal} {Phys.
  Solid State}\ }\textbf {\bibinfo {volume} {39}},\ \bibinfo {pages} {584}
  (\bibinfo {year} {1997})}\BibitemShut {NoStop}%
\bibitem [{\citenamefont {Wu}\ \emph {et~al.}(2017)\citenamefont {Wu},
  \citenamefont {Nikitin}, \citenamefont {Frontzek}, \citenamefont
  {Kolesnikov}, \citenamefont {Ehlers}, \citenamefont {Lumsden}, \citenamefont
  {Shaykhutdinov}, \citenamefont {Guo}, \citenamefont {Savici}, \citenamefont
  {Gai}, \citenamefont {Sefat},\ and\ \citenamefont {Podlesnyak}}]{Wu2017}%
  \BibitemOpen
  \bibfield  {author} {\bibinfo {author} {\bibfnamefont {L.~S.}\ \bibnamefont
  {Wu}}, \bibinfo {author} {\bibfnamefont {S.~E.}\ \bibnamefont {Nikitin}},
  \bibinfo {author} {\bibfnamefont {M.}~\bibnamefont {Frontzek}}, \bibinfo
  {author} {\bibfnamefont {A.~I.}\ \bibnamefont {Kolesnikov}}, \bibinfo
  {author} {\bibfnamefont {G.}~\bibnamefont {Ehlers}}, \bibinfo {author}
  {\bibfnamefont {M.~D.}\ \bibnamefont {Lumsden}}, \bibinfo {author}
  {\bibfnamefont {K.~A.}\ \bibnamefont {Shaykhutdinov}}, \bibinfo {author}
  {\bibfnamefont {E.~J.}\ \bibnamefont {Guo}}, \bibinfo {author} {\bibfnamefont
  {A.~T.}\ \bibnamefont {Savici}}, \bibinfo {author} {\bibfnamefont
  {Z.}~\bibnamefont {Gai}}, \bibinfo {author} {\bibfnamefont {A.~S.}\
  \bibnamefont {Sefat}}, \ and\ \bibinfo {author} {\bibfnamefont
  {A.}~\bibnamefont {Podlesnyak}},\ }\href@noop {} {\bibfield  {journal}
  {\bibinfo  {journal} {Physical Review B}\ }\textbf {\bibinfo {volume} {96}},\
  \bibinfo {pages} {144407} (\bibinfo {year} {2017})}\BibitemShut {NoStop}%
\bibitem [{\citenamefont {Zvezdin}\ and\ \citenamefont
  {Mukhin}(2009)}]{Zvezdin2009}%
  \BibitemOpen
  \bibfield  {author} {\bibinfo {author} {\bibfnamefont {A.~K.}\ \bibnamefont
  {Zvezdin}}\ and\ \bibinfo {author} {\bibfnamefont {A.~A.}\ \bibnamefont
  {Mukhin}},\ }\href@noop {} {\bibfield  {journal} {\bibinfo  {journal} {JETP
  Letters}\ }\textbf {\bibinfo {volume} {88}},\ \bibinfo {pages} {505}
  (\bibinfo {year} {2009})}\BibitemShut {NoStop}%
\bibitem [{\citenamefont {Rajeswaran}\ \emph {et~al.}(2013)\citenamefont
  {Rajeswaran}, \citenamefont {Sanyal}, \citenamefont {Mahuya}, \citenamefont
  {Sundarayya}, \citenamefont {Sundaresan},\ and\ \citenamefont
  {Rao}}]{Rajeswaran2013}%
  \BibitemOpen
  \bibfield  {author} {\bibinfo {author} {\bibfnamefont {B.}~\bibnamefont
  {Rajeswaran}}, \bibinfo {author} {\bibfnamefont {D.}~\bibnamefont {Sanyal}},
  \bibinfo {author} {\bibfnamefont {C.}~\bibnamefont {Mahuya}}, \bibinfo
  {author} {\bibfnamefont {Y.}~\bibnamefont {Sundarayya}}, \bibinfo {author}
  {\bibfnamefont {A.}~\bibnamefont {Sundaresan}}, \ and\ \bibinfo {author}
  {\bibfnamefont {C.~N.~R.}\ \bibnamefont {Rao}},\ }\href@noop {} {\bibfield
  {journal} {\bibinfo  {journal} {Euro Physics Letters}\ }\textbf {\bibinfo
  {volume} {101}},\ \bibinfo {pages} {17001} (\bibinfo {year}
  {2013})}\BibitemShut {NoStop}%
\bibitem [{\citenamefont {Ankita}\ \emph {et~al.}(2017)\citenamefont {Ankita},
  \citenamefont {Jain}, \citenamefont {Avijeet}, \citenamefont {Padmanabhan},
  \citenamefont {Ruchika}, \citenamefont {Vivian}, \citenamefont {Sajid},
  \citenamefont {Yusuf}, \citenamefont {Maitra},\ and\ \citenamefont
  {Malik}}]{Ankita2017}%
  \BibitemOpen
  \bibfield  {author} {\bibinfo {author} {\bibfnamefont {S.}~\bibnamefont
  {Ankita}}, \bibinfo {author} {\bibfnamefont {A.}~\bibnamefont {Jain}},
  \bibinfo {author} {\bibfnamefont {R.}~\bibnamefont {Avijeet}}, \bibinfo
  {author} {\bibfnamefont {B.}~\bibnamefont {Padmanabhan}}, \bibinfo {author}
  {\bibfnamefont {Y.}~\bibnamefont {Ruchika}}, \bibinfo {author} {\bibfnamefont
  {N.}~\bibnamefont {Vivian}}, \bibinfo {author} {\bibfnamefont
  {H.}~\bibnamefont {Sajid}}, \bibinfo {author} {\bibfnamefont {S.~M.}\
  \bibnamefont {Yusuf}}, \bibinfo {author} {\bibfnamefont {T.}~\bibnamefont
  {Maitra}}, \ and\ \bibinfo {author} {\bibfnamefont {V.~K.}\ \bibnamefont
  {Malik}},\ }\href@noop {} {\bibfield  {journal} {\bibinfo  {journal}
  {Physical review B}\ }\textbf {\bibinfo {volume} {96}},\ \bibinfo {pages}
  {144420} (\bibinfo {year} {2017})}\BibitemShut {NoStop}%
\bibitem [{\citenamefont {Tokunaga}\ \emph {et~al.}(2014)\citenamefont
  {Tokunaga}, \citenamefont {Taguchi}, \citenamefont {Arima},\ and\
  \citenamefont {Tokura}}]{Tokunaga2014}%
  \BibitemOpen
  \bibfield  {author} {\bibinfo {author} {\bibfnamefont {Y.}~\bibnamefont
  {Tokunaga}}, \bibinfo {author} {\bibfnamefont {Y.}~\bibnamefont {Taguchi}},
  \bibinfo {author} {\bibfnamefont {T.}~\bibnamefont {Arima}}, \ and\ \bibinfo
  {author} {\bibfnamefont {Y.}~\bibnamefont {Tokura}},\ }\href {\doibase
  10.1103/PhysRevLett.112.037203} {\bibfield  {journal} {\bibinfo  {journal}
  {Phys. Rev. Lett.}\ }\textbf {\bibinfo {volume} {112}},\ \bibinfo {pages}
  {037203} (\bibinfo {year} {2014})}\BibitemShut {NoStop}%
\bibitem [{\citenamefont {M.Mihalik}\ \emph {et~al.}(2013)\citenamefont
  {M.Mihalik}, \citenamefont {Mihalik}, \citenamefont {Fitta}, \citenamefont
  {Ba{\l}anda}, \citenamefont {Vavra}, \citenamefont {Gab{\'{a}}ni},
  \citenamefont {Zentkov{\'{a}}},\ and\ \citenamefont
  {Brian{\v{c}}in}}]{MIHALIK2013}%
  \BibitemOpen
  \bibfield  {author} {\bibinfo {author} {\bibnamefont {M.Mihalik}}, \bibinfo
  {author} {\bibfnamefont {M.}~\bibnamefont {Mihalik}}, \bibinfo {author}
  {\bibfnamefont {M.}~\bibnamefont {Fitta}}, \bibinfo {author} {\bibfnamefont
  {M.}~\bibnamefont {Ba{\l}anda}}, \bibinfo {author} {\bibfnamefont
  {M.}~\bibnamefont {Vavra}}, \bibinfo {author} {\bibfnamefont
  {S.}~\bibnamefont {Gab{\'{a}}ni}}, \bibinfo {author} {\bibfnamefont
  {M.}~\bibnamefont {Zentkov{\'{a}}}}, \ and\ \bibinfo {author} {\bibfnamefont
  {J.}~\bibnamefont {Brian{\v{c}}in}},\ }\href@noop {} {\bibfield  {journal}
  {\bibinfo  {journal} {Journal of Magnetism and Magnetic Materials}\ }\textbf
  {\bibinfo {volume} {345}},\ \bibinfo {pages} {125} (\bibinfo {year}
  {2013})}\BibitemShut {NoStop}%
\bibitem [{\citenamefont {Chakraborty}\ and\ \citenamefont
  {Elizabeth}(2018{\natexlab{a}})}]{CHAKRABORTY2018}%
  \BibitemOpen
  \bibfield  {author} {\bibinfo {author} {\bibfnamefont {T.}~\bibnamefont
  {Chakraborty}}\ and\ \bibinfo {author} {\bibfnamefont {S.}~\bibnamefont
  {Elizabeth}},\ }\href@noop {} {\bibfield  {journal} {\bibinfo  {journal}
  {Journal of Magnetism and Magnetic Materials}\ }\textbf {\bibinfo {volume}
  {462}},\ \bibinfo {pages} {78} (\bibinfo {year}
  {2018}{\natexlab{a}})}\BibitemShut {NoStop}%
\bibitem [{\citenamefont {Wu}\ \emph {et~al.}(2014)\citenamefont {Wu},
  \citenamefont {Cao}, \citenamefont {Liu}, \citenamefont {Cao}, \citenamefont
  {Kang}, \citenamefont {Zhang},\ and\ \citenamefont {Ren}}]{Wu2014}%
  \BibitemOpen
  \bibfield  {author} {\bibinfo {author} {\bibfnamefont {H.}~\bibnamefont
  {Wu}}, \bibinfo {author} {\bibfnamefont {S.}~\bibnamefont {Cao}}, \bibinfo
  {author} {\bibfnamefont {M.}~\bibnamefont {Liu}}, \bibinfo {author}
  {\bibfnamefont {Y.}~\bibnamefont {Cao}}, \bibinfo {author} {\bibfnamefont
  {B.}~\bibnamefont {Kang}}, \bibinfo {author} {\bibfnamefont {J.}~\bibnamefont
  {Zhang}}, \ and\ \bibinfo {author} {\bibfnamefont {W.}~\bibnamefont {Ren}},\
  }\href@noop {} {\bibfield  {journal} {\bibinfo  {journal} {Physical review
  B}\ }\textbf {\bibinfo {volume} {90}},\ \bibinfo {pages} {144415} (\bibinfo
  {year} {2014})}\BibitemShut {NoStop}%
\bibitem [{\citenamefont {Lazurova}\ \emph {et~al.}(2015)\citenamefont
  {Lazurova}, \citenamefont {Mihalik}, \citenamefont {jr.}, \citenamefont
  {Vavra1}, \citenamefont {Zentkova1}, \citenamefont {Briancin}, \citenamefont
  {Perovic}, \citenamefont {Kusigerski}, \citenamefont {Schneeweiss},
  \citenamefont {Roupcova}, \citenamefont {Kamenev}, \citenamefont {Misek},\
  and\ \citenamefont {Jaglicic}}]{Lazurova2015}%
  \BibitemOpen
  \bibfield  {author} {\bibinfo {author} {\bibfnamefont {J.}~\bibnamefont
  {Lazurova}}, \bibinfo {author} {\bibfnamefont {M.}~\bibnamefont {Mihalik}},
  \bibinfo {author} {\bibfnamefont {M.~M.}\ \bibnamefont {jr.}}, \bibinfo
  {author} {\bibfnamefont {M.}~\bibnamefont {Vavra1}}, \bibinfo {author}
  {\bibfnamefont {M.}~\bibnamefont {Zentkova1}}, \bibinfo {author}
  {\bibfnamefont {J.}~\bibnamefont {Briancin}}, \bibinfo {author}
  {\bibfnamefont {M.}~\bibnamefont {Perovic}}, \bibinfo {author} {\bibfnamefont
  {V.}~\bibnamefont {Kusigerski}}, \bibinfo {author} {\bibfnamefont
  {O.}~\bibnamefont {Schneeweiss}}, \bibinfo {author} {\bibfnamefont
  {P.}~\bibnamefont {Roupcova}}, \bibinfo {author} {\bibfnamefont {K.~V.}\
  \bibnamefont {Kamenev}}, \bibinfo {author} {\bibfnamefont {M.}~\bibnamefont
  {Misek}}, \ and\ \bibinfo {author} {\bibfnamefont {Z.}~\bibnamefont
  {Jaglicic}},\ }\href@noop {} {\bibfield  {journal} {\bibinfo  {journal}
  {Journal of Physics: Conference Series}\ }\textbf {\bibinfo {volume} {592}},\
  \bibinfo {pages} {012117} (\bibinfo {year} {2015})}\BibitemShut {NoStop}%
\bibitem [{\citenamefont {Chakraborty}\ \emph
  {et~al.}(2016{\natexlab{b}})\citenamefont {Chakraborty}, \citenamefont
  {Yadav}, \citenamefont {Elizabeth},\ and\ \citenamefont {Bhat}}]{Tirtha2016}%
  \BibitemOpen
  \bibfield  {author} {\bibinfo {author} {\bibfnamefont {T.}~\bibnamefont
  {Chakraborty}}, \bibinfo {author} {\bibfnamefont {R.}~\bibnamefont {Yadav}},
  \bibinfo {author} {\bibfnamefont {S.}~\bibnamefont {Elizabeth}}, \ and\
  \bibinfo {author} {\bibfnamefont {H.~L.}\ \bibnamefont {Bhat}},\ }\href@noop
  {} {\bibfield  {journal} {\bibinfo  {journal} {Physical Chemistry and
  Chemical Physics}\ }\textbf {\bibinfo {volume} {18}},\ \bibinfo {pages} {536}
  (\bibinfo {year} {2016}{\natexlab{b}})}\BibitemShut {NoStop}%
\bibitem [{\citenamefont {Nair}\ \emph {et~al.}(2016)\citenamefont {Nair},
  \citenamefont {Chatterji}, \citenamefont {Kumar}, \citenamefont {Hansen},
  \citenamefont {Nhalil}, \citenamefont {Elizabeth},\ and\ \citenamefont
  {M.Strydom}}]{Harikrishnan2016}%
  \BibitemOpen
  \bibfield  {author} {\bibinfo {author} {\bibfnamefont {H.~S.}\ \bibnamefont
  {Nair}}, \bibinfo {author} {\bibfnamefont {T.}~\bibnamefont {Chatterji}},
  \bibinfo {author} {\bibfnamefont {C.~M.~N.}\ \bibnamefont {Kumar}}, \bibinfo
  {author} {\bibfnamefont {T.}~\bibnamefont {Hansen}}, \bibinfo {author}
  {\bibfnamefont {H.}~\bibnamefont {Nhalil}}, \bibinfo {author} {\bibfnamefont
  {S.}~\bibnamefont {Elizabeth}}, \ and\ \bibinfo {author} {\bibfnamefont
  {A.}~\bibnamefont {M.Strydom}},\ }\href@noop {} {\bibfield  {journal}
  {\bibinfo  {journal} {Journal of Applied Physics}\ }\textbf {\bibinfo
  {volume} {119}},\ \bibinfo {pages} {053901} (\bibinfo {year}
  {2016})}\BibitemShut {NoStop}%
\bibitem [{\citenamefont {Chakraborty}\ and\ \citenamefont
  {Elizabeth}(2018{\natexlab{b}})}]{Tirtha2018}%
  \BibitemOpen
  \bibfield  {author} {\bibinfo {author} {\bibfnamefont {T.}~\bibnamefont
  {Chakraborty}}\ and\ \bibinfo {author} {\bibfnamefont {S.}~\bibnamefont
  {Elizabeth}},\ }\href@noop {} {\bibfield  {journal} {\bibinfo  {journal}
  {Journal of Magnetism and Magnetic Materials}\ }\textbf {\bibinfo {volume}
  {462}},\ \bibinfo {pages} {78} (\bibinfo {year}
  {2018}{\natexlab{b}})}\BibitemShut {NoStop}%
\bibitem [{\citenamefont {Rietveld}(1969)}]{rietveld1969profile}%
  \BibitemOpen
  \bibfield  {author} {\bibinfo {author} {\bibfnamefont {H.}~\bibnamefont
  {Rietveld}},\ }\href@noop {} {\bibfield  {journal} {\bibinfo  {journal}
  {Journal of Applied Crystallography}\ }\textbf {\bibinfo {volume} {2}},\
  \bibinfo {pages} {65} (\bibinfo {year} {1969})}\BibitemShut {NoStop}%
\bibitem [{\citenamefont {Rodriguez-Carvajal}(1990)}]{rodriguez1990fullprof}%
  \BibitemOpen
  \bibfield  {author} {\bibinfo {author} {\bibfnamefont {J.}~\bibnamefont
  {Rodriguez-Carvajal}},\ }in\ \href@noop {} {\emph {\bibinfo {booktitle}
  {satellite meeting on powder diffraction of the XV congress of the IUCr}}},\
  Vol.\ \bibinfo {volume} {127}\ (\bibinfo {organization} {Toulouse,
  France:[sn]},\ \bibinfo {year} {1990})\BibitemShut {NoStop}%
\bibitem [{\citenamefont {Hovestreydt}\ \emph {et~al.}(1992)\citenamefont
  {Hovestreydt}, \citenamefont {Aroyo}, \citenamefont {Sattler},\ and\
  \citenamefont {Wondratschek}}]{Hovestreydt92}%
  \BibitemOpen
  \bibfield  {author} {\bibinfo {author} {\bibfnamefont {E.}~\bibnamefont
  {Hovestreydt}}, \bibinfo {author} {\bibfnamefont {M.}~\bibnamefont {Aroyo}},
  \bibinfo {author} {\bibfnamefont {S.}~\bibnamefont {Sattler}}, \ and\
  \bibinfo {author} {\bibfnamefont {H.}~\bibnamefont {Wondratschek}},\
  }\href@noop {} {\bibfield  {journal} {\bibinfo  {journal} {Journal of Applied
  Crystallography}\ }\textbf {\bibinfo {volume} {25}},\ \bibinfo {pages} {544}
  (\bibinfo {year} {1992})}\BibitemShut {NoStop}%
\bibitem [{\citenamefont {Kresse}\ and\ \citenamefont
  {Furthm{\"u}ller}(1996)}]{kresse1996efficient}%
  \BibitemOpen
  \bibfield  {author} {\bibinfo {author} {\bibfnamefont {G.}~\bibnamefont
  {Kresse}}\ and\ \bibinfo {author} {\bibfnamefont {J.}~\bibnamefont
  {Furthm{\"u}ller}},\ }\href@noop {} {\bibfield  {journal} {\bibinfo
  {journal} {Physical review B}\ }\textbf {\bibinfo {volume} {54}},\ \bibinfo
  {pages} {11169} (\bibinfo {year} {1996})}\BibitemShut {NoStop}%
\bibitem [{\citenamefont {Perdew}\ \emph {et~al.}(1996)\citenamefont {Perdew},
  \citenamefont {Burke},\ and\ \citenamefont {Ernzerhof}}]{perdew1996}%
  \BibitemOpen
  \bibfield  {author} {\bibinfo {author} {\bibfnamefont {J.~P.}\ \bibnamefont
  {Perdew}}, \bibinfo {author} {\bibfnamefont {K.}~\bibnamefont {Burke}}, \
  and\ \bibinfo {author} {\bibfnamefont {M.}~\bibnamefont {Ernzerhof}},\
  }\href@noop {} {\bibfield  {journal} {\bibinfo  {journal} {Physical Review
  Letters}\ }\textbf {\bibinfo {volume} {77}},\ \bibinfo {pages} {3865}
  (\bibinfo {year} {1996})}\BibitemShut {NoStop}%
\bibitem [{\citenamefont {Anisimov}\ \emph {et~al.}(1993)\citenamefont
  {Anisimov}, \citenamefont {Solovyev}, \citenamefont {Korotin}, \citenamefont
  {Czyzyk},\ and\ \citenamefont {Sawatzky}}]{anisimov}%
  \BibitemOpen
  \bibfield  {author} {\bibinfo {author} {\bibfnamefont {V.~I.}\ \bibnamefont
  {Anisimov}}, \bibinfo {author} {\bibfnamefont {I.~V.}\ \bibnamefont
  {Solovyev}}, \bibinfo {author} {\bibfnamefont {M.~A.}\ \bibnamefont
  {Korotin}}, \bibinfo {author} {\bibfnamefont {M.~T.}\ \bibnamefont {Czyzyk}},
  \ and\ \bibinfo {author} {\bibfnamefont {G.~A.}\ \bibnamefont {Sawatzky}},\
  }\href@noop {} {\bibfield  {journal} {\bibinfo  {journal} {Physical Review
  B}\ }\textbf {\bibinfo {volume} {48}},\ \bibinfo {pages} {16929} (\bibinfo
  {year} {1993})}\BibitemShut {NoStop}%
\bibitem [{\citenamefont {Kumar}\ \emph {et~al.}(2015)\citenamefont {Kumar},
  \citenamefont {Xiao}, \citenamefont {Lunkenheimer}, \citenamefont {Loidl},\
  and\ \citenamefont {Ohl}}]{Naveen2015}%
  \BibitemOpen
  \bibfield  {author} {\bibinfo {author} {\bibfnamefont {C.~M.~N.}\
  \bibnamefont {Kumar}}, \bibinfo {author} {\bibfnamefont {Y.}~\bibnamefont
  {Xiao}}, \bibinfo {author} {\bibfnamefont {P.}~\bibnamefont {Lunkenheimer}},
  \bibinfo {author} {\bibfnamefont {A.}~\bibnamefont {Loidl}}, \ and\ \bibinfo
  {author} {\bibfnamefont {M.}~\bibnamefont {Ohl}},\ }\href@noop {} {\bibfield
  {journal} {\bibinfo  {journal} {Physical Review B}\ }\textbf {\bibinfo
  {volume} {91}},\ \bibinfo {pages} {235149} (\bibinfo {year}
  {2015})}\BibitemShut {NoStop}%
\bibitem [{\citenamefont {Shen}\ \emph {et~al.}(2013)\citenamefont {Shen},
  \citenamefont {Cheng}, \citenamefont {Hong}, \citenamefont {Jiayue},
  \citenamefont {Yuan}, \citenamefont {Cao},\ and\ \citenamefont
  {Wang}}]{Huishen2013}%
  \BibitemOpen
  \bibfield  {author} {\bibinfo {author} {\bibfnamefont {H.}~\bibnamefont
  {Shen}}, \bibinfo {author} {\bibfnamefont {Z.}~\bibnamefont {Cheng}},
  \bibinfo {author} {\bibfnamefont {F.}~\bibnamefont {Hong}}, \bibinfo {author}
  {\bibfnamefont {X.}~\bibnamefont {Jiayue}}, \bibinfo {author} {\bibfnamefont
  {S.}~\bibnamefont {Yuan}}, \bibinfo {author} {\bibfnamefont {S.}~\bibnamefont
  {Cao}}, \ and\ \bibinfo {author} {\bibfnamefont {X.}~\bibnamefont {Wang}},\
  }\href@noop {} {\bibfield  {journal} {\bibinfo  {journal} {Applied Physics
  Letters}\ }\textbf {\bibinfo {volume} {103}},\ \bibinfo {pages} {192404}
  (\bibinfo {year} {2013})}\BibitemShut {NoStop}%
\bibitem [{\citenamefont {Zhang}\ \emph {et~al.}(2019)\citenamefont {Zhang},
  \citenamefont {Xia}, \citenamefont {Ke}, \citenamefont {Zhang}, \citenamefont
  {Cheng}, \citenamefont {Ouyang}, \citenamefont {Wang}, \citenamefont {Huang},
  \citenamefont {Yang}, \citenamefont {Song}, \citenamefont {Xiao},
  \citenamefont {Deng},\ and\ \citenamefont {Jiang}}]{Zhang2019}%
  \BibitemOpen
  \bibfield  {author} {\bibinfo {author} {\bibfnamefont {X.~X.}\ \bibnamefont
  {Zhang}}, \bibinfo {author} {\bibfnamefont {Z.~C.}\ \bibnamefont {Xia}},
  \bibinfo {author} {\bibfnamefont {Y.~J.}\ \bibnamefont {Ke}}, \bibinfo
  {author} {\bibfnamefont {X.~Q.}\ \bibnamefont {Zhang}}, \bibinfo {author}
  {\bibfnamefont {Z.~H.}\ \bibnamefont {Cheng}}, \bibinfo {author}
  {\bibfnamefont {Z.~W.}\ \bibnamefont {Ouyang}}, \bibinfo {author}
  {\bibfnamefont {J.~F.}\ \bibnamefont {Wang}}, \bibinfo {author}
  {\bibfnamefont {S.}~\bibnamefont {Huang}}, \bibinfo {author} {\bibfnamefont
  {F.}~\bibnamefont {Yang}}, \bibinfo {author} {\bibfnamefont {Y.~J.}\
  \bibnamefont {Song}}, \bibinfo {author} {\bibfnamefont {G.~L.}\ \bibnamefont
  {Xiao}}, \bibinfo {author} {\bibfnamefont {H.}~\bibnamefont {Deng}}, \ and\
  \bibinfo {author} {\bibfnamefont {D.~Q.}\ \bibnamefont {Jiang}},\ }\href@noop
  {} {\bibfield  {journal} {\bibinfo  {journal} {Physical review B}\ }\textbf
  {\bibinfo {volume} {100}},\ \bibinfo {pages} {054418} (\bibinfo {year}
  {2019})}\BibitemShut {NoStop}%
\bibitem [{\citenamefont {Epstein}\ and\ \citenamefont
  {Shaked}(1969)}]{Epstein1969}%
  \BibitemOpen
  \bibfield  {author} {\bibinfo {author} {\bibfnamefont {A.}~\bibnamefont
  {Epstein}}\ and\ \bibinfo {author} {\bibfnamefont {H.}~\bibnamefont
  {Shaked}},\ }\href@noop {} {\bibfield  {journal} {\bibinfo  {journal}
  {Physical Letters}\ }\textbf {\bibinfo {volume} {29A}},\ \bibinfo {pages}
  {659} (\bibinfo {year} {1969})}\BibitemShut {NoStop}%
\bibitem [{\citenamefont {Singh}\ \emph {et~al.}(2020)\citenamefont {Singh},
  \citenamefont {Rajput}, \citenamefont {Balasubramanian}, \citenamefont
  {Anas}, \citenamefont {Damay}, \citenamefont {Kumar}, \citenamefont {Eguchi},
  \citenamefont {Jain}, \citenamefont {Yusuf}, \citenamefont {Maitra},\ and\
  \citenamefont {Malik}}]{Ankita2019b}%
  \BibitemOpen
  \bibfield  {author} {\bibinfo {author} {\bibfnamefont {A.}~\bibnamefont
  {Singh}}, \bibinfo {author} {\bibfnamefont {S.}~\bibnamefont {Rajput}},
  \bibinfo {author} {\bibfnamefont {P.}~\bibnamefont {Balasubramanian}},
  \bibinfo {author} {\bibfnamefont {M.}~\bibnamefont {Anas}}, \bibinfo {author}
  {\bibfnamefont {F.}~\bibnamefont {Damay}}, \bibinfo {author} {\bibfnamefont
  {C.~M.~N.}\ \bibnamefont {Kumar}}, \bibinfo {author} {\bibfnamefont
  {G.}~\bibnamefont {Eguchi}}, \bibinfo {author} {\bibfnamefont
  {A.}~\bibnamefont {Jain}}, \bibinfo {author} {\bibfnamefont {S.~M.}\
  \bibnamefont {Yusuf}}, \bibinfo {author} {\bibfnamefont {T.}~\bibnamefont
  {Maitra}}, \ and\ \bibinfo {author} {\bibfnamefont {V.~K.}\ \bibnamefont
  {Malik}},\ }\href@noop {} {\bibfield  {journal} {\bibinfo  {journal} {Phys.
  Rev. B}\ }\textbf {\bibinfo {volume} {102}},\ \bibinfo {pages} {144432}
  (\bibinfo {year} {2020})}\BibitemShut {NoStop}%
\bibitem [{\citenamefont {Bertaut}(1967)}]{Bertaut1968}%
  \BibitemOpen
  \bibfield  {author} {\bibinfo {author} {\bibfnamefont {E.~F.}\ \bibnamefont
  {Bertaut}},\ }\href@noop {} {\bibfield  {journal} {\bibinfo  {journal} {Acta
  Crystallographica}\ }\textbf {\bibinfo {volume} {A24}},\ \bibinfo {pages}
  {217} (\bibinfo {year} {1967})}\BibitemShut {NoStop}%
\bibitem [{\citenamefont {Saito}\ \emph {et~al.}(2001)\citenamefont {Saito},
  \citenamefont {Yamamuraa}, \citenamefont {Mayerb}, \citenamefont {Kobayashi},
  \citenamefont {Miyazakia}, \citenamefont {Ensling}, \citenamefont
  {G{\"u}tlich}, \citenamefont {Le\'{s}niewska},\ and\ \citenamefont
  {Sorai}}]{Saito2001}%
  \BibitemOpen
  \bibfield  {author} {\bibinfo {author} {\bibfnamefont {K.}~\bibnamefont
  {Saito}}, \bibinfo {author} {\bibfnamefont {Y.}~\bibnamefont {Yamamuraa}},
  \bibinfo {author} {\bibfnamefont {J.}~\bibnamefont {Mayerb}}, \bibinfo
  {author} {\bibfnamefont {H.}~\bibnamefont {Kobayashi}}, \bibinfo {author}
  {\bibfnamefont {Y.}~\bibnamefont {Miyazakia}}, \bibinfo {author}
  {\bibfnamefont {J.}~\bibnamefont {Ensling}}, \bibinfo {author} {\bibfnamefont
  {P.}~\bibnamefont {G{\"u}tlich}}, \bibinfo {author} {\bibfnamefont
  {B.}~\bibnamefont {Le\'{s}niewska}}, \ and\ \bibinfo {author} {\bibfnamefont
  {M.}~\bibnamefont {Sorai}},\ }\href@noop {} {\bibfield  {journal} {\bibinfo
  {journal} {Journal of Magnetism and Magnetic Materials}\ }\textbf {\bibinfo
  {volume} {225}},\ \bibinfo {pages} {381} (\bibinfo {year}
  {2001})}\BibitemShut {NoStop}%
\bibitem [{\citenamefont {Faulhaber}\ \emph {et~al.}(1967)\citenamefont
  {Faulhaber}, \citenamefont {H\u"fner}, \citenamefont {Orlich},\ and\
  \citenamefont {Schuchert}}]{FaulhaberEr1967}%
  \BibitemOpen
  \bibfield  {author} {\bibinfo {author} {\bibfnamefont {R.}~\bibnamefont
  {Faulhaber}}, \bibinfo {author} {\bibfnamefont {E.}~\bibnamefont {H\u"fner}},
  \bibinfo {author} {\bibfnamefont {E.}~\bibnamefont {Orlich}}, \ and\ \bibinfo
  {author} {\bibfnamefont {H.}~\bibnamefont {Schuchert}},\ }\href@noop {}
  {\bibfield  {journal} {\bibinfo  {journal} {Zeitschrift fur Physik}\ }\textbf
  {\bibinfo {volume} {204}},\ \bibinfo {pages} {101} (\bibinfo {year}
  {1967})}\BibitemShut {NoStop}%
\bibitem [{\citenamefont {Schuchert}\ and\ \citenamefont
  {H\u"fner}(1969)}]{FaulhaberDy1969}%
  \BibitemOpen
  \bibfield  {author} {\bibinfo {author} {\bibfnamefont {H.}~\bibnamefont
  {Schuchert}}\ and\ \bibinfo {author} {\bibfnamefont {R.}~\bibnamefont
  {H\u"fner}, \bibfnamefont {E.~Faulhaber}},\ }\href@noop {} {\bibfield
  {journal} {\bibinfo  {journal} {Zeitschrift fur Physik}\ }\textbf {\bibinfo
  {volume} {220}},\ \bibinfo {pages} {273} (\bibinfo {year}
  {1969})}\BibitemShut {NoStop}%
\bibitem [{\citenamefont {Hasson}\ \emph {et~al.}(1975)\citenamefont {Hasson},
  \citenamefont {Hornreich}, \citenamefont {Komett}, \citenamefont {Wanklyn},\
  and\ \citenamefont {Yaeger}}]{Hasson1975}%
  \BibitemOpen
  \bibfield  {author} {\bibinfo {author} {\bibfnamefont {A.}~\bibnamefont
  {Hasson}}, \bibinfo {author} {\bibfnamefont {R.~M.}\ \bibnamefont
  {Hornreich}}, \bibinfo {author} {\bibfnamefont {Y.}~\bibnamefont {Komett}},
  \bibinfo {author} {\bibfnamefont {B.~M.}\ \bibnamefont {Wanklyn}}, \ and\
  \bibinfo {author} {\bibfnamefont {I.}~\bibnamefont {Yaeger}},\ }\href@noop {}
  {\bibfield  {journal} {\bibinfo  {journal} {Physical Review B}\ }\textbf
  {\bibinfo {volume} {12}},\ \bibinfo {pages} {5051} (\bibinfo {year}
  {1975})}\BibitemShut {NoStop}%
\bibitem [{\citenamefont {Wood}\ \emph {et~al.}(1969)\citenamefont {Wood},
  \citenamefont {Holmes},\ and\ \citenamefont {Remeika}}]{Wood1969}%
  \BibitemOpen
  \bibfield  {author} {\bibinfo {author} {\bibfnamefont {D.~L.}\ \bibnamefont
  {Wood}}, \bibinfo {author} {\bibfnamefont {L.~M.}\ \bibnamefont {Holmes}}, \
  and\ \bibinfo {author} {\bibfnamefont {J.~P.}\ \bibnamefont {Remeika}},\
  }\href@noop {} {\bibfield  {journal} {\bibinfo  {journal} {Physical review
  B}\ }\textbf {\bibinfo {volume} {185}},\ \bibinfo {pages} {689} (\bibinfo
  {year} {1969})}\BibitemShut {NoStop}%
\bibitem [{\citenamefont {Weingart}\ \emph {et~al.}(2012)\citenamefont
  {Weingart}, \citenamefont {Spaldin},\ and\ \citenamefont
  {Bousquet}}]{Spaldin2012}%
  \BibitemOpen
  \bibfield  {author} {\bibinfo {author} {\bibfnamefont {C.}~\bibnamefont
  {Weingart}}, \bibinfo {author} {\bibfnamefont {N.}~\bibnamefont {Spaldin}}, \
  and\ \bibinfo {author} {\bibfnamefont {E.}~\bibnamefont {Bousquet}},\
  }\href@noop {} {\bibfield  {journal} {\bibinfo  {journal} {Physical Review
  B}\ }\textbf {\bibinfo {volume} {86}},\ \bibinfo {pages} {094413} (\bibinfo
  {year} {2012})}\BibitemShut {NoStop}%
\bibitem [{\citenamefont {Chen}\ \emph {et~al.}(2011)\citenamefont {Chen},
  \citenamefont {Wu},\ and\ \citenamefont {Selloni}}]{Jiachen2011}%
  \BibitemOpen
  \bibfield  {author} {\bibinfo {author} {\bibfnamefont {J.}~\bibnamefont
  {Chen}}, \bibinfo {author} {\bibfnamefont {X.}~\bibnamefont {Wu}}, \ and\
  \bibinfo {author} {\bibfnamefont {A.}~\bibnamefont {Selloni}},\ }\href@noop
  {} {\bibfield  {journal} {\bibinfo  {journal} {Physical Review B}\ }\textbf
  {\bibinfo {volume} {83}},\ \bibinfo {pages} {245204} (\bibinfo {year}
  {2011})}\BibitemShut {NoStop}%
\bibitem [{\citenamefont {Horneich}\ \emph {et~al.}(1975)\citenamefont
  {Horneich}, \citenamefont {Komet}, \citenamefont {Nolan},\ and\ \citenamefont
  {Wanklyn}}]{Horneich1975}%
  \BibitemOpen
  \bibfield  {author} {\bibinfo {author} {\bibfnamefont {R.~M.}\ \bibnamefont
  {Horneich}}, \bibinfo {author} {\bibfnamefont {Y.}~\bibnamefont {Komet}},
  \bibinfo {author} {\bibfnamefont {R.}~\bibnamefont {Nolan}}, \ and\ \bibinfo
  {author} {\bibfnamefont {I.}~\bibnamefont {Wanklyn}, \bibfnamefont {B.~M.
  an~Yaeger}},\ }\href@noop {} {\bibfield  {journal} {\bibinfo  {journal}
  {Physical Review B}\ }\textbf {\bibinfo {volume} {12}},\ \bibinfo {pages}
  {5094} (\bibinfo {year} {1975})}\BibitemShut {NoStop}%
\bibitem [{\citenamefont {Schuchert}\ and\ \citenamefont
  {H{\"u}fner}(1969)}]{SchuchertDy1969}%
  \BibitemOpen
  \bibfield  {author} {\bibinfo {author} {\bibfnamefont {H.}~\bibnamefont
  {Schuchert}}\ and\ \bibinfo {author} {\bibfnamefont {R.}~\bibnamefont
  {H{\"u}fner}, \bibfnamefont {E.~Faulhaber}},\ }\href@noop {} {\bibfield
  {journal} {\bibinfo  {journal} {Zeitschrift fur Physik}\ }\textbf {\bibinfo
  {volume} {222}},\ \bibinfo {pages} {105} (\bibinfo {year}
  {1969})}\BibitemShut {NoStop}%
\end{thebibliography}%
\end{document}